\documentclass[twocolumn,showpacs,preprintnumbers,amsmath,pra,amssymb,superscriptaddress]{revtex4-1}

\usepackage{amsfonts}
\usepackage{amsmath}
\usepackage[makeroom]{cancel}
\usepackage[english]{babel}
\usepackage[T1]{fontenc}
\usepackage{times}
\usepackage{mathrsfs}
\usepackage{graphicx}
\usepackage{dcolumn}
\usepackage{bm}
\usepackage{wasysym}
\usepackage[colorlinks,bookmarks=true,citecolor=blue,linkcolor=red,urlcolor=blue]{hyperref}
\usepackage[tight, FIGTOPCAP, hang, raggedright, nooneline]{subfigure}
\usepackage{hyperref}
\usepackage{xcolor}
\usepackage{epstopdf}
\usepackage{epsfig}
\usepackage{amssymb}
\usepackage{lipsum}

\newcommand{\subfigimg}[3][,]{%
  \setbox1=\hbox{\includegraphics[#1]{#3}}
  \leavevmode\rlap{\usebox1}
  \rlap{\hspace*{0pt}\raisebox{\dimexpr\ht1-2\baselineskip}{#2}}
  \phantom{\usebox1}
  }
  
 \definecolor{Green}{RGB}{80,182,0}

\usepackage{color}

\newcommand{\la}{\langle}
\newcommand{\ra}{\rangle}

\newcommand{\q}{\mathbf{q}}

\begin{document}
\title{Majorana spectroscopy of 3D  Kitaev spin-liquids}
\author{A.~Smith}
\affiliation{T.C.M. Group, Cavendish Laboratory, J. J. Thomson Avenue, Cambridge CB3 OHE, United Kingdom}
\author{J.~Knolle}
\affiliation{T.C.M. Group, Cavendish Laboratory, J. J. Thomson Avenue, Cambridge CB3 OHE, United Kingdom}
\author{D.~L.~Kovrizhin}
\affiliation{T.C.M. Group, Cavendish Laboratory, J. J. Thomson Avenue, Cambridge CB3 OHE, United Kingdom}
\affiliation{National Research Centre Kurchatov Institute, 1 Kurchatov Square, Moscow 123182, Russia}
\author{J.~T.~Chalker}
\affiliation{Theoretical Physics, Oxford University, 1, Keble Road, Oxford OX1 3NP, United Kingdom}
\author{R.~Moessner}
\affiliation{Max Planck Institute for the Physics of Complex Systems, D-01187 Dresden, Germany}
\date{\today}

\begin{abstract}
We analyse the dynamical response of a range of 3D Kitaev quantum spin-liquids,
using lattice models chosen to explore the different possible low-energy spectra for gapless Majorana
fermions, with either Fermi surfaces, nodal lines or Weyl points. We find that the 
behaviour of the dynamical structure factor is distinct in all three cases, reflecting
the quasiparticle density of states
in two fundamentally 
different ways. First, the low-energy response is either straightforwardly related to the power with which 
the low-energy density of states vanishes; or for a non-vanishing density of states, 
to the phase shifts encountered in the corresponding X-ray edge problem, whose phenomenology we
extend to the case of Majorana fermions. Second, at higher energies,  
there is a rich fine-structure,
determined by microscopic features of the Majorana spectrum. 
Our theoretical results test the usefulness of inelastic neutron scattering as a probe of
these quantum spin liquids: we find that although spin flips fractionalise, 
the main features of the dynamical spin response
nevertheless admit straightforward interpretations in terms of Majorana and flux loop excitations.
\end{abstract}

\maketitle


\section{Introduction}
Space dimensionality is known to radically change the character  of a physical system, as was already evident from the works of 
Ising and Onsager on the classical Ising model.  In addition, dimensionality greatly influences the tractability of a problem -- 
in non-trivial quantum systems, exact solutions exist primarily in one dimension. With the recognition
that exotic magnetic quantum phases are available -- and can be fundamentally distinct -- away from 
 one dimension, instances of tractable models in higher dimension are most valuable. 

The Kitaev spin model~\cite{Kitaev} is uniquely useful in this respect, combining the following three properties. First, its 
phenomenology is very rich -- 
it provides an example of a quantum spin liquid (QSL) hosting fractionalized quasiparticles: Majorana fermions and flux excitations. Second, the model allows for an exact solution. This is true not only as originally formulated for a two-dimensional quantum spin system, but also in three dimensions -- it can be naturally extended from the honeycomb lattice to other tricoordinated lattices~\cite{MandalKitaev}, allowing variation not only of dimensionality
but also of the nature of the low-energy spectrum of Majorana fermions~\cite{Kimchi,HermannsZoo}. 
Third, the model is simple enough to be approximately realizable in physical systems; in fact, a search for materials with dominant Kitaev-like interactions have recently become the subject of intensive experimental work~\cite{Jackeli,Singh2010,Singh2012,Plumb2014,Sears2015,Majumdar2015,Sandilands2015,Banerjee2015,Modic,Takayama,Kim}. In particular, 
the synthesis of 3D materials $\beta$- and $\gamma$-Li$_2$IrO$_3$~\cite{Modic,Takayama,Kim} has stimulated interest in theoretical studies of the whole class of systems~\cite{HermannsZoo}, which includes the harmonic honeycomb series~\cite{Kimchi}. However, so far all materials eventually do form a long range magnetically ordered state at low temperatures and in a strict sense do not realise Kitaev QSL ground states. Nevertheless, there is growing evidence that the high energy (temperature) features (above the scale of non-Kitaev interactions inducing the residual magnetism) of spectroscopic experiments can still be interpreted in terms of the fractionalised quasiparticles of the unperturbed Kitaev models~\cite{Plumb2014,Banerjee2015,Yamaji2016,Nasu2016}. This is one of the motivations to look for distinct signatures in the dynamical properties of different Kitaev QSL phases.

Ever since P.W.~Anderson's original proposal of the RVB QSL~\cite{AndersonRVB}, a central obstacle to probing experimentally 
the physics of QSLs -- and topologically-ordered states more generally~\cite{WenTopological}  -- 
has been the featureless nature of their ground states. 
As a possible remedy, it has been recognised for a long time that the fingerprints of liquidity and fractionalisation are more accessible in the excited state spectrum, even though the coupling of experimental probes to fractionalised quasiparticles may itself be rather non-trivial. 

In this paper we build upon methods developed in our previous work on the 2D honeycomb Kitaev model~\cite{PRL,Knolle2015} to study the dynamical response of Kitaev QSLs in 3D~\cite{HermannsZoo} in order to 
investigate the effect of varying spatial dimensionality and low-energy spectrum. We study the dynamical structure factor for the full range of varieties of gapless Kitaev 3D QSLs defined on the hyperoctagon~\cite{HermannsQSL}, hyperhoneycomb~\cite{MandalKitaev,Lee,Takayama}, and hyperhexagon ((8,3)b from Ref.~\cite{HermannsZoo}) lattices. These models have excitations with, respectively, Majorana Fermi surfaces,  nodal lines, and Weyl points, and provide a characteristic set of 3D gapless Kitaev QSLs. Results for the hyperhoneycomb lattice were presented in a recent Rapid Communications~\cite{Smith} and are reproduced here for direct comparison.


\begin{figure*}[htb]
\centering
\subfigimg[width=0.31\textwidth]{\large  \hspace*{-10pt} (a)}{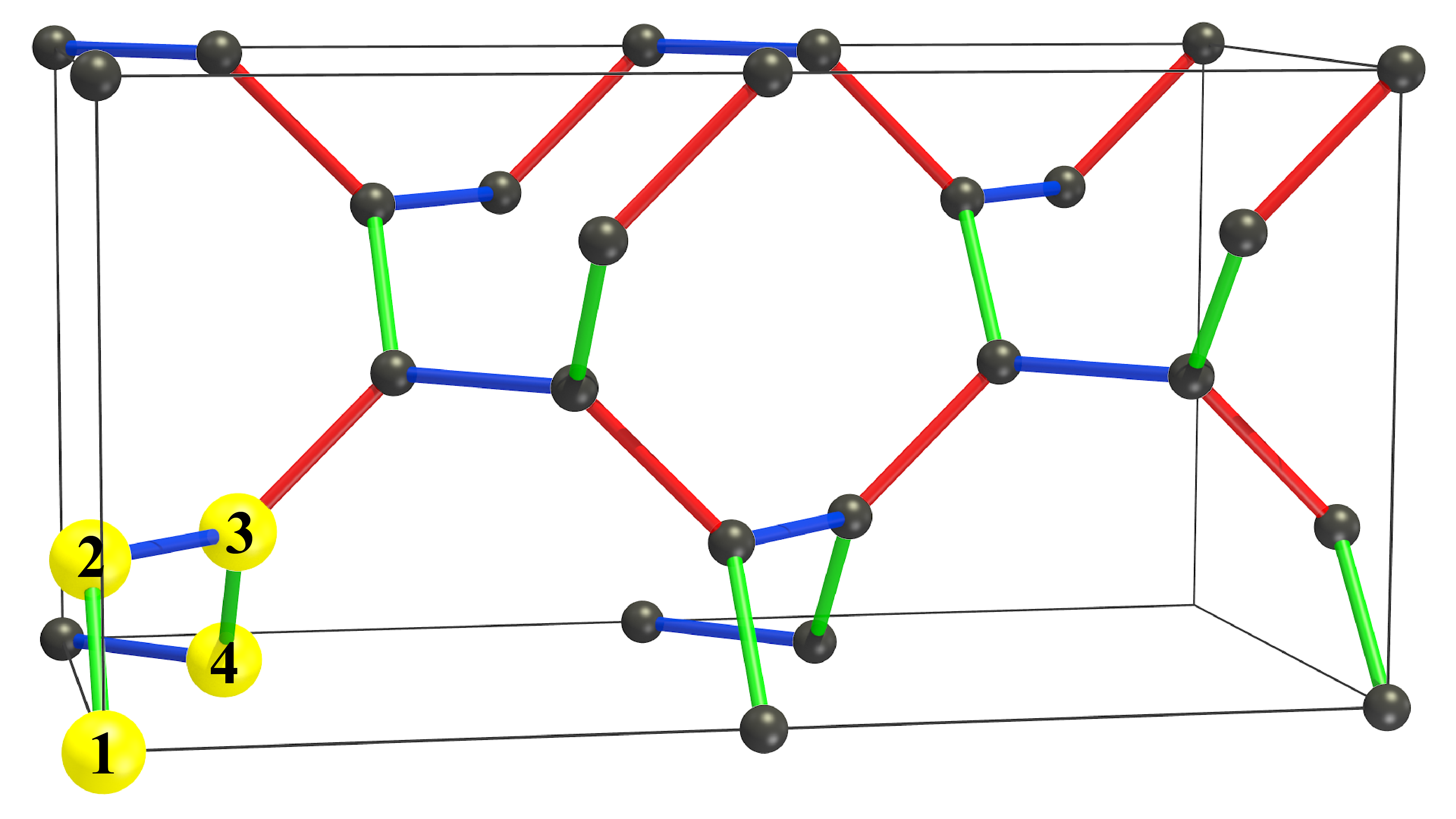}
\quad\subfigimg[width=0.31\textwidth]{\large \hspace*{-10pt} (b)}{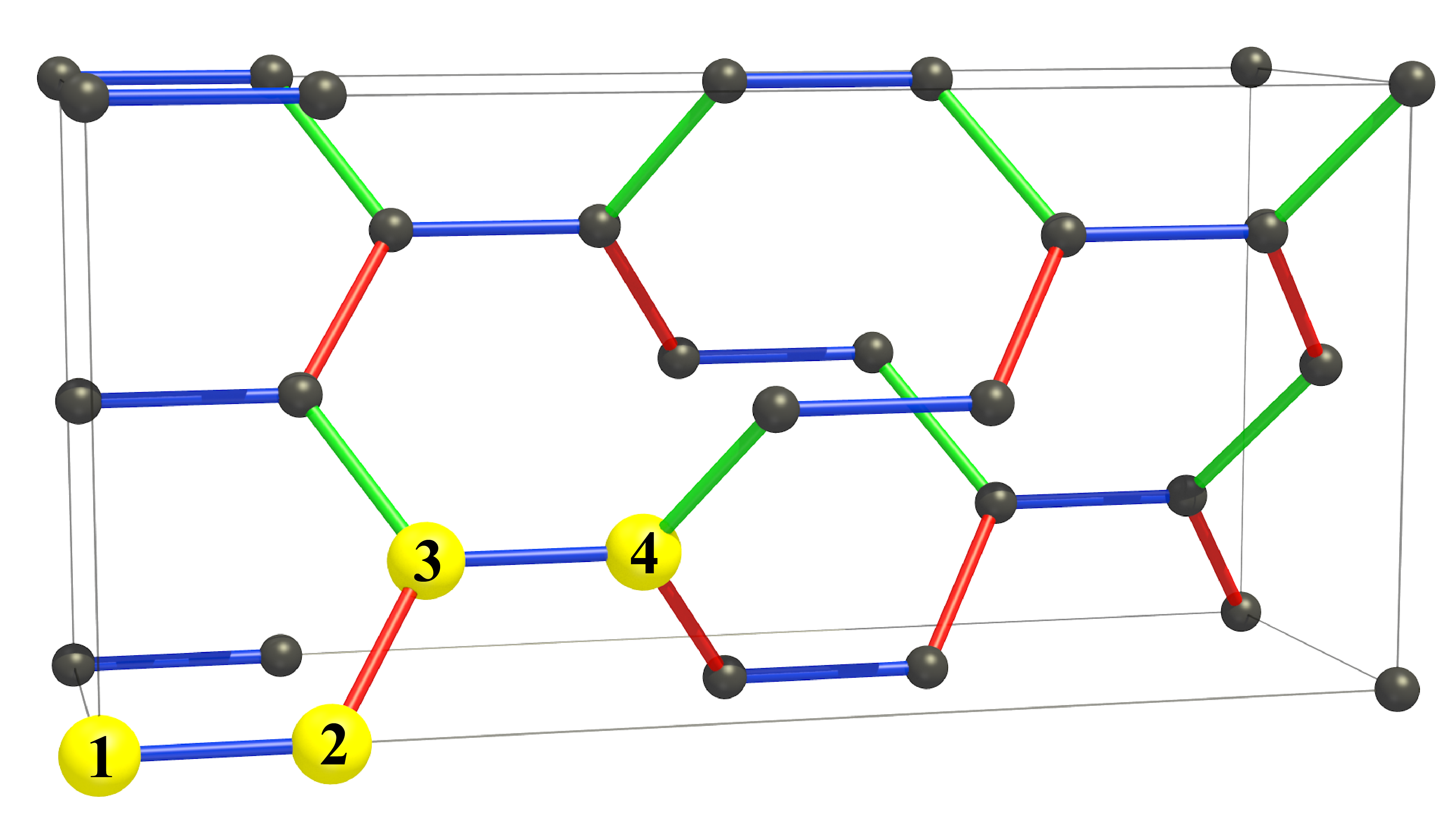}
\subfigimg[width=0.31\textwidth]{\large (c)}{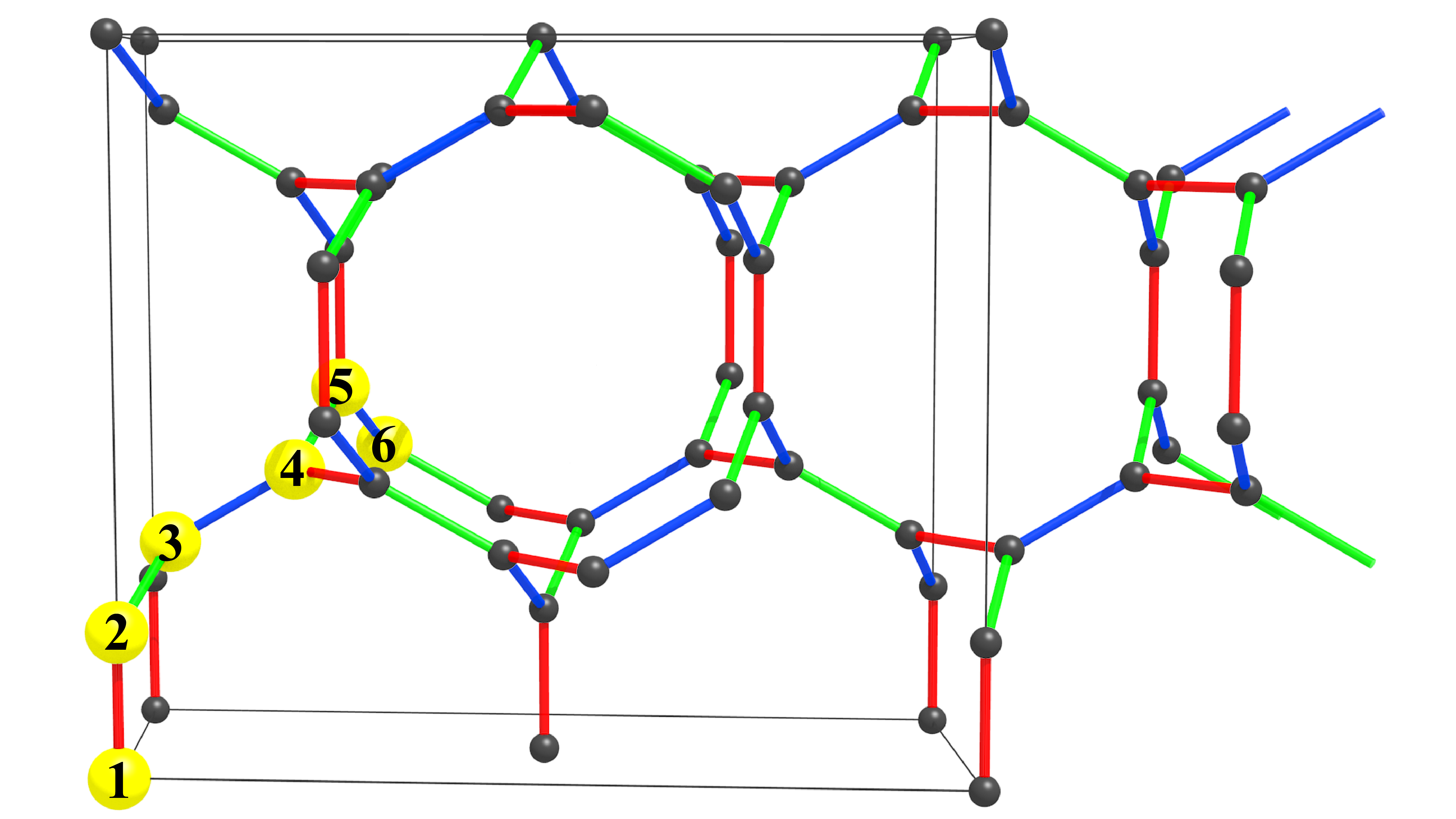}
\caption{Unit cells for (a) the hyperoctagon, (b) hyperhoneycomb, and (c) hyperhexagon lattices. The numbered yellow sites form primitive unit cells. The x,y, and z-bonds are shown in red, green, and blue respectively.}
\label{fig: lattices}
\end{figure*}

\begin{figure}[b]
\centering
\includegraphics[width=0.23\textwidth]{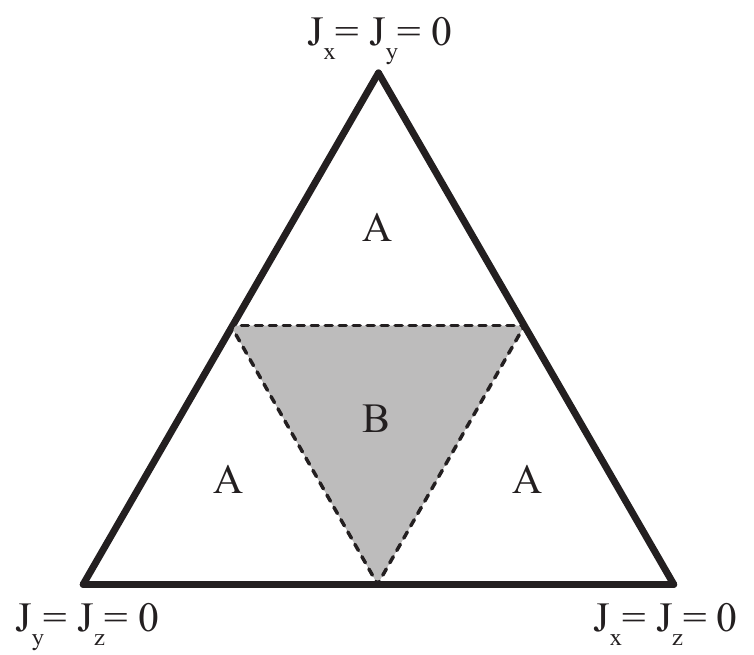}
\includegraphics[width=0.23\textwidth]{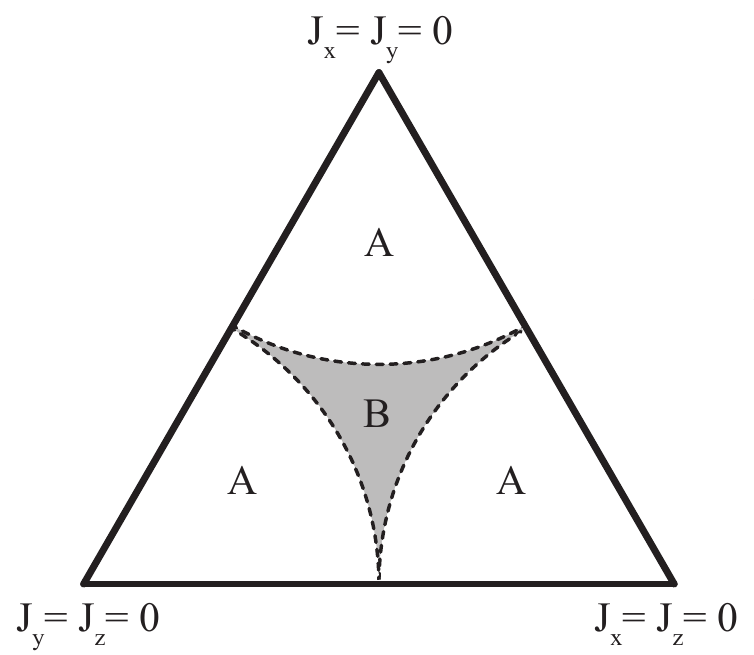}
\caption{Phase diagram of the Kitaev model through a cut in the parameter space defined by the relation $J_x + J_y + J_z =1$ for the hyperhoneycomb and hyperoctagon lattice (left), and hyperhexagon lattice (right). The gapped regions are indicated by ``A''. The central shaded ``B'' regions correspond to gapless phases.} 
\label{fig: phase diagram}
\end{figure}
 
We show that the dynamical structure factor (DSF) provides two types of complementary information about these spin liquids. At low energies there is no response, even in QSL regimes with gapless Majorana fermions, because a spin flip necessarily introduces gapped flux excitations~\cite{PRL}.
At energies $\omega$ just above the flux gap $\Delta$, the dynamical spin response falls into different categories, depending on the nature of the low energy Majorana density of states (DOS). If the latter vanishes at small energy $\varepsilon$ as $\varepsilon^x$ with $x>0$ (as for Weyl and Dirac points, and for nodal lines), the low energy response follows the same power-law, with $S(\omega)\propto (\omega-\Delta)^x$. Alternatively, if the DOS is constant at low energies (as in the presence of a Majorana Fermi-surface), then $S(\omega)\propto(\omega-\Delta)^{-\alpha}$, where $\alpha$ is an X-ray edge exponent. This universality has its origin in the presence of dynamic Majorana fermions and static point-like gauge fluxes -- into which spin flips fractionalize, which make the measurement of the dynamical structure factor a perfect tool to probe the \textit{local} DOS of Majorana fermions. 
In addition, away from the low-energy limit, the DSF reflects in considerable detail the band structure of the fermionic excitations. This allows the identification of Majorana fermion physics well away from the `universal' low-energy behaviour, which itself may be be fragile in the presence of additional terms that lead to collective instabilities~\cite{Hermanns2015}, higher-dimensional ordering, or the destruction of integrability (and solubility). 

Taking these two items together, the DSF as probed through inelastic neutron scattering (INS) experiments provides a handle on Majorana excitations in Kitaev quantum spin-liquids that is more direct than might have been expected in view of their fractionalized character.

The paper is organized as follows. In Section~\ref{sec: Kitaev Model} we outline the exact solution of the Kitaev model using Majorana fermion representation, and highlight some key aspects of our approach. Defining three extensions of the Kitaev model to 3D lattices we use the conserved loop operators and a Majorana representation of spins, as in 2D, to recast the Hamiltonian in terms of a Majorana tight-binding model coupled to a static $\mathbb{Z}_2$ gauge field. In Section~\ref{sec: spin correlators} we define the dynamical structure factor
and present the results of calculations for the three lattices. We also discuss the ways in which features of the dynamical response are characteristic of the Majorana fermion spectrum, both asymptotically at low-energy, and at higher energies and across the Brillouin zone. We close with an summary in Section~\ref{sec: Summary}. Details of the calculations are deferred to the Appendices.

\section{Kitaev Model}\label{sec: Kitaev Model}

The Kitaev model, which can be defined on any tri-coordinated lattice, describes spin-1/2 degrees of freedom interacting via bond-dependent, nearest-neighbour Ising exchange $J_a$. Below we concentrate on the dynamics of the Kitaev model for the cases of the hyperoctagon, hyperhoneycomb and hyperhexagon lattices~\cite{HermannsZoo}. We label three types of lattice bond $a=x,y,z$ referring to the components of spins involved in the Ising interaction (see Fig.~\ref{fig: lattices}). Using the notation $\la j k \ra_a$ for a pair of sites $j,k$ connected by bond $a$, the Hamiltonian is 
\begin{equation}\label{eq: H Kitaev}
\hat{H} = -\frac{J_x}{2} \sum_{\la j k \ra_x} \hat{\sigma}^x_j \hat{\sigma}^x_k - \frac{J_y}{2} \sum_{\la j k \ra_y} \hat{\sigma}^y_j \hat{\sigma}^y_k - \frac{J_z}{2} \sum_{\la j k \ra_z} \hat{\sigma}^z_j \hat{\sigma}^z_k,
\end{equation}
with Pauli matrices $\hat{\sigma}_j^a$. Ground states of Kitaev model are gapped and gapless quantum spin liquids: see the phase diagrams in Fig.~\ref{fig: phase diagram}. The phase diagrams are identical for the 2D honeycomb~\cite{Kitaev}, hyperhoneycomb~\cite{HermannsQSL,MandalKitaev}, and hyperoctagon lattices~\cite{HermannsQSL} (and also for the harmonic honeycomb series~\cite{Kimchi}), but different in the hyperhexagon case (Fig.~\ref{fig: phase diagram}, right).

\subsection{Loop Operators and Flux Sectors}

\begin{figure*}[!t]
\subfigimg[width=0.32\textwidth]{\large  \hspace*{-0pt} (a)}{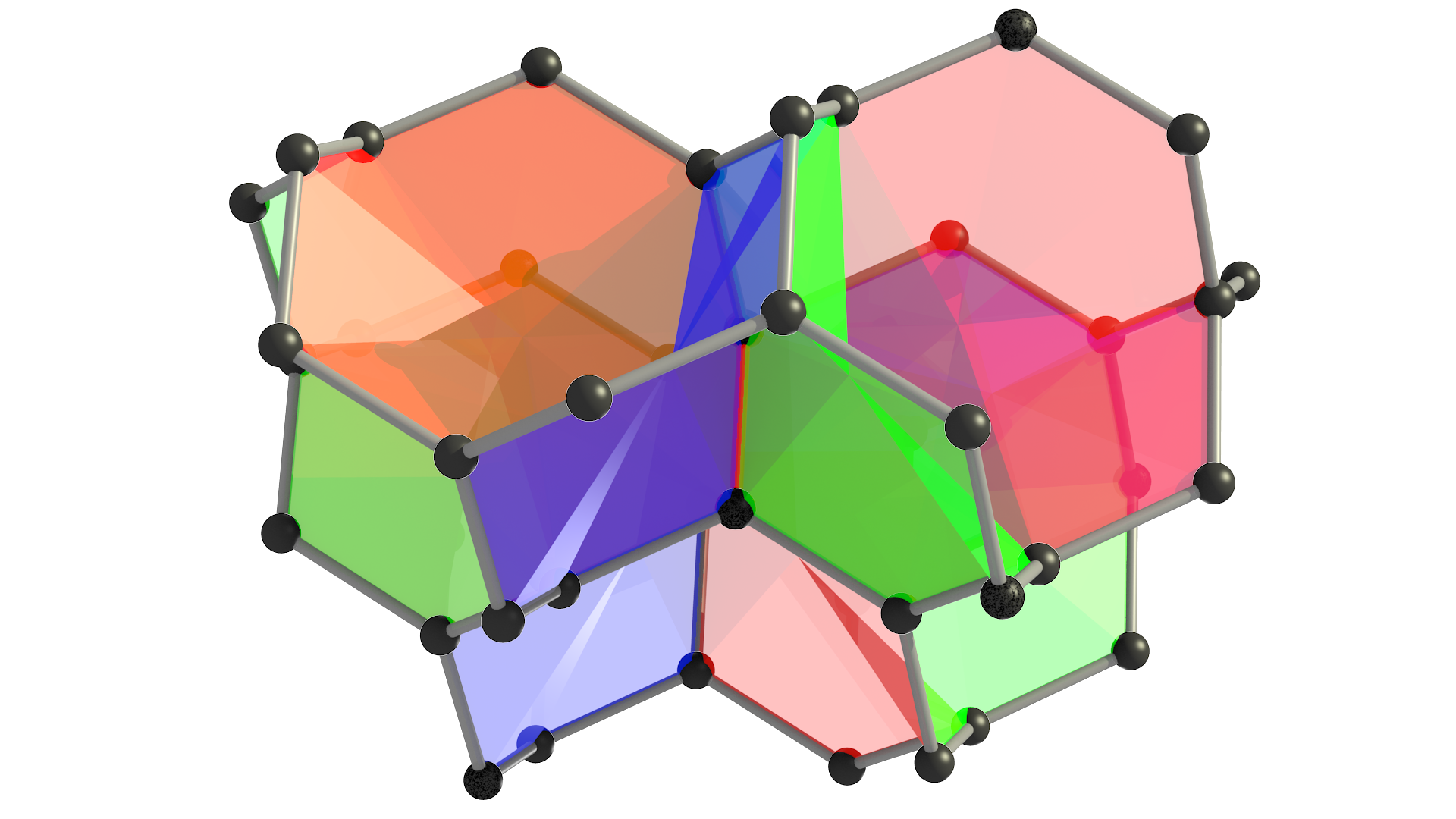}
\subfigimg[width=0.32\textwidth]{\large \hspace*{-10pt} (b)}{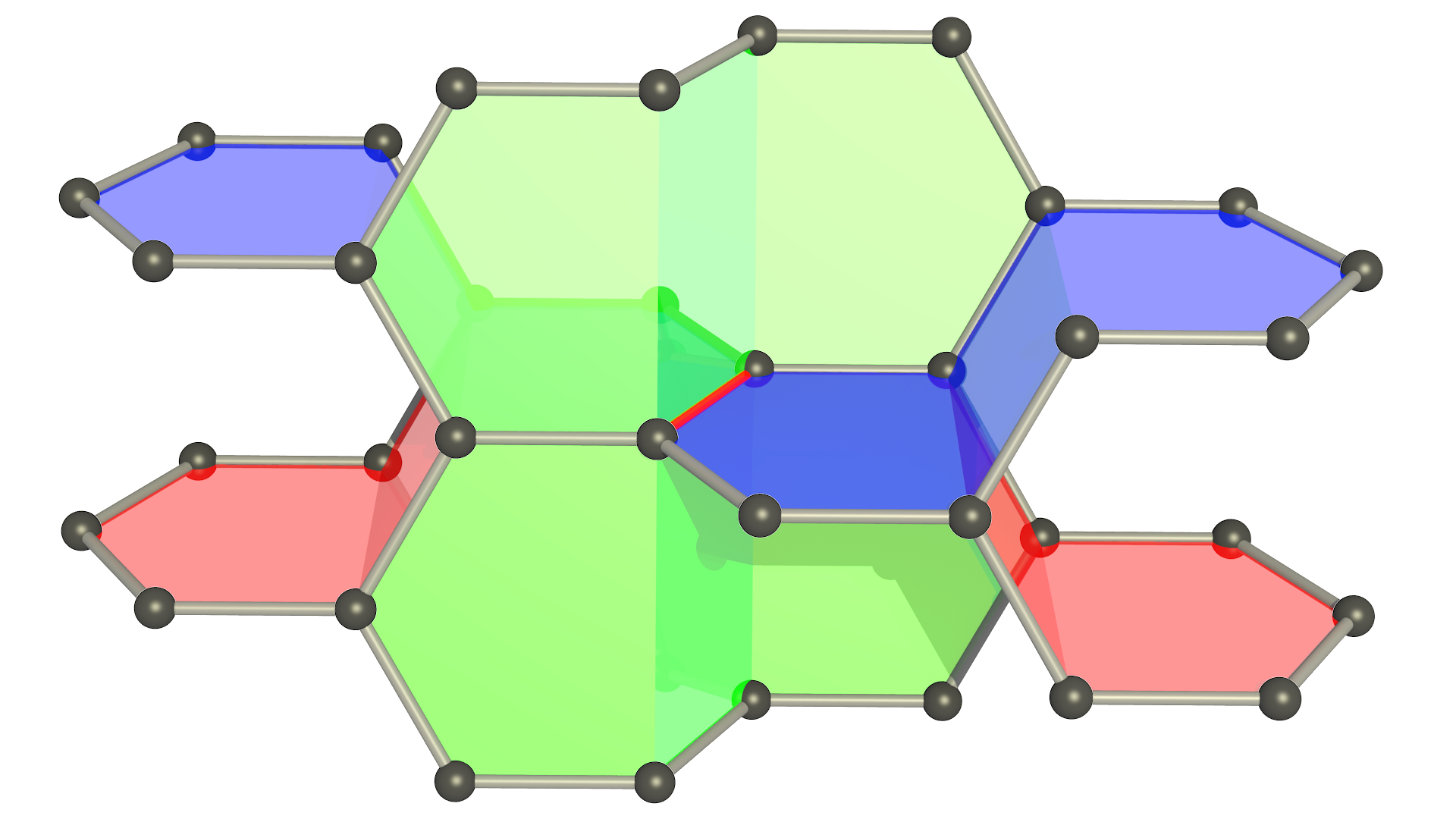}
\subfigimg[width=0.32\textwidth]{\large \hspace*{-0pt} (c)}{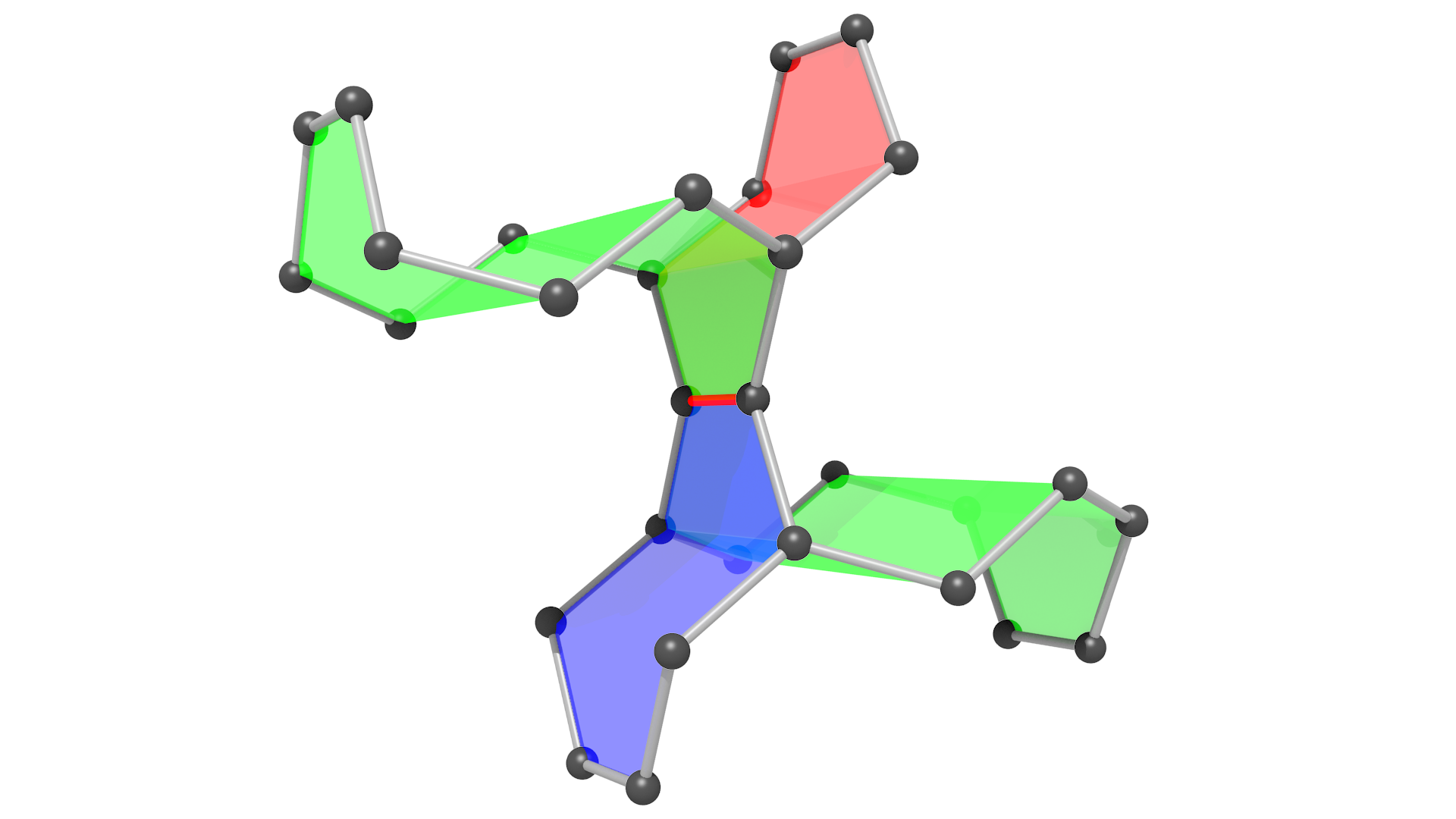}
\caption{Examples of flux loops created by flipping a single bond presented for the three lattice models discussed in the text. The flipped bond is shown in red and the flux loops are shown as coloured surfaces.}
\label{fig: loops}
\end{figure*}

One of the key insights of Kitaev~\cite{Kitaev} was that there exists an extensive number of conserved quantities defined on each plaquette of the 2D honeycomb lattice, which play a crucial role in the exact solution of the model. In three dimensions one can identify similar operators $\hat{\mathcal{W}}_l$ which define fluxes through the loops living on the bonds of the lattice. 
These loop operators can be written as
\begin{equation}\label{eq: loop definition}
\hat{\mathcal{W}}_\gamma = \prod_{l\in\gamma} K_l, \qquad\qquad K_l = \hat{\sigma}_j^\alpha \hat{\sigma}_k^\alpha,
\end{equation}
where $l$ denotes the $\alpha$-bond connecting $j$ and $k$ sites.

Loops in two and three dimensions differ in a fundamental way. Increased dimensionality allows for a possibility of e.g.~knotted loops. However, for all the 3D lattices which we study here, the irreducible loops arising in the calculations are simple, and we can treat them in a way similar to the 2D case. Note that different loops are not all independent, since the product of flux operators for a set of loops enclosing a volume is equal to identity, see Refs.~\cite{HermannsQSL,MandalKitaev,HermannsZoo}. 

The operators $\hat{\mathcal{W}}_l$ have eigenvalues $\pm 1$, and we identify $\hat{\mathcal{W}}_l$ with a $\mathbb{Z}_2$-flux through the loop $l$.  We say that a loop with eigenvalue $+1$ is flux-free, and otherwise has a $\pi$-flux. Since  $[\hat{\mathcal{W}}_l,\hat{H}] = 0$, and $[\hat{\mathcal{W}}_{l},\hat{\mathcal{W}}_{l'} ] =0$, the Hilbert space $\mathcal{H}$ of the Hamiltonian $\hat{H}$ can be separated into flux sectors $\mathcal{H}_{\{\mathcal{W}_l\}}$ classified by the eigenvalues of $\{\mathcal{W}_l\}$. The full Hilbert space in the Majorana representation is a direct product of `flux' $|F\ra$ and `matter' $|M \ra$ sectors, and we denote the ground state by $| 0 \ra = | F_0 \ra \otimes | M_0 \ra$. 

In the 2D honeycomb case the ground state flux configuration can be identified using a theorem due to Lieb~\cite{Lieb}. This theorem is not generally applicable in 3D, but the authors of Ref.~\cite{HermannsZoo} have determined a set of lattices for which it applies and used numerics to find the configuration for the others. Using their numerical results we fix the ground state flux configurations in the hyperhoneycomb and hyperoctagon cases such that all irreducible loops are flux-free, and Lieb's theorem gives $\pi$-flux in the hyperhexagon case.

\subsection{Majorana Representation}\label{sec: Majorana}

The approach originally taken by Kitaev~\cite{Kitaev}, and the first step in our calculation, is to represent spins using four Majorana fermions  $\hat{b}_j^x, \hat{b}_j^y, \hat{b}_j^z, \hat{c}_j$ at each lattice site. These have the commutation relations
$\{\hat{b}^\alpha_j,\hat{b}^\beta_k\} = 2\delta_{\alpha\beta}\delta_{j k}$ and $\{\hat{c}_j,\hat{c}_k\} = 2\delta_{jk}$. 
Spin operators can be written in terms of Majorana fermions as $\hat{\sigma}^\alpha_j = i\hat{b}^\alpha_j \hat{c}_j$. The Hilbert space of the Majorana fermions is larger than that of the  spins, and the physical Hilbert space is defined via constraints that the eigenvalues of the operators $\hat{D}_j = \hat{b}^x_j \hat{b}^y_j \hat{b}^z_j \hat{c}_j$ are equal to $+1$.

Using the Majorana fermion representation of spins one can recast a general Kitaev Hamiltonian which we study here in the form
\begin{equation}\label{eq: H Majorana}
\hat{H} = \frac{i}{2} \sum_{\la jk \ra}  J_{\alpha_{jk}} \hat{u}_{jk} \;\hat{c}_{j} \hat{c}_{k},
\end{equation}
where $\hat{u}_{jk} \equiv i\hat{b}^{\alpha_{jk}}_j \hat{b}^{\alpha_{jk}}_k$. The notation $\la j k \ra$ indicates the sum is over nearest neighbour sites $j$ and $k$. The bond operators $\hat{u}_{jk} = i \hat{b}^{a_{jk}}_j \hat{b}^{a_{jk}}_k$  commute with the Hamiltonian and amongst themselves: $[\hat{u}_{jk},\hat{H}]=0  =[\hat{u}_{ij},\hat{u}_{kl}]$. Their eigenvalues are given by $\pm 1$, and can be associated with the `direction' of a bond (indeed, $\hat{u}_{jk} = - \hat{u}_{kj}$).

\begin{figure*}[tb]
\centering
\subfigimg[width=0.33\textwidth]{\large (a)}{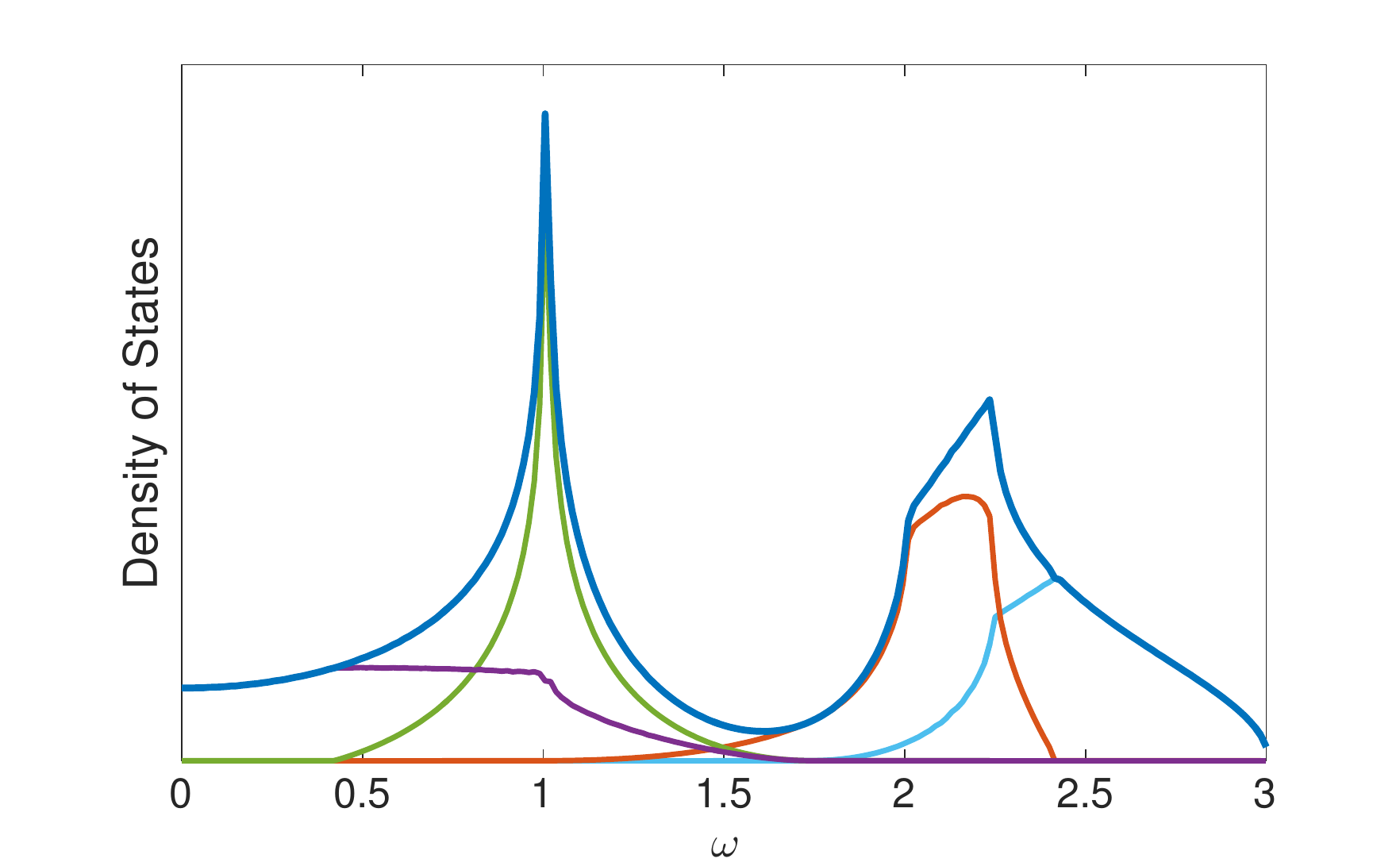}
\!\!\subfigimg[width=0.33\textwidth]{\large (b)}{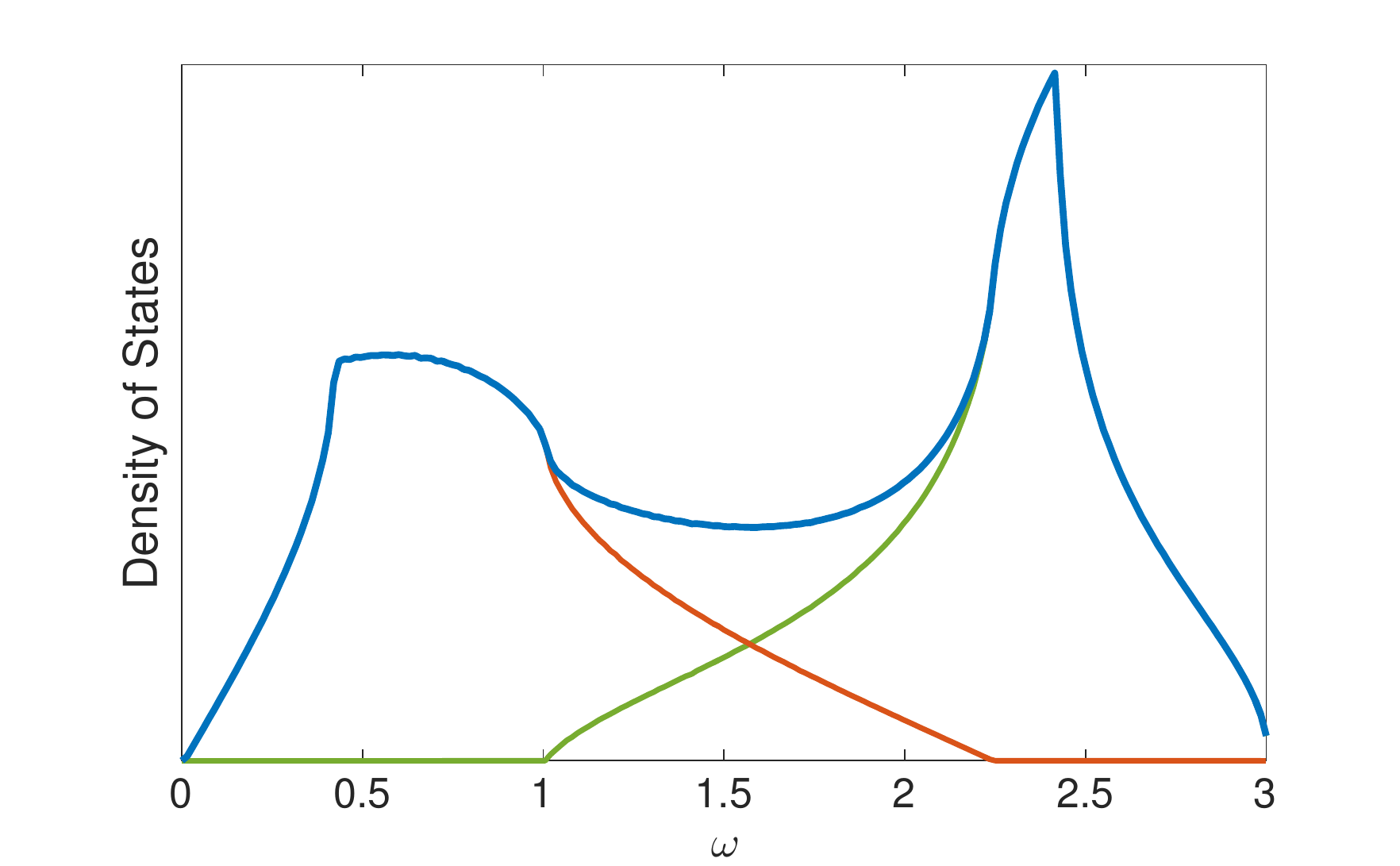}
\!\!\subfigimg[width=0.33\textwidth]{\large (c)}{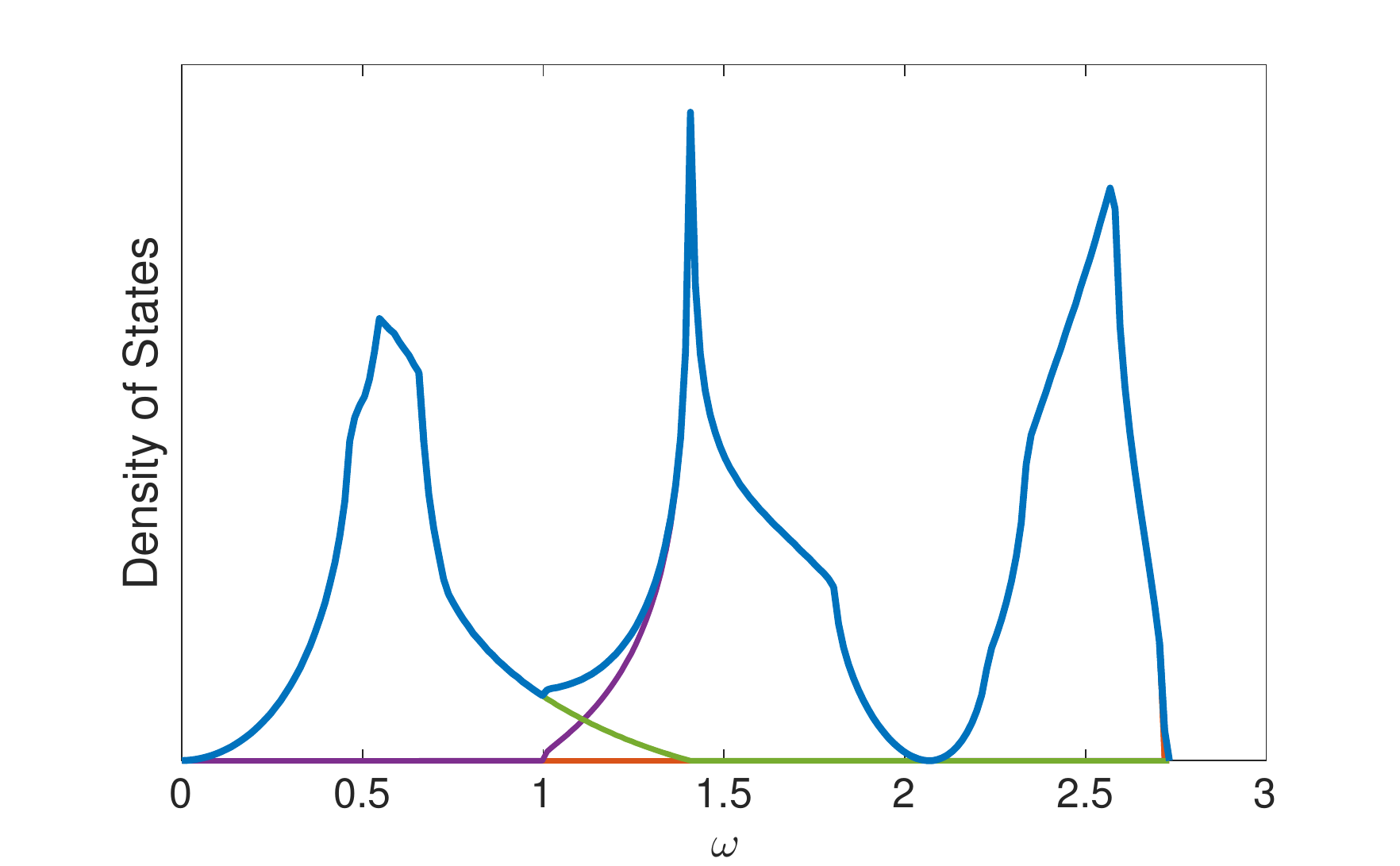}
\!\!\subfigimg[width=0.33\textwidth]{\large (a)}{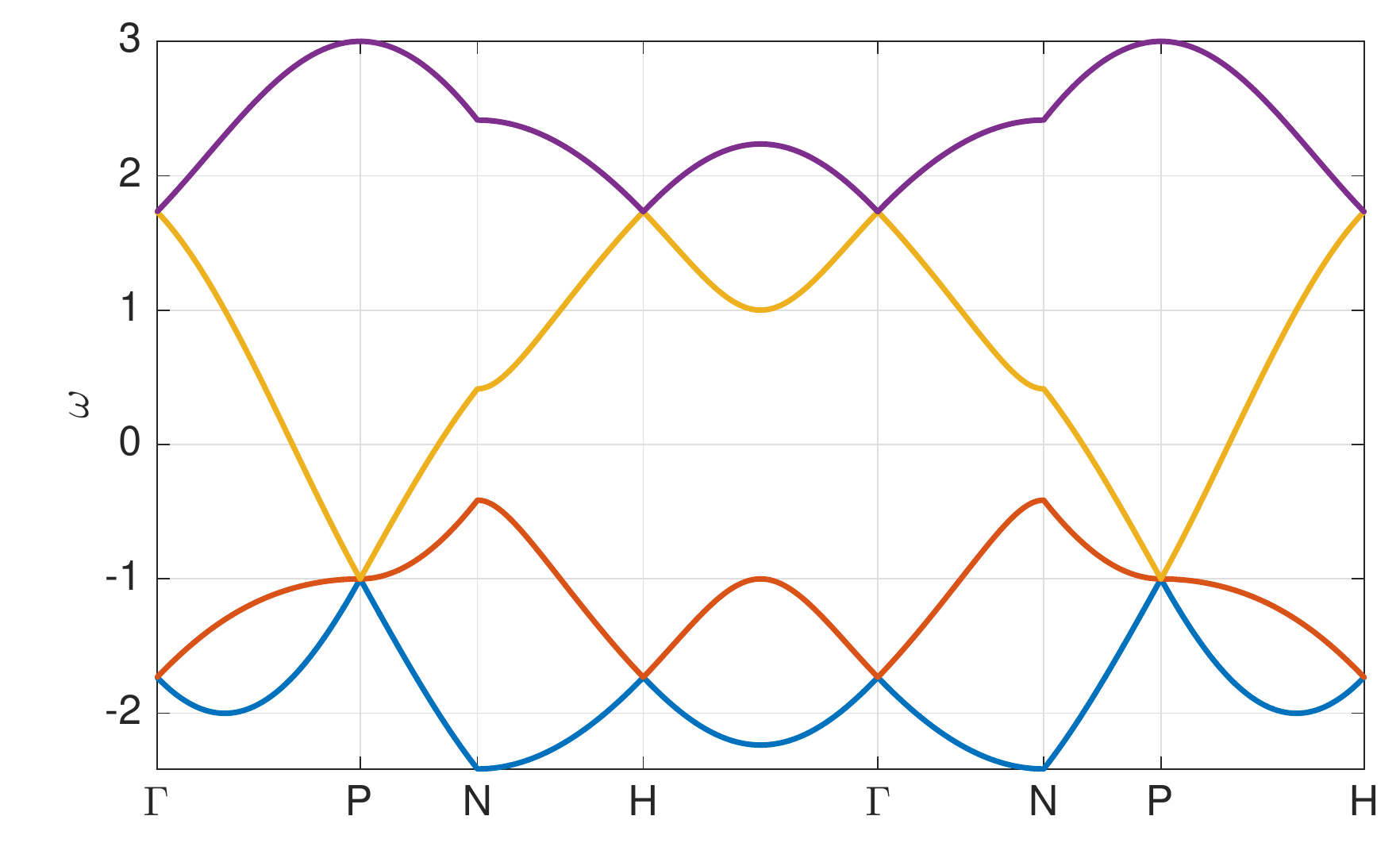}
\subfigimg[width=0.33\textwidth]{\large \hspace*{-3pt}(b)}{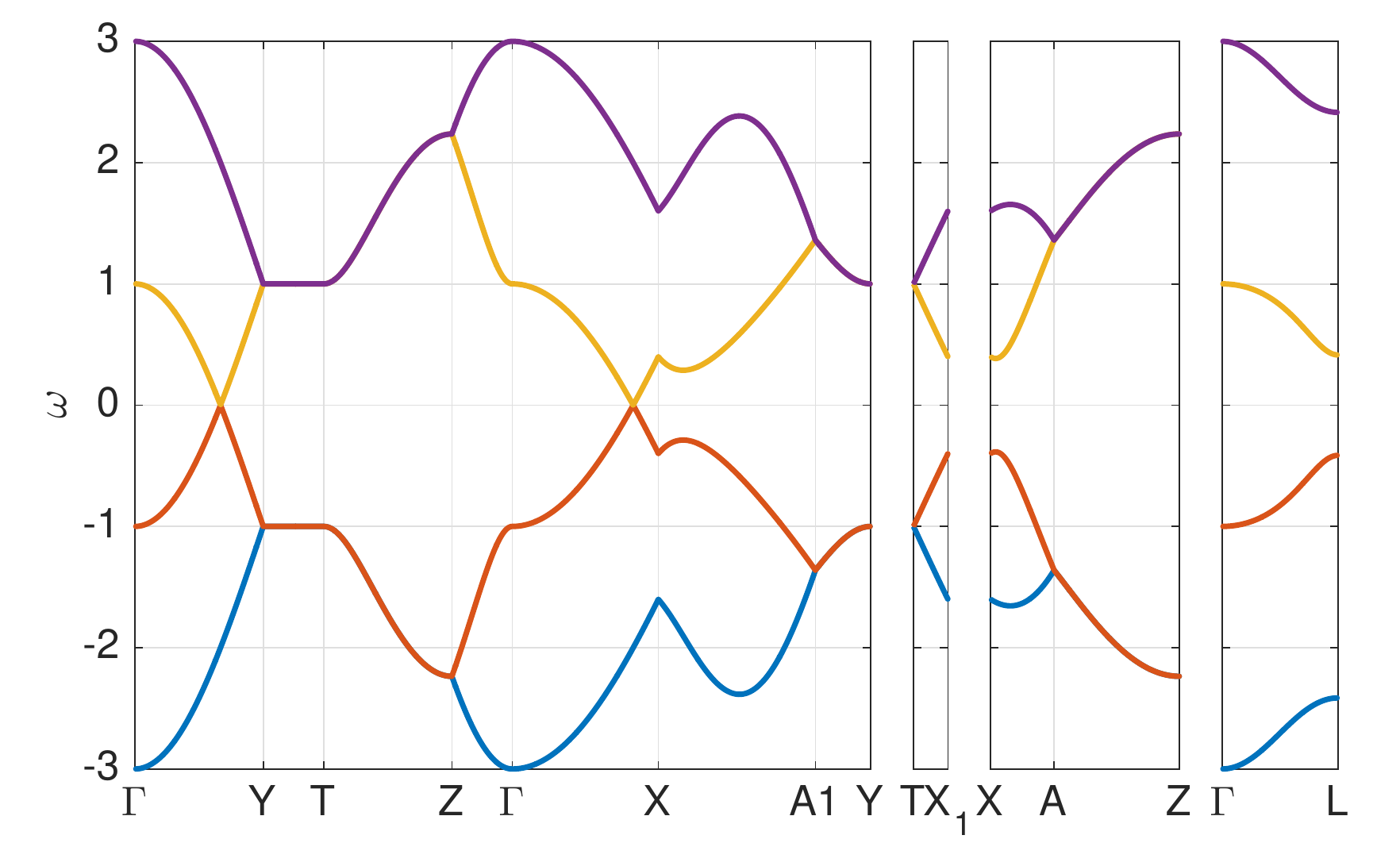}
\!\!\subfigimg[width=0.33\textwidth]{\large (c)}{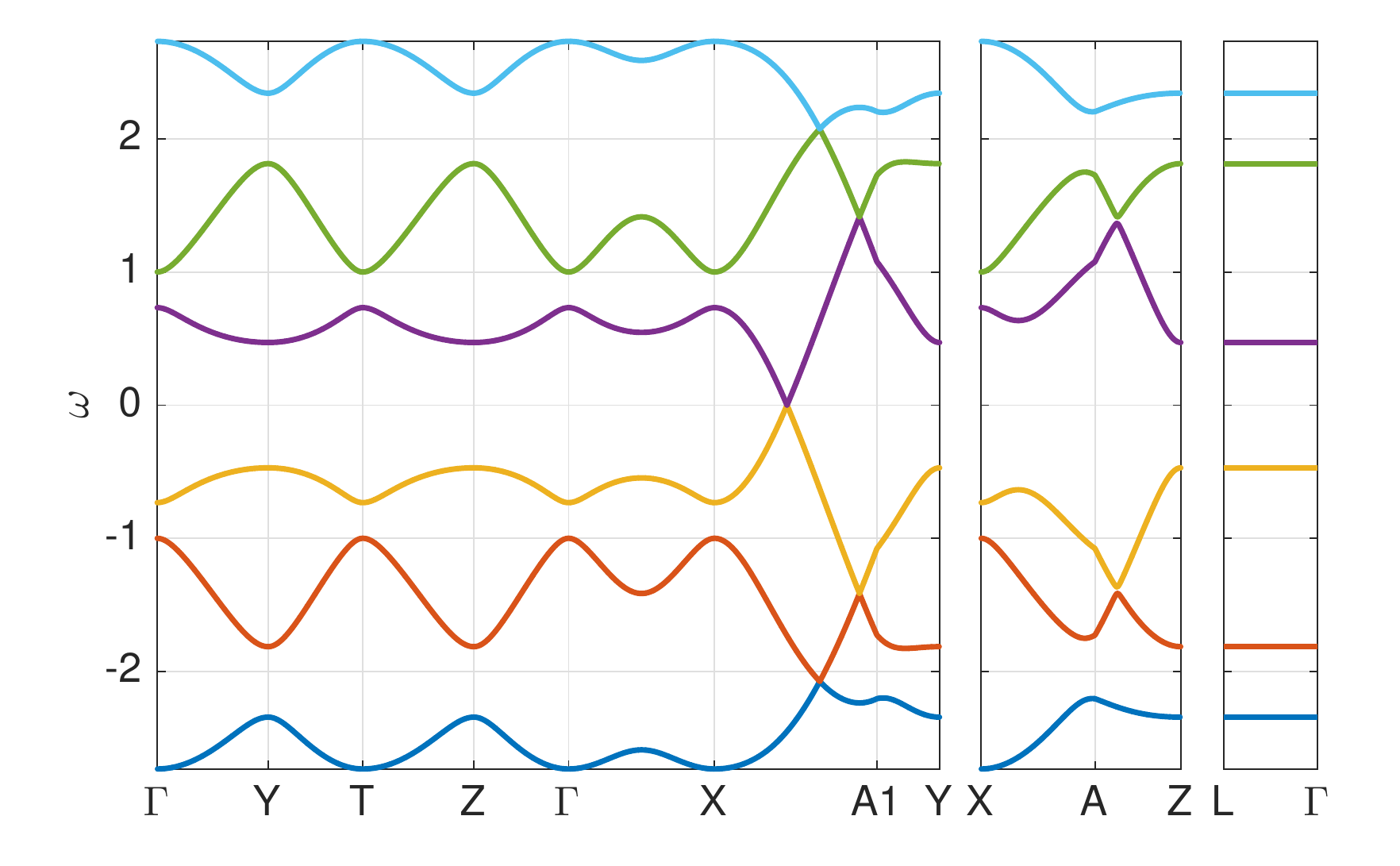}
\caption{Band resolved density of states (top row) and the Majorana dispersion relations (bottom row) for the hyperoctagon (a), hyperhoneycomb (b) and hyperhexagon (c). The dispersion relations are plotted along high symmetry directions in the Brillouin zone.}
\label{fig: spectra}
\end{figure*}

Using the definitions given above one can write loop operators in terms of bond fermions 
\begin{equation}\label{eq: u loop}
\hat{\mathcal{W}}_\gamma = \prod_{l\in\gamma} \tilde{K}_l,
\end{equation}
where $\tilde{K}_l = -i\hat{u}_{jk}$, bond $l$ connects sites $j$ and $k$, and the labels $j$ and $k$ appear in the order traversed around the loop. Using this representation one can obtain a flux through a loop by traversing it (in either direction) and multiplying by $\pm 1$ for each bond that is traversed in the correct or opposite way. Fig.~\ref{fig: loops} shows irreducible loops for all three lattices which we study here, and in particular the fluxes which change sign after flipping direction of a single bond.

Physical observables depend only on the eigenvalues of the flux operators, but clearly, as can be seen from Eq.~\eqref{eq: u loop}, there are many distinct sets $\{u_{jk}\}$ that give rise to the same set of eigenvalues. The extra dimensionality of the Hilbert space for Majorana fermions compared to that of spins can be associated with the gauge freedom of the $\mathbb{Z}_2$ fluxes. We can fix the gauge, and hence the flux sector, by choosing a particular set of $\{u_{jk}\}$. This gives a hopping problem for the matter Majorana fermions $\hat{c}$, as can be seen by replacing $\hat{u}_{jk}$ by their eigenvalues $\pm1$ in Eq.~\eqref{eq: H Majorana}.

\subsection{Majorana Spectrum and Density of States}

For a given flux sector we have rephrased the Kitaev model in terms of a Majorana hopping problem. The Hamiltonian of the latter is quadratic and can be diagonalised to obtain the Majorana spectrum, and the DOS. Details can be found in Appendix~\ref{ap: H momentum diagonalisation}. One of the striking observations in Refs.~\cite{Knolle2015,Smith} was that the dynamical response at low energies is primarily determined by the Majorana DOS, and more generally by the Green function obtained within the adiabatic approximation. Here we wish to extend this phenomenology to a characteristic set of 3D lattices. Compared to the results in 2D, the hyperoctagon lattice provides a qualitatively distinct example, having a Majorana Fermi-surface.

In Fig.~\ref{fig: spectra} we present the spectrum and the DOS of ``matter'' Majorana fermions. A difference between the three lattices which is important for the further discussion is in the low-energy behaviour of the DOS. The latter is finite for the hyperoctagon, linear in energy for the hyperhoneycomb, and quadratic in the hyperhexagon case. The consequences of this different behaviour for the dynamical spin correlation functions are discussed in Sec.~\ref{sec: results}. While the physical quantities such as the DOS and the spin correlation functions are invariant under $\mathbb{Z}_2$ gauge transformations, the Majorana dispersion relation depends on a chosen gauge.
The spectrum in Fig.~\ref{fig: spectra} is shown for gauge choices in which the gapless points, lines or surfaces intersect standard high symmetry cuts in the Brillouin zone. The most striking feature, which makes the hyperoctagon case distinct from the two other lattices is that its Majorana dispersion relation is not ``particle-hole" symmetric, i.e. $E(\mathbf{k}) \neq - \tilde{E}(\mathbf{k})$ (but of course $E(\mathbf{k}) = -\tilde{E}(-\mathbf{k})$ holds) because the symmetry-related excitations with positive and negative frequencies occur at different points in the Brillouin zone (rather than at the same $k$-vector). The latter is the consequence of the hyperoctagon lattice being non-bipartite, whereas the other two lattices are bipartite~\cite{HermannsQSL,HermannsZoo}.

\begin{figure*}[tb]
\centering
\subfigimg[width=0.34\textwidth]{\large \hspace*{-3pt}(a)}{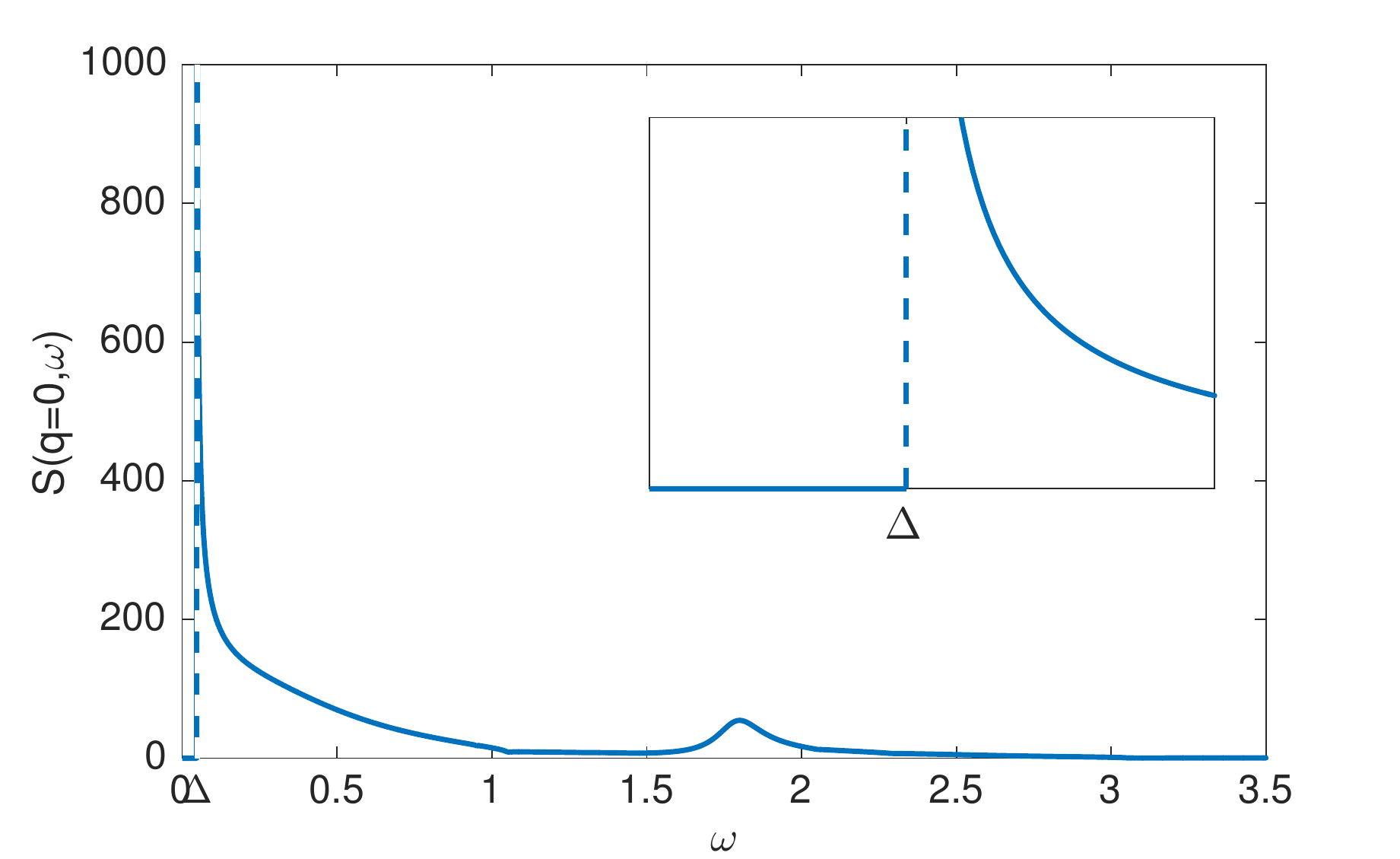}
\!\!\!\!\!\!\subfigimg[width=0.34\textwidth]{\large \hspace*{-3pt}(b)}{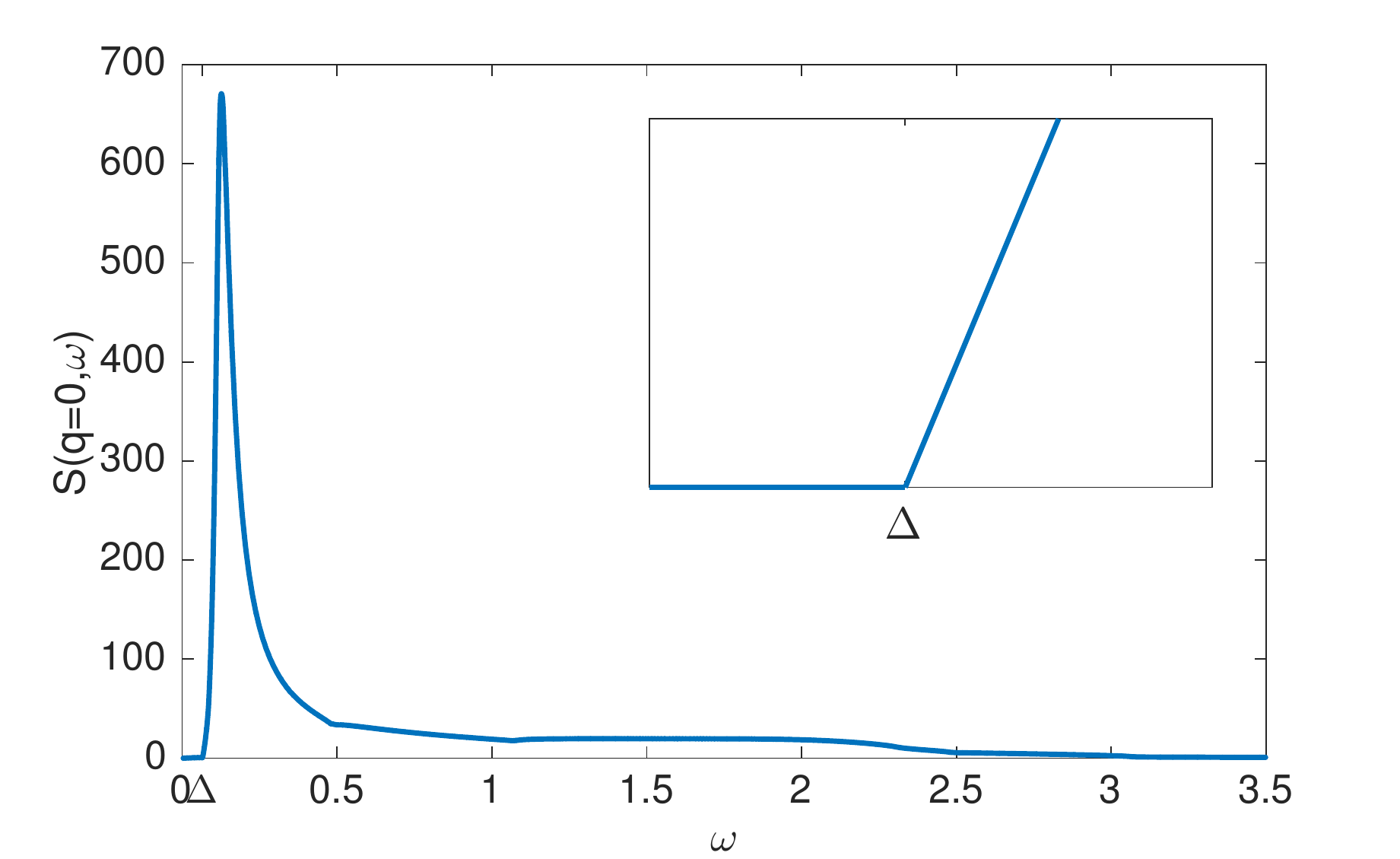}
\!\!\!\!\!\subfigimg[width=0.34\textwidth]{\large \hspace*{-5pt}(c)}{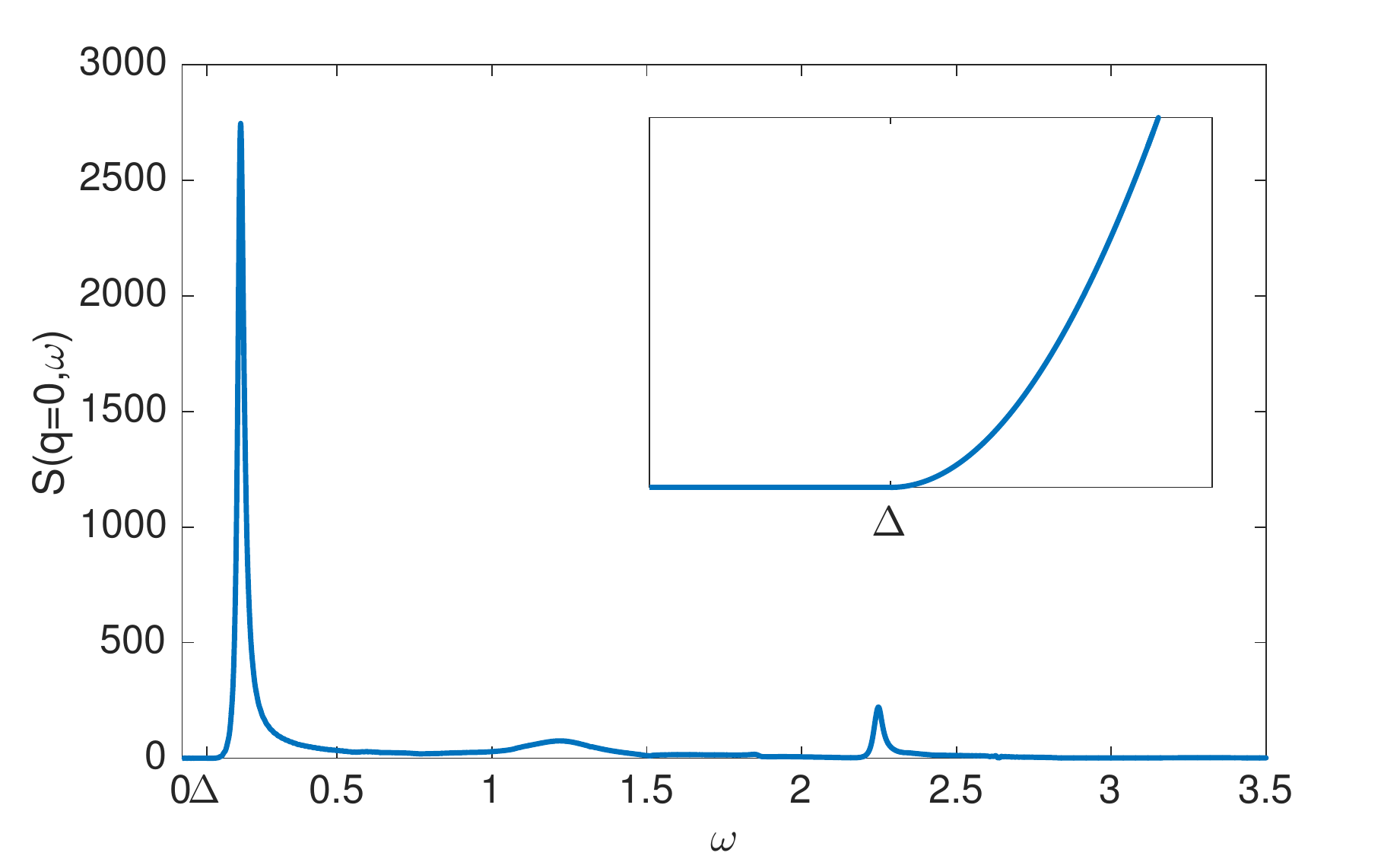}
\subfigimg[width=0.345\textwidth]{\large (a)}{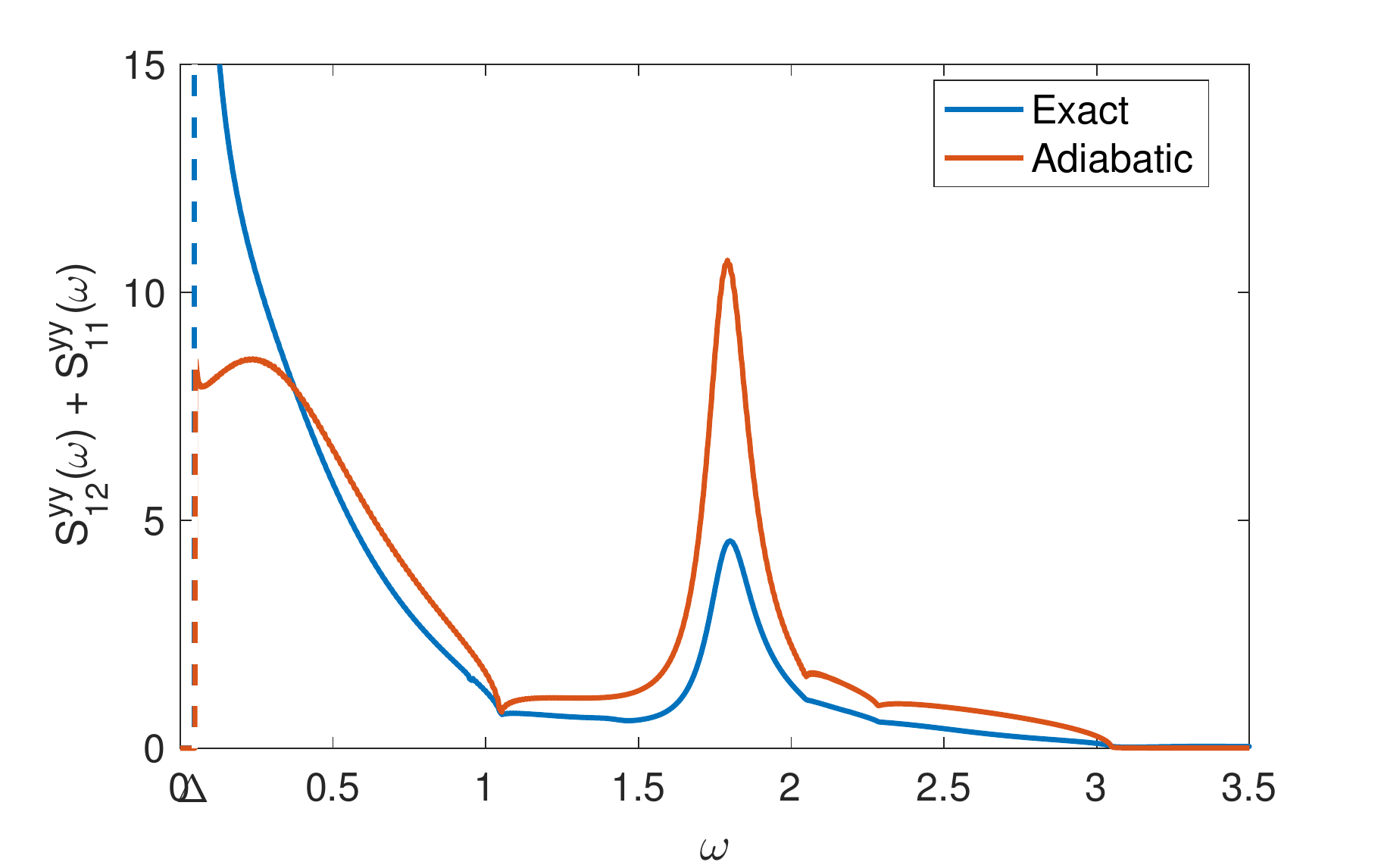}
\!\!\!\!\!\!\!\!\subfigimg[width=0.345\textwidth]{\large (b)}{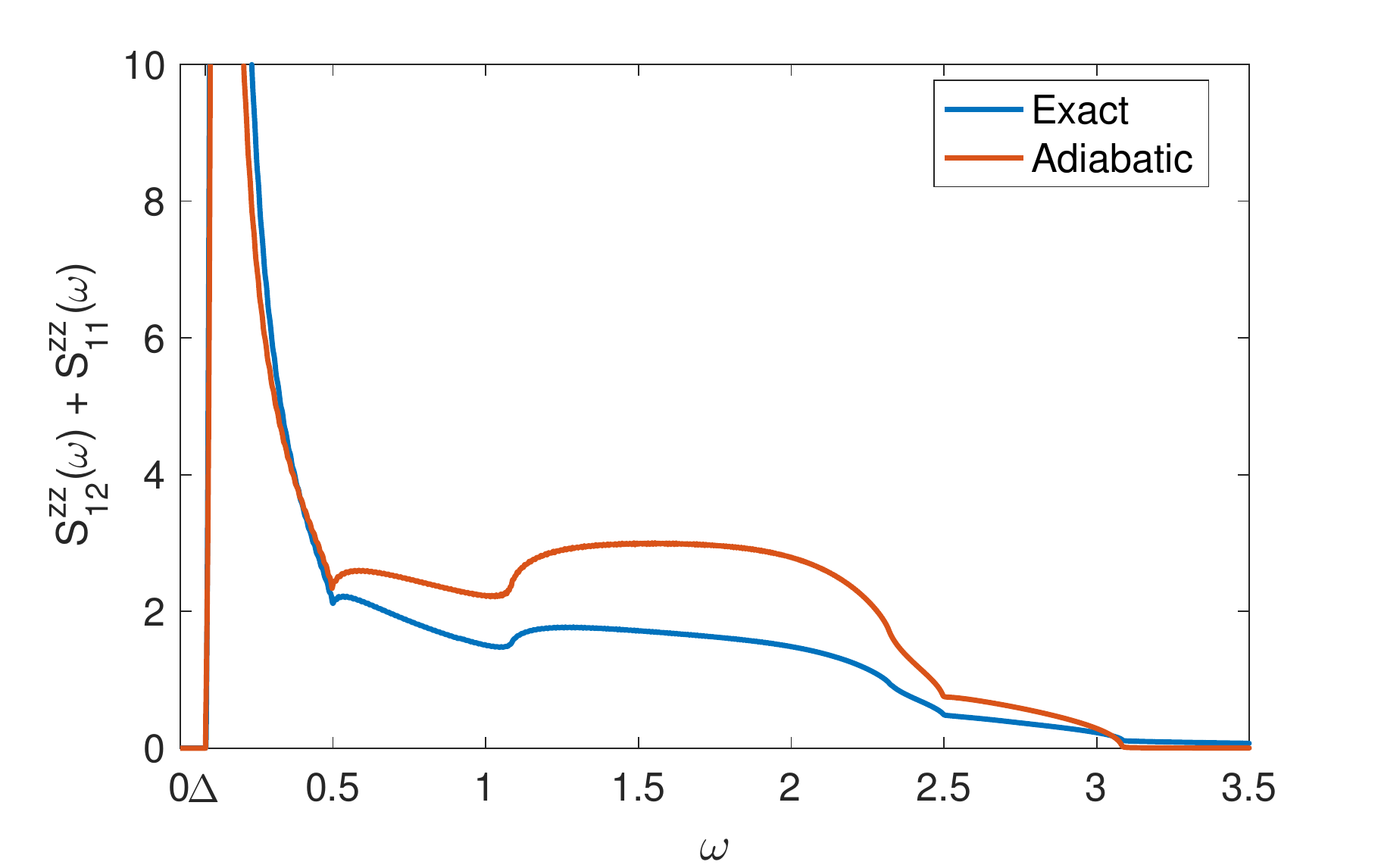}
\!\!\!\!\!\!\!\!\subfigimg[width=0.345\textwidth]{\large (c)}{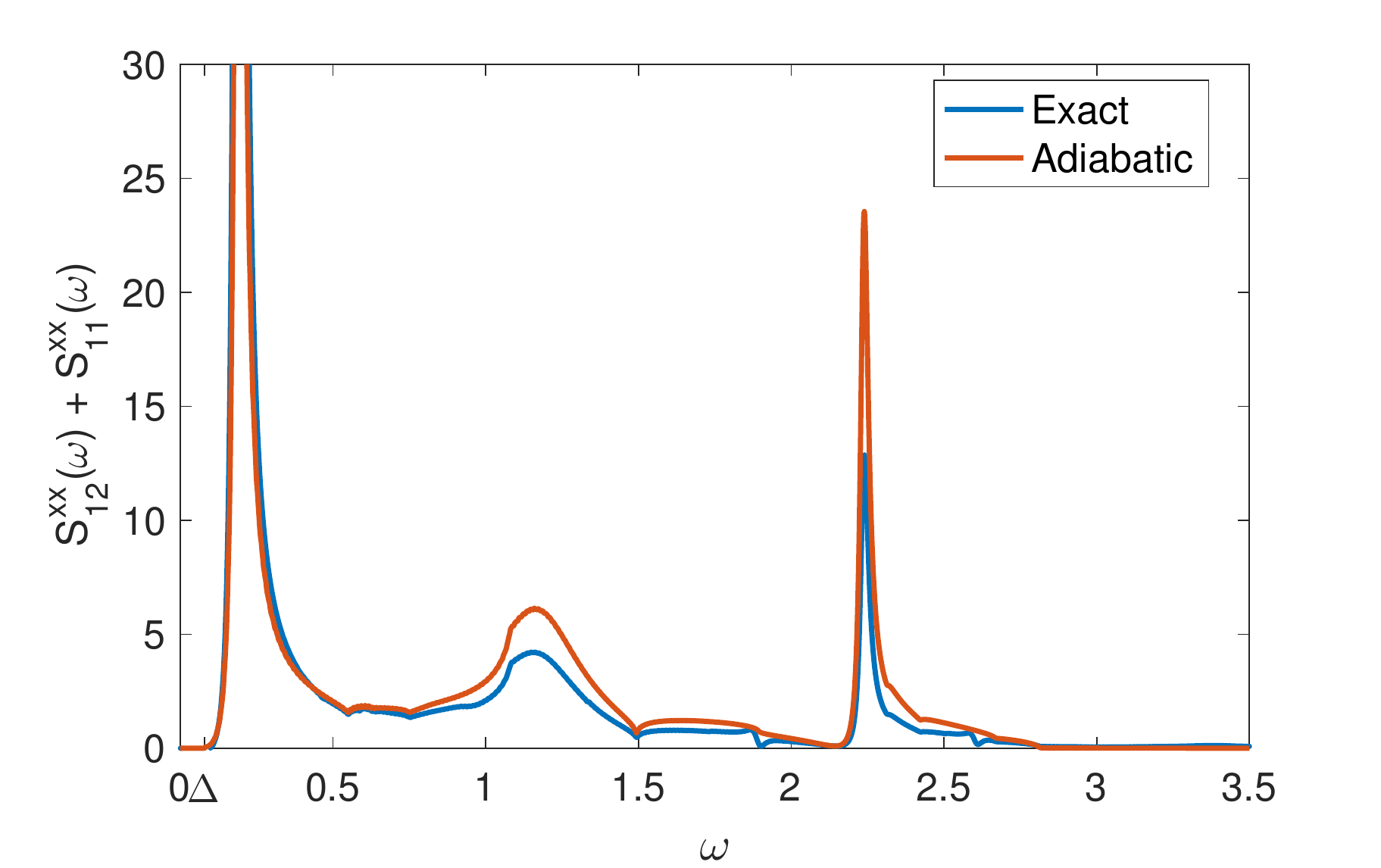}
\caption{(Top) Dynamical structure factor for (a) hyperoctagon, (b) hyperhoneycomb, and (c) hyperhexagon lattices at $\q=0$ with the inset showing schematically the low energy behaviour. (Bottom) Comparison of the exact response and the adiabatic approximation for particular bond contributions. The adiabatic response is rescaled by a numerical factor to obey the sum rule.}
\label{fig: q=0}
\end{figure*}

\begin{figure*}[tb]
\centering
\!\subfigimg[width=0.335\textwidth]{\color{white}\large \hspace*{15pt}(a)}{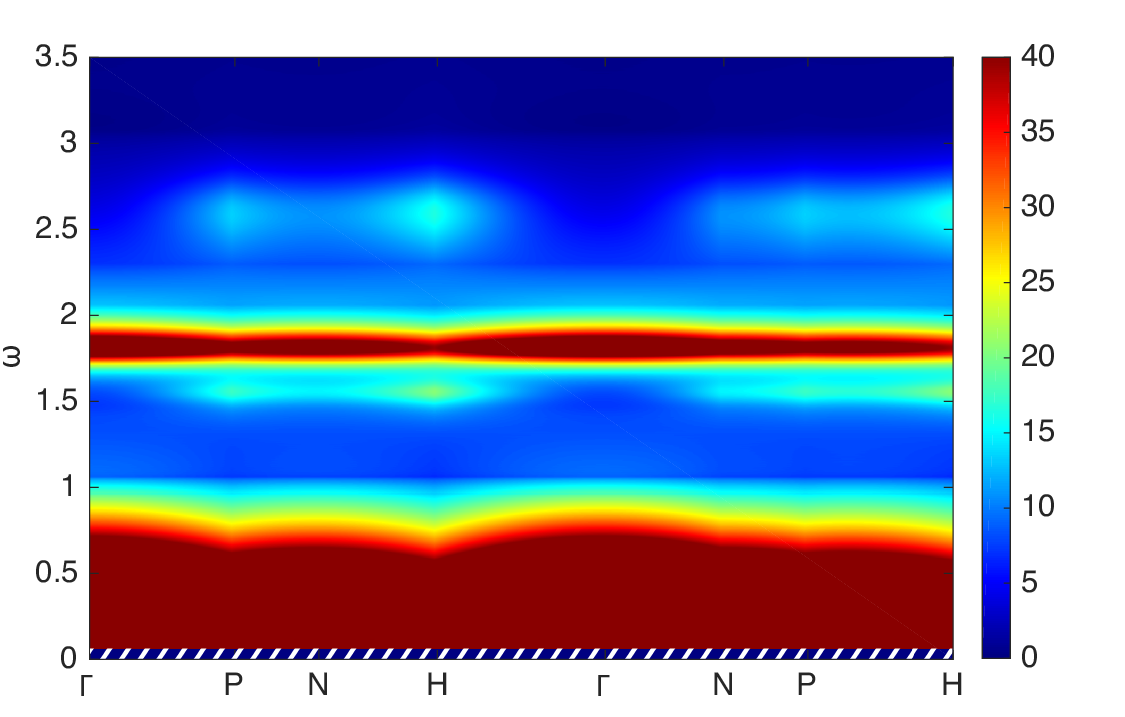}
\!\!\!\!\!\subfigimg[width=0.335\textwidth]{\color{white}\large \hspace*{15pt}(b)}{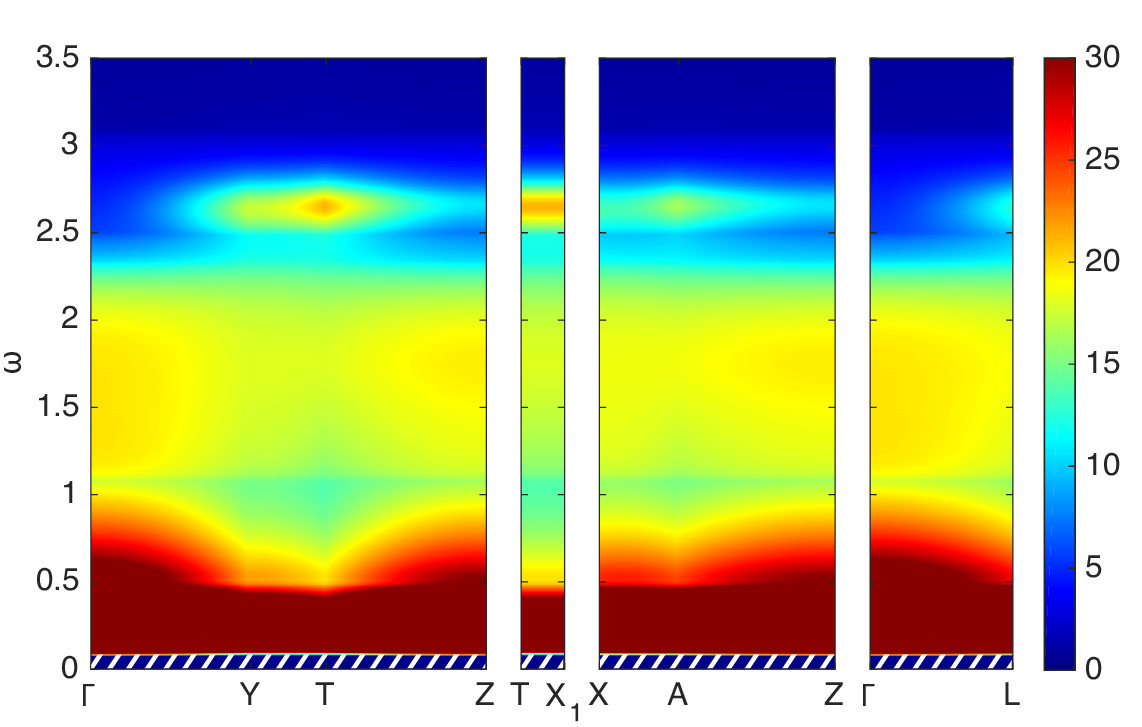}
\!\subfigimg[width=0.335\textwidth]{\color{white}\large \hspace*{15pt}(c)}{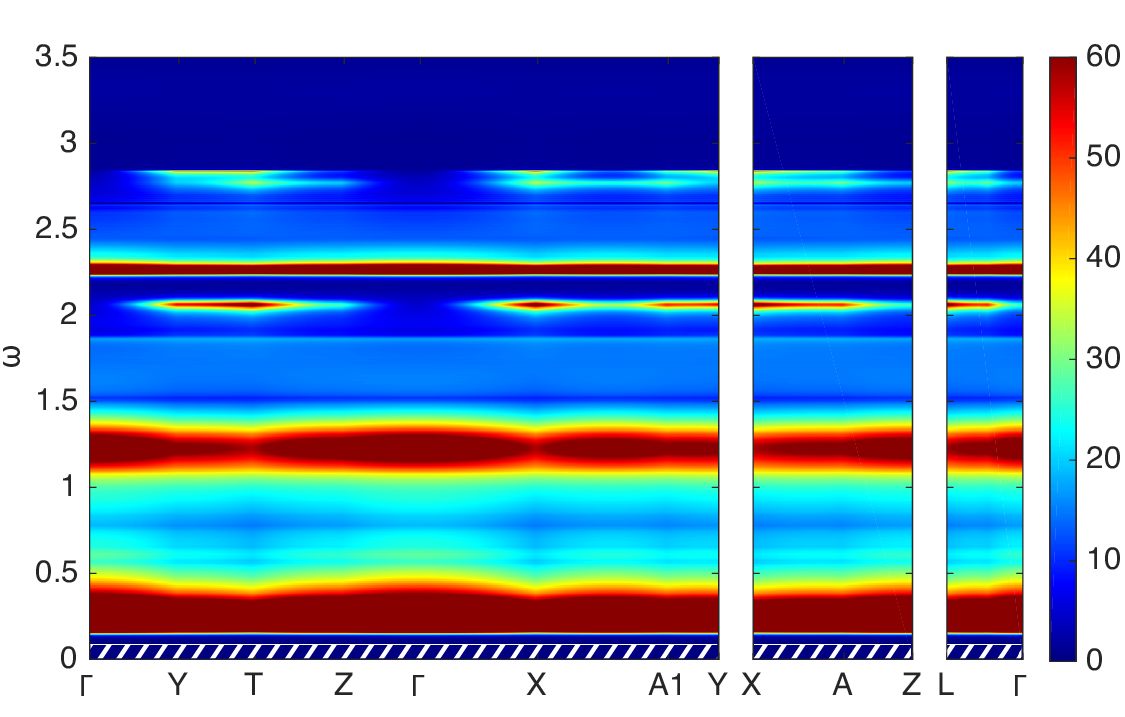}
\caption{Dynamical structure factor for (a) the hyperoctagon, (b) hyperhoneycomb and (c) hyperhexagon lattices. All intensity values above $40$, $30$ and $60$ respectively (in arb.~units) are shown in dark red. Shaded region represents the frequencies below the flux gap.}\label{fig: HO  Weyl DSF}
\end{figure*}

\begin{figure*}[htb]
\centering
\subfigimg[width=0.345\textwidth]{\large (a)}{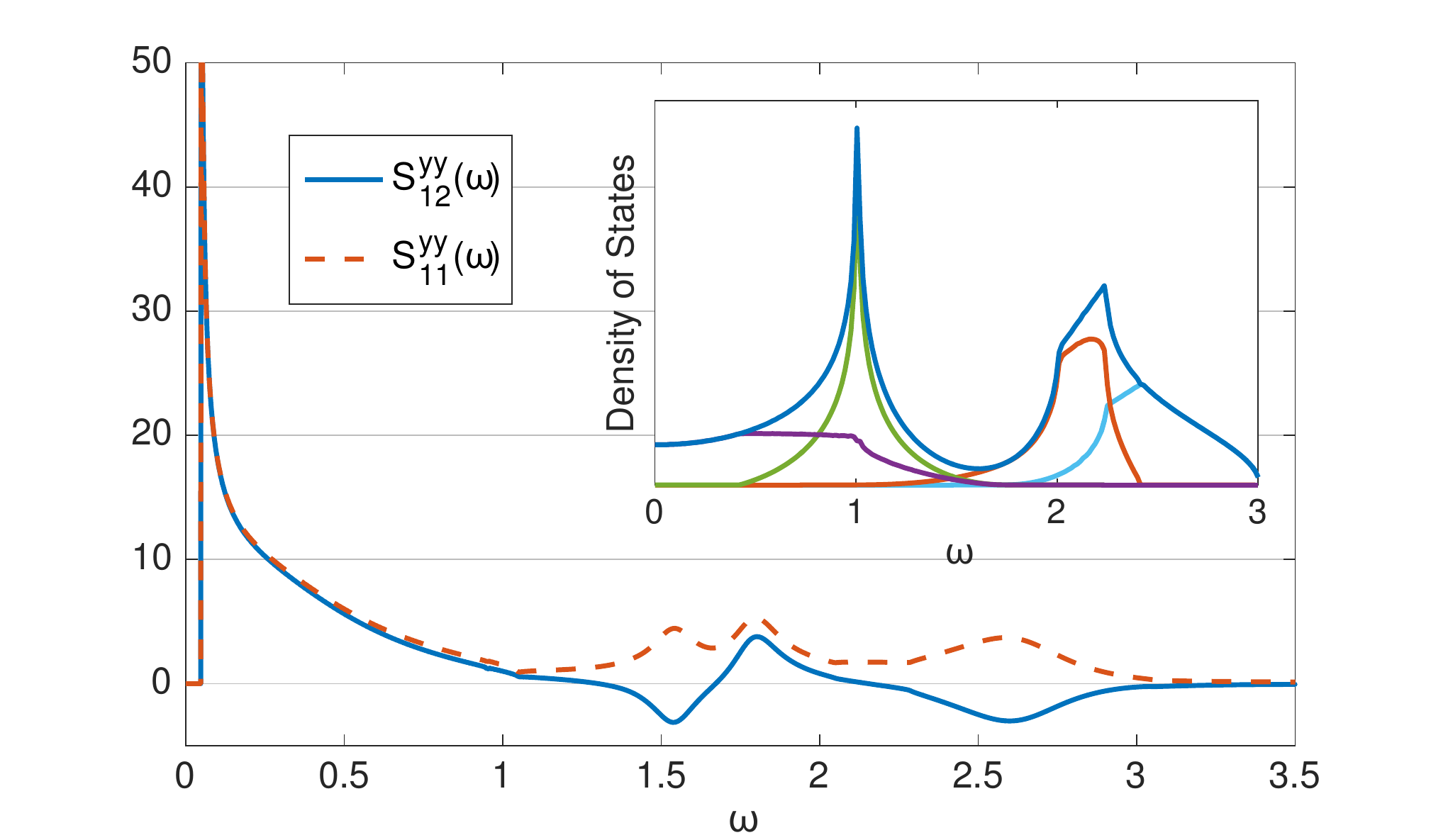}
\!\!\!\!\!\!\!\!\subfigimg[width=0.345\textwidth]{\large (b)}{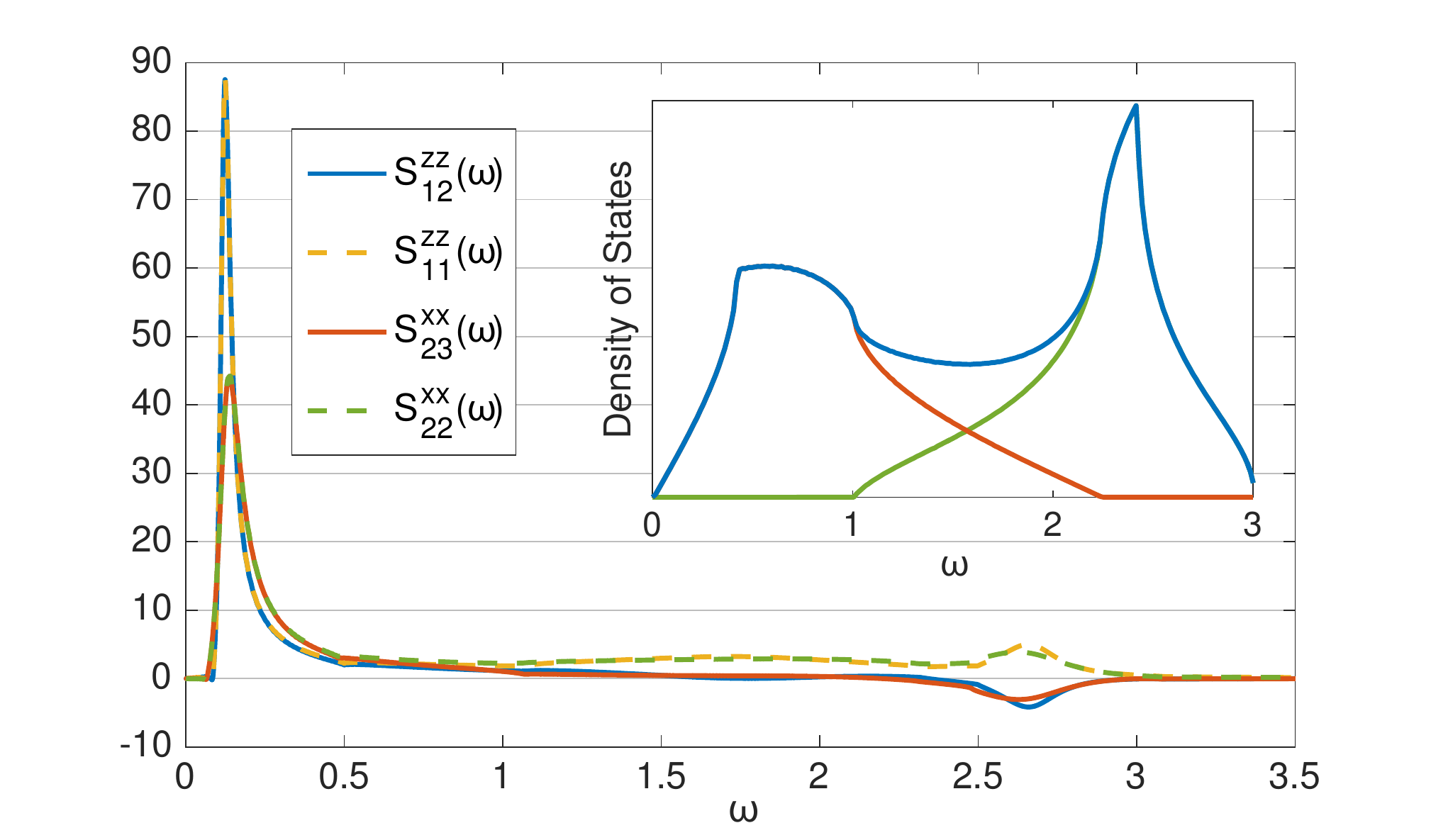}
\!\!\!\!\!\!\!\!\subfigimg[width=0.345\textwidth]{\large \hspace*{-1pt}(c)}{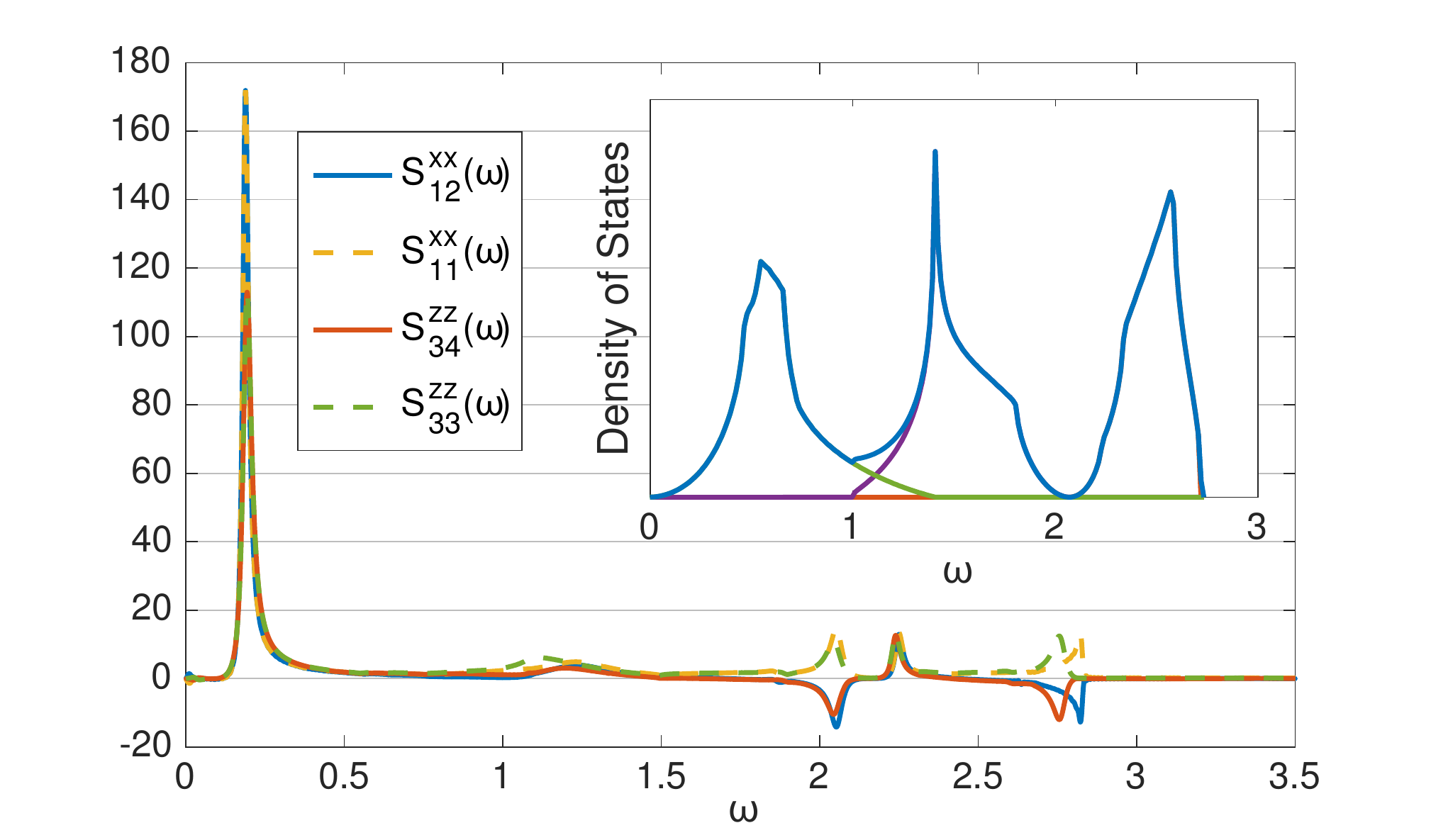}
\caption{Inequivalent components of the dynamic structure factor for (a) the hyperoctagon, (b) hyperhoneycomb and (c) hyperhexagon  lattices. The hyperoctagon has two inequivalent components, whereas the hyperhoneycomb and hyperhexagon have four. (Inset) Band resolved density of states.}\label{fig: components}
\end{figure*}

\section{Dynamical Structure Factor} \label{sec: spin correlators}\label{sec: results}

Our central task is the calculation of time-dependent spin correlators, and the corresponding dynamic structure factor. 
This problem can be mapped to a non-equilibrium problem in which Majorana fermions propagate in the presence of a suddenly inserted flux excitation, due to the action of a spin operator on the ground state~\cite{BaskaranExact}. In previous work~\cite{PRL,Knolle2015} we developed a method for calculating dynamical response in the 2D Kitaev model, which allows one to obtain exact results for the dynamical response in the thermodynamic limit. These ideas have been further applied in Ref.~\cite{Smith} to a three-dimensional case. The details of the calculations are presented in Appendix~\ref{ap: spin correlators}; see also~\cite{PRL,Knolle2015}. 

The dynamical spin structure factor is defined as
\begin{equation}\label{eq: DSF definition}
S(\mathbf{q},\omega) = \sum_{a,b,j,k} e^{-i\mathbf{q}\cdot (\mathbf{r}_j - \mathbf{r}_k)} \int^\infty_{-\infty} dt \; e^{i\omega t} S^{ab}_{jk}(t),
\end{equation} 
where $S^{ab}_{jk}(t) = \la 0 | \hat{\sigma}^a_j(t) \hat{\sigma}^b_k(0) | 0 \ra$ is the time-dependent spin correlation function. The DSF is directly related to cross sections measured in inelastic neutron scatting (INS) experiments~\cite{Lovesey} and at $\mathbf{q}=0$ to the signal obtained in electron spin resonance (ESR) experiments.

In the following we will focus on the results at the isotropic point $J_x = J_y = J_z$, which is representative of the gapless Kitaev QSL phases, and is also relevant for experiments, which indicate that e.g.~the hyperhoneycomb material $\beta$-Li$_2$IrO$_3$ lies in proximity to the isotropic point~\cite{Kim}. 
\begin{figure*}[t]
\centering
\subfigimg[width=0.33\textwidth]{\large (a)}{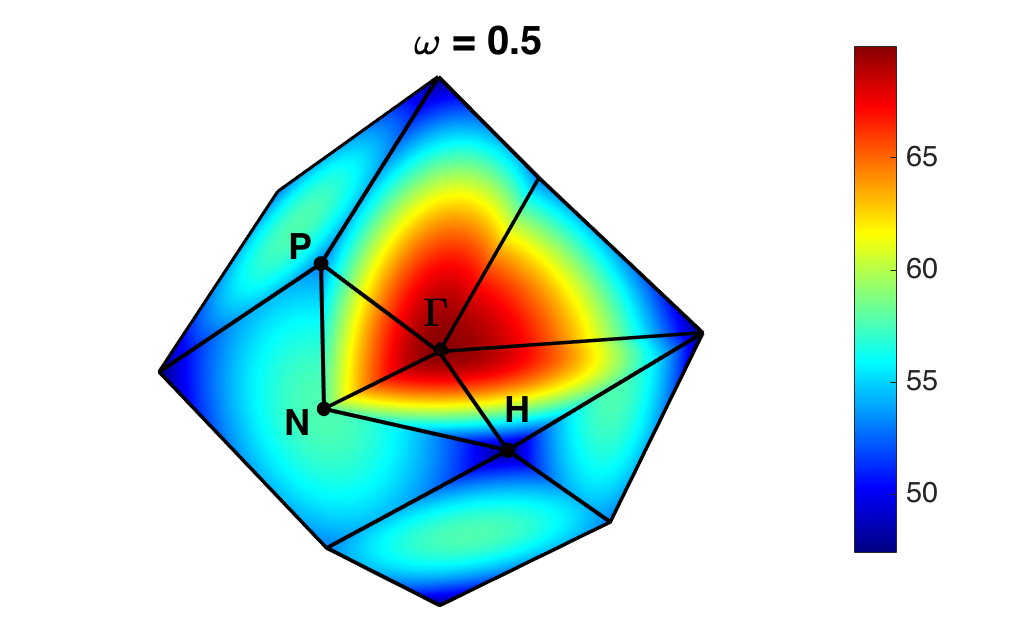}
\subfigimg[width=0.33\textwidth]{\large (b)}{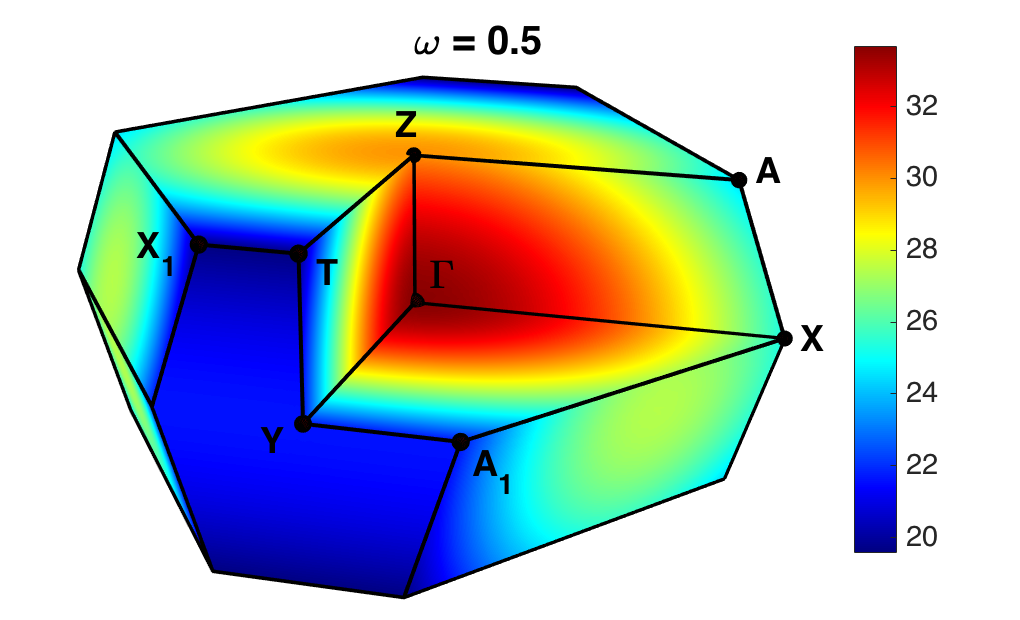}
\subfigimg[width=0.33\textwidth]{\large (c)}{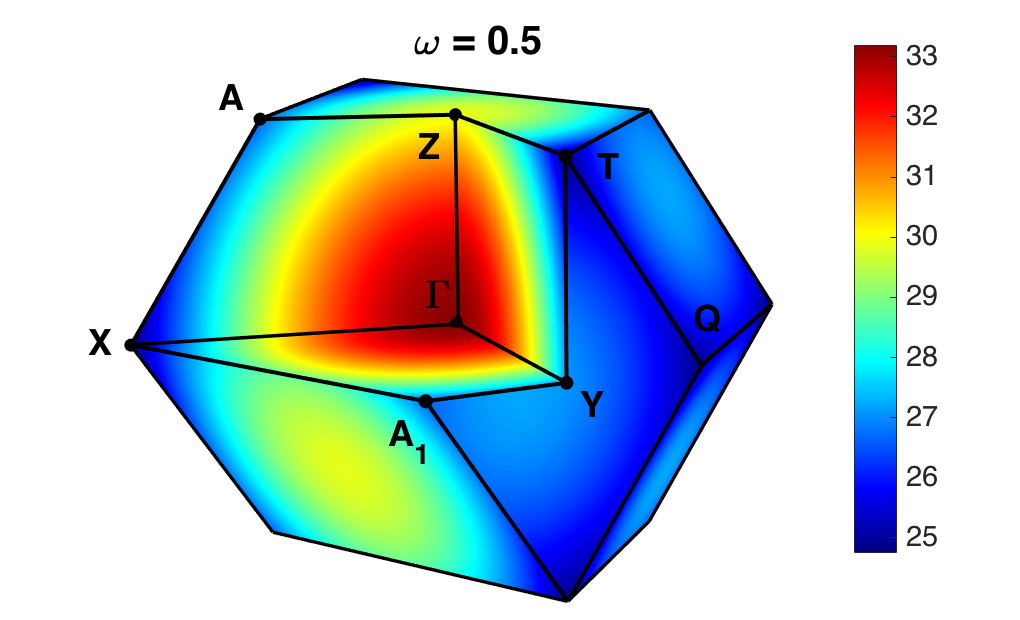}
\subfigimg[width=0.33\textwidth]{\large (a)}{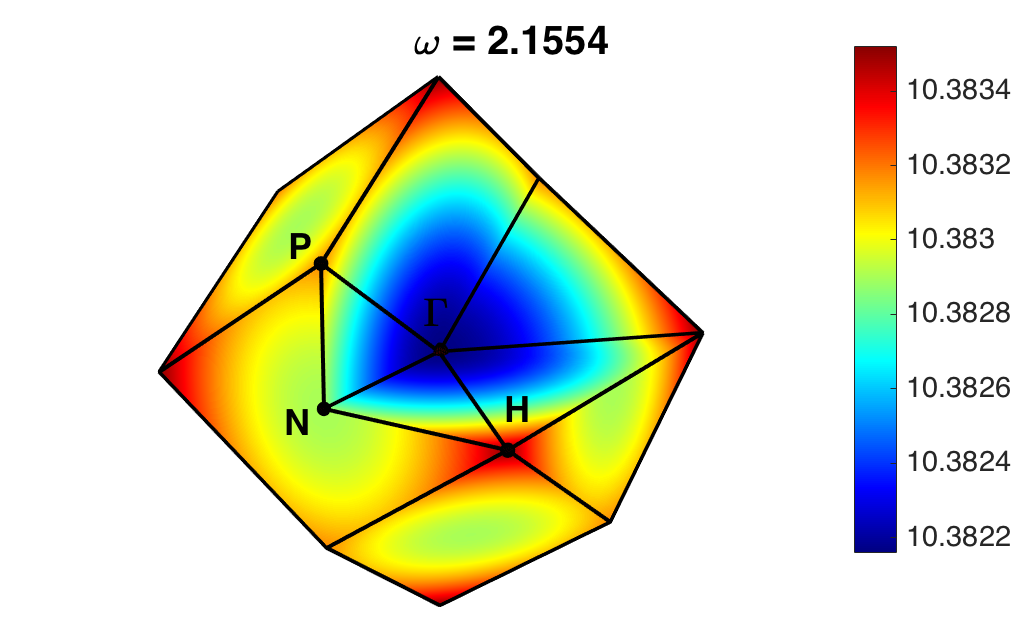}
\subfigimg[width=0.33\textwidth]{\large (b)}{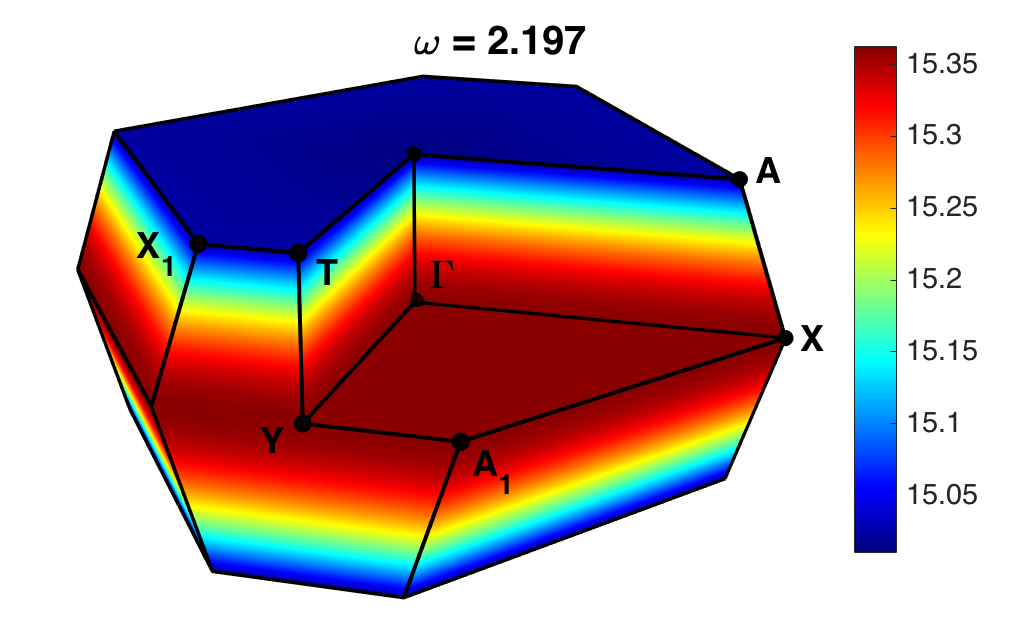}
\subfigimg[width=0.33\textwidth]{\large (c)}{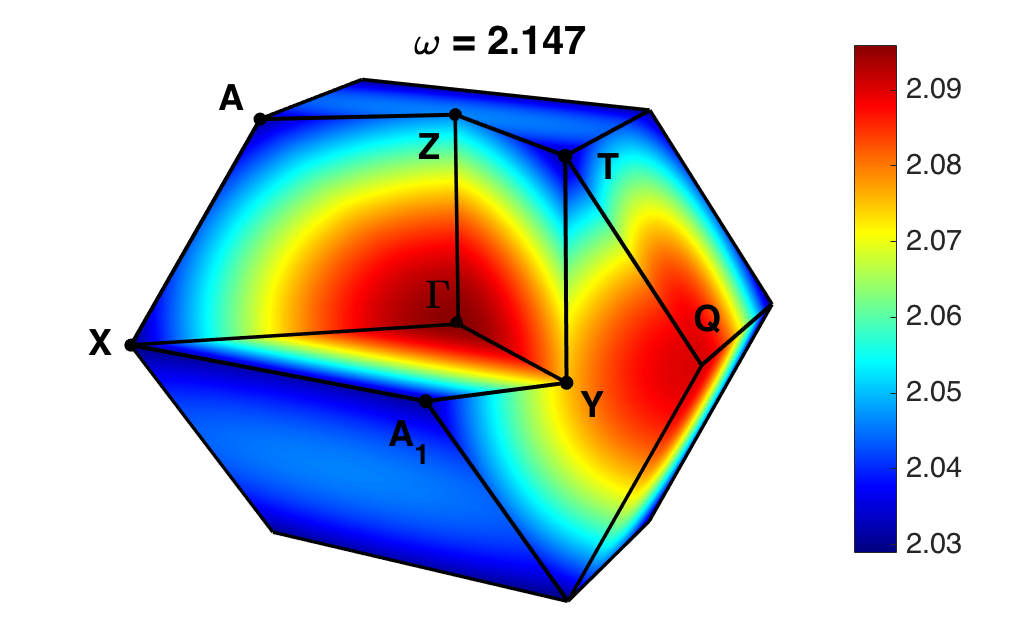}
\caption{Dynamical structure factor for (a) hyperoctagon, (b) hyperhoneycomb, and (c) hyperhexagon at fixed values of $\omega$. Intensity is shown across high symmetry planes in the Brillouin zone. (Top) Representative momentum dependence in the middle of the spectrum. (Bottom) Momentum dependence in the vicinity of inflexion points in frequency. Note the frequency scale.}\label{fig: 3D BZ}
\end{figure*}

Figure~\ref{fig: q=0} shows results for the dynamic structure factor at $\q=0$ for the three lattices studied in the text. In all cases the response vanishes at energies below a threshold, which is given precisely by the value of the flux gap. Distinct behaviour for different lattices is apparent just above threshold, reflecting their low-energy Majorana DOS. For the hyperhoneycomb lattice $S(0,\omega)$ increases linearly with energy as also found in the 2D honeycomb case. This is a direct consequence of the linear behaviour of the Majorana DOS. For the hyperhexagon lattice $S(0,\omega)$ increases quadratically with energy due to quadratic dependence of the DOS. By contrast, the constant low-energy DOS for the hyperoctagon lattice results in a divergence of $S(0,\omega)$ at the threshold.

Because of this non-vanishing DOS the low-energy behaviour of the correlators for the hyperoctagon lattice can be obtained using standard methods developed for the X-ray edge problem~\cite{Bosonization,Nozieres}, from which we can read off the value of the X-ray edge exponent $\alpha=2g-g^2$, %
where $g = \delta_0/\pi$, and $\delta_0$ is the phase shift related to the strength of the local potential (which in our case corresponds to coupling between Majorana fermions and the flux excitation), see Ref.~\cite{Bosonization}. This phase shift in the case of Kitaev model is given by the equation
\begin{equation}\label{eq: general delta0}
\delta_0 = - \arctan \left( \frac{2 J\Im m[G^R_0(0)] }{1+\tilde{J}\Re e[G^R_0(0)]}  \right),
\end{equation}
where $\tilde{J}=2 u_{jk}J$, and $u_{jk}$ is the ground-state flux on the measured bond.
The exponent $\alpha$ obtained from our numerical solution agrees with this asymptotic behaviour.

We note that the mapping of dynamical correlators in the 2D Kitaev honeycomb model to the X-ray edge problem was suggested by Baskaran et.al.~in Ref.~\cite{BaskaranExact}. However, as we showed previously~\cite{PRL} while this mapping is useful, the physics, in the 2D case is different from that of the X-ray edge problem. Notably, there is no singularity in the dynamical correlators of the 2D Kitaev model because of the vanishing low-energy Majorana DOS. By contrast, for 3D lattices there is a possibility for a Majorana Fermi surface (hyperoctagon), and thus for non-vanishing DOS, which allows one to extend the phenomenology developed for the X-ray edge problem directly onto the case of Majorana fermions. The physics of the Fermi-edge singularity for a system with a Fermi sea of Majorana excitations was also used in Ref.~\cite{Tikhonov2010} to obtain the long-time dynamical response of the Kitaev model on a decorated honeycomb lattice.

Figure~\ref{fig: q=0} shows a comparison of the results obtained using exact non-equilibrium calculation, and the adiabatic approximation. The latter, which was introduced in~\cite{PRL,JohannesThesis} provides an insight into the origin of the fine-structure in the response. The basic assumption of this approximation is that the non-equilibrium correlators can be approximated by equilibrium ones. The latter correspond to adiabaticaly introducing the fluxes that are generated by the action of spin operators on the ground state.  
The local Green's function (GF), see Eqs.~\eqref{eq: GF def}--\eqref{eq: spin GF ss}, in the adiabatic approximation can be written in a particularly transparent form, e.g.~the expression for the advanced GF $\tilde{G}^A_0$ reads
\begin{equation}\label{eq: adiab}
\tilde{G}^A_0(\omega)=\frac{G^A_0(\omega)}{1+\tilde{J} G^A_0(\omega)}.
\end{equation} 
For small density of states, i.e.~small $\Im m [G_0^A(\omega)]$, zeros of $[1+\tilde{J}\Re e[G_0^A(\omega)]]$ produce  peaks in the response.

Although the result of adiabatic approximation does not agree quantitatively with the exact solution, both do exhibit a similar qualitative behaviour. In fact, it can be shown analytically that the low-energy dynamical structure factor is exact in the adiabatic approximation, provided that the Majorana DOS vanishes at the threshold which is the case for the hyperhoneycomb and the hyperhexagon lattices. In contrast, for the hyperoctagon lattice, the adiabatic approximation fails to provide a correct description of the low-energy response. The reason for this true non-equilibrium effect is the non-vanishing low energy DOS which leads to the divergence of the response at the threshold, similarly to the classic X-ray edge problem. However, at higher energies, the adiabatic approximation follows the shape of the exact response remarkably well even in the hyperoctagon case. By combining this approximation with the X-ray edge approach one can obtain a good qualitative description of the response across the whole frequency region.

The frequency dependence of the dynamical structure factor along high symmetry planes in the Brillouin zone is shown in Fig.~\ref{fig: HO  Weyl DSF}. As in the other cases~\cite{PRL,Smith} the response above the flux gap is a continuous function of frequency that falls off rapidly above the energy of a single particle Majorana bandwidth. Beyond that, the main features of the response are a series of peaks and almost flat bands, with some of these bands showing respectively inverted dispersion. These features are also apparent from the behaviour of inequivalent spin correlators shown in Fig.~\ref{fig: components} in which different correlators show (anti)aligned peaks. Depending on whether these correlators add up constructively/destructively at zero momentum leads to two qualitatively different types of response with complementary momentum dependence.

In Fig.~\ref{fig: 3D BZ} we show the momentum dependence of the dynamic structure factor in the Brillouin zone for fixed values of $\omega$. The top row is representative of the behaviour in the middle of the spectrum. Depending on whether the zero-frequency correlators add up destructively/constructively, the intensity in the centre of the BZ shows minimum/maximum. In the bottom row of Fig.~\ref{fig: 3D BZ} we present the response in the vicinity of a transition separating bands with opposite momentum dependence. For the hyperoctagon lattice one can see a spherically symmetric inversion of the momentum dependence when crossing the inflexion point. In the case of other two lattices this inversion is more complicated due to the anisotropic momentum dependence. For the hyperhoneycomb lattice we find that near the inflexion point there exist narrow regions of $\omega$ where we observe flat bands dispersing only along $\Gamma-Z$ direction. In the hyperhexagon case we find that there is no dependence on the $\Gamma-Q$ direction, and the momentum dependence is rotationally symmetric about this axis. 

This behaviour is a consequence of the fact that all correlators beyond nearest neighbour vanish, and only nearest neighbour correlators, but not on-site ones, contribute to the momentum dependence of the response. The sign of the $\la j k \ra$ spin-components of the correlators thus determines the positions of the maxima/minima along the $r_j - r_k$ direction. For the hyperoctagon lattice nearest-neighbour correlators for all the three bonds are equal, which produces a spherically symmetric momentum dependence of the response, or no momentum dependence at all when nearest-neighbour correlators vanish simultaneously.  For the hyperhoneycomb and hyperhexagon lattices different spin-components of the correlators vanish at slightly different values of $\omega$. This explains a smooth transition across the inflexion point, where the momentum dependence of the response becomes flat in different directions for different $\omega$.

\section{Summary, discussion and outlook}\label{sec: Summary}
We have presented a systematic study  of the dynamical spin response in 3D Kitaev quantum spin liquids 
exhibiting fractionalized gapless Majorana fermion excitations with low-energy behaviour represented by Weyl points, nodal lines and Majorana Fermi surfaces. One of the main features of the dynamic structure factor is that its low energy behaviour is defined by the corresponding Majorana DOS. Here there are two distinct possibilities. In the case of vanishing DOS the response vanishes at low energies with the same power-law as the DOS, as illustrated by the hyperhexagon and hyperhoneycomb lattices. Alternatively, if the Majorana DOS is constant at low energy, as for the hyperoctagon lattice, 
the dynamical structure factor at energies just above threshold is governed by the true non-equilibrium physics of the X-ray edge problem. This allows one to extend the X-ray edge phenomenology to the case of Majorana fermions, and obtain the results for the response using standard methods. Here the response shows a power-law behaviour at the threshold. The exponent can be related to the strength of the local scattering potential for Majorana fermions which arises from a sudden insertion of fluxes as a result of fractionalization of spin. At high energies the response is broad in all three lattices showing a fine-structure which is governed by the respective Majorana DOS. If one knows in other ways that one has a Kitaev spin liquid, then the INS reveals the Majorana excitations more simply than one might have expected.

From a methodological perspective, the adiabatic approximation, which neglects the non-equilibrium features of the problem, provides a good qualitative understanding of the dynamical response. Even in the case of the hyperoctagon lattice, which has the divergent response at low energies, it captures well the gross features beyond this contribution. Thus by combining the knowledge of the Majorana DOS with the adiabatic approximation one can easily find a qualitative behaviour of the INS response for other Kitaev models (close to the isotropic point).

One has to emphasize that the integrability of the Kitaev models comes at the usual price of fine-tuning the Hamiltonian. Adding integrability breaking terms to the Hamiltonian turns out to be less deleterious than one might imagine, as some rather natural perturbations -- e.g.\ a Heisenberg exchange -- do not necessarily destroy central features such as the conservation of Fermion parity or the gaplessness of the Majorana spectrum. However, some details do change: while general considerations imply that the gapfulness of the flux excitations is perturbatively stable, their non-dynamical nature is not. Therefore, results depending on this feature in detail, such as the strict vanishing of the response below the gap for creating a flux pair will not in general hold. However, our results provide a good starting point to investigate the integrability breaking terms perturbatively.

We have thus presented a general phenomenology amongst Kitaev QSLs which should extend to the harmonic honeycomb series, as well as to the whole zoo of lattices studied in Ref.~\cite{HermannsZoo}, and beyond. The methods developed here can be applied generally to a full range of Kitaev models which can be represented in terms of itinerant Majorana fermions coupled to static flux degrees of freedom.

Overall, in this work we obtained a detailed set of predictions for the inelastic neutron scattering, and electron spin resonance experiments for a wide range of potential Kitaev QSLs in 3D, which may be useful for identifying fractionalised quantum spin liquids in three-dimensional materials. 

\textbf{Acknowledgements} The collaboration was supported by the Helmholtz Virtual Institute ``New States of Matter and their Excitations'' and the German Science Foundation under SFB 1143. The work of J.K. is supported by a Fellowship within the Postdoc-Program of the German Academic Exchange Service (DAAD). J.T.C. is supported by EPSRC Grant No. EP/I032487/1, D.K. is supported by EPSRC Grant No. EP/M007928/1. 
JK acknowledges helpful discussions with M. Hermanns, and S. Trebst.

\appendix

\section{Momentum Space Diagonalisation} \label{ap: H momentum diagonalisation}

Since it is possible to choose a translationally invariant gauge in the ground state flux sector, we can simplify the Hamiltonian \eqref{eq: H Majorana} by Fourier transform, writing it in the form
\begin{equation}\label{eq: Fourier H}
\hat{H} = \sum_\q \Psi_{-\q}^T H_\q \Psi_\q
\end{equation}
where $\Psi_\q^T = (\hat{c}_{1,\q},\ldots,\hat{c}_{n,\q})$, $H_\q$ is a $n \times n$ matrix for each $\q$, and $n$ is the number of sites in the primitive unit cell.

To rewrite this in terms of complex fermions we need to define a linear transformation between the $n$ Majorana fermions and $n/2$ complex fermions. For the hyperoctagon and hyperhoneycomb we can thus define two types of complex fermions at once which we call $\hat{f}$ and $\hat{g}$, and a third for the hyperhexagon, $\hat{h}$ and we denote the transformation by a matrix $\Gamma$. The Hamiltonian then becomes
\begin{equation}\label{eq: Gamma H}
\hat{H} = \sum_\q \Phi_\q^\dag \Gamma^\dag H_\q \Gamma \Phi_\q \equiv \sum_{\q \in B.Z.} \Phi_\q^\dag \tilde{H}_\q \Phi_\q
\end{equation} 
where $\Phi_\q^T = (\hat{f}_\q, \hat{f}^\dag_{-\q}, \hat{g}_\q, \hat{g}^\dag_{-\q})$ for the hyperhoneycomb and hyperoctagon, and $\Phi_\q^T = (\hat{f}_\q, \hat{f}^\dag_{-\q}, \hat{g}_\q, \hat{g}^\dag_{-\q}, \hat{h}_\q, \hat{h}^\dag_{-\q})$ in the hyperhexagon case.

In momentum space, the pairs of Majorana fermions that span two primitive cells introduce phases to the $\Gamma$-matrices. If we take the $\langle14\rangle_x$ and $\langle23\rangle_y$ bonds for the hyperhoneycomb as an example, we define $\hat{f}_\mathbf{r} = \frac{1}{2} (\hat{c}_{1,\mathbf{r}+\mathbf{a}_1} + i \hat{c}_{4,\mathbf{r}} )$
\begin{equation}
\hat{c}_{1,\q}  = e^{-i\q\cdot\mathbf{a}_1} ( \hat{f}_\q + \hat{f}^\dag_{-\q} ),\ \ 
\hat{c}_{4,\q} = i (\hat{f}^\dag_{-\q} - \hat{f}_\q).
\end{equation}
Similarly, defining $\hat{g}_\mathbf{r} = \frac{1}{2} (\hat{c}_{2,\mathbf{r}+\mathbf{a}_3} + i \hat{c}_{3,\mathbf{r}} )$ gives
\begin{equation}
\hat{c}_{2,\q} = e^{-i\q\cdot\mathbf{a}_3} ( \hat{g}_\q + \hat{g}^\dag_{-\q} ), \ \ \hat{c}_{3,\q} = i (\hat{g}^\dag_{-\q} - \hat{g}_\q).
\end{equation}
Hence, the corresponding $\Gamma$-matrix is
\begin{equation}
\Gamma = \left(\begin{array}{cccc}e^{-i\q\cdot\mathbf{a}_1} & e^{-i\q\cdot\mathbf{a}_1} & 0 & 0 \\0 & 0 & e^{-i\q\cdot\mathbf{a}_3}  & e^{-i\q\cdot\mathbf{a}_3}  \\0 & 0 & -i & i \\-i & i & 0 & 0\end{array}\right).
\end{equation}
In a similar way we can obtain a $\Gamma$-matrix for a pair (or triple for hyperhexagon) of bonds for other lattices. Note also that the $\Gamma$-matrices in Eq.~\eqref{eq: Gamma H} assume implicit momentum dependence.

\subsection{Diagonalization of the Majorana Hamiltonian}

One of the subtleties of using the exact integral approach that was not an issue for the 2D honeycomb is that our definitions of the complex fermions must be such that the anomalous Green's functions vanish. Because of this we must redefine the matter fermions for each correlator we calculate. We will now show that we can work around this apparent complication and diagonalize directly the Majorana Hamiltonian as a matrix once for each $\q$ and use transformation matrices to calculate all of the different spin correlators.

In equation \eqref{eq: Gamma H} we make it explicit that the sum is over all momenta in the Brillouin zone. To diagonalise the Hamiltonian we then split it into two halves over $\mathbf{q} > 0$ and $\mathbf{q} < 0$ separately. By $\mathbf{q} > 0$ we simply mean any half of the Brillouin zone that does not overlap with its inversion through the origin (which we denote $\mathbf{q}<0$). 
\begin{widetext}
\begin{equation}\label{eq: block H}
\begin{aligned}
\hat{H} &= \sum_{\q>0} \Phi^\dag_\q \tilde{H}_\q \Phi_\q + \sum_{\q > 0} \Phi^\dag_{-\q} \tilde{H}_{-\q} \Phi_{-\q}  = \sum_{\q >0} \left(\begin{array}{cc}\Phi_\q^\dag & \Phi^\dag_{-\q}\end{array}\right) \left(\begin{array}{cc}\tilde{H}_\q & 0 \\0 & \tilde{H}_{-\q}\end{array}\right) \left(\begin{array}{c}\Phi_\q \\\Phi_{-\q}\end{array}\right).\\
& = \sum_{\q >0} \left(\begin{array}{cc}\Phi_\q^\dag & \Phi^\dag_{-\q}\end{array}\right) \left(\begin{array}{cc}\Gamma^\dag_\mathbf{q} & 0 \\0 & \Gamma^\dag_\mathbf{-q}\end{array}\right)\left(\begin{array}{cc}H_\q & 0 \\0 & H_{-\q}\end{array}\right) \left(\begin{array}{cc}\Gamma_\mathbf{q} & 0 \\0 & \Gamma_\mathbf{-q}\end{array}\right)\left(\begin{array}{c}\Phi_\q \\\Phi_{-\q}\end{array}\right).
\end{aligned}
\end{equation}
\end{widetext}
The block matrix in the last line of \eqref{eq: block H} is of the form
\begin{equation}\label{eq: BR form}
\left(\begin{array}{cc}A & B \\-B^* & -A^*\end{array}\right).
\end{equation}
This allows us to use the results of Blaizot and Ripka~\cite{BlaizotRipka}, namely that we can write the Hamiltonian as
\begin{equation}\label{eq: mom diag}
\begin{aligned}
\hat{H} = \sum_{\q > 0} \beta_\q^\dag \Omega_\q \beta_\q &\equiv \sum_{\q>0} \left(\begin{array}{cc}\hat{b}_\q^\dag & \hat{b}_\q\end{array}\right) \left(\begin{array}{cc}\omega_\q & 0 \\0 & -\omega_\q\end{array}\right) \left(\begin{array}{c}\hat{b}_\q \\\hat{b}_\q^\dag\end{array}\right),\\
& = \sum_{\q>0} \left( \hat{b}^\dag \omega \hat{b} - \hat{b}\omega \hat{b}^\dag \right)\\
& = \sum_{\q>0} \left( 2\hat{b}^\dag \omega \hat{b} - \text{tr}[\omega] \right).
\end{aligned}
\end{equation}
where $\omega_\q$ is a $n\times n$ diagonal positive semi-definite matrix for each $\q$. From \eqref{eq: mom diag} we find the time dependence of $b_i(t)$ via $i \partial_t b_i(t) = [b_i,\hat{H}]$ which gives
\begin{equation}\label{eq: fermion t dependence}
\hat{b}_i(t) = \hat{b}_i e^{-2i\omega_i t}.
\end{equation}

The Hamiltonian \eqref{eq: mom diag} is in diagonal form and our goal now is to find a transformation between $\Phi$ and $\beta$. To do this we first symmetrize \eqref{eq: block H} and write the Hamiltonian as
\begin{equation}\label{eq: H symmetrized}
\hat{H} = \frac{1}{2}\sum_{\text{all }\q} \left(\begin{array}{cc}\Phi_\q^\dag & \Phi^\dag_{-\q}\end{array}\right) \left(\begin{array}{cc}\tilde{H}_\q & 0 \\0 & \tilde{H}_{-\q}\end{array}\right) \left(\begin{array}{c}\Phi_\q \\\Phi_{-\q}\end{array}\right),
\end{equation}
where the sum is now over the entire Brillouin zone. As mentioned above we are now double counting the fermions, but since have already acquired the correct time dependence this will cause us no further problems.

We can then diagonalise the Hamiltonian by diagonalising the sub-matrices $\tilde{H}_\q$ and $\tilde{H}_{-\q}$:
\begin{equation}\label{eq: diagonalising U}
\begin{aligned}
\tilde{H}_\q &= U_+ \Lambda U_+^\dag \\
\tilde{H}_{-\q} &= U_- (-C\Lambda C) U^\dag_-
\end{aligned}
\end{equation}
where $\Lambda$ has the eigenvalues ascending on its diagonal and $C$ is the matrix with ones along the diagonal from bottom left to top right. We write the diagonalization in this way so that both $\Lambda$ and $(-C\Lambda C)$ are in ascending order.
We can then write the Hamiltonian in a form similar to \eqref{eq: mom diag}:
\begin{widetext}
\begin{equation}
\hat{H} = \frac{1}{2} \sum_{\text{all }\mathbf{q}} \left(\begin{array}{cc}\Phi_\q^\dag & \Phi^\dag_{-\q}\end{array}\right) \left(\begin{array}{cc}U_+ & 0 \\0 & U_- C \end{array}\right) \left(\begin{array}{cc}\Lambda & 0 \\0 & -\Lambda \end{array}\right) \left(\begin{array}{cc}U^\dag_+ & 0 \\0 & C U^\dag_-  \end{array}\right)\left(\begin{array}{c}\Phi_\q \\\Phi_{-\q}\end{array}\right).
\end{equation}
\end{widetext}
Although this is now in diagonal form, the matrices $\Lambda$ are not positive semi-definite and thus we need an extra rotation $\beta \rightarrow \tilde{\beta}$ to relate $\Phi$ and $\beta$ with the help of which we can make the identification
\begin{equation}\label{eq: identification}
\left(\begin{array}{c}\Phi_\q \\\Phi_{-\q}\end{array}\right) = \left(\begin{array}{cc}U_+ & 0 \\0 & U_- C \end{array}\right) \tilde{\beta}.
\end{equation}

\subsection{Calculating $U_+$ and $U_-$}

To calculate the matrices $U_+$ and $U_-$ we use the definition $\tilde{H}_\q = \Gamma^\dag H_\q \Gamma$ and the diagonalised form of $H_\q$ to get
\begin{equation}
\begin{aligned}
\tilde{H}_\q &= \Gamma^\dag \left(P \frac{1}{2}\Lambda P^\dag\right) \Gamma\\
&= \left( \frac{1}{\sqrt{2}}\Gamma^\dag P \right) \Lambda \left( \frac{1}{\sqrt{2}}\Gamma^\dag P \right)^\dag.
\end{aligned}
\end{equation}
We write the diagonal matrix $\frac{1}{2}\Lambda$ in this form, with the factor of a half, because the matrices $\frac{1}{\sqrt{2}}\Gamma$ are unitary and thus we can make the identification
\begin{equation}
U_+ = \frac{1}{\sqrt{2}}\Gamma^\dag P.
\end{equation}
Since we have symmetrized the Hamiltonian we only need to calculate $U_+$. For completeness, the corresponding calculation for $U_-$ uses $H_{-\q} = -H_\q^*$ to give
\begin{equation}
\begin{aligned}
\tilde{H}_{-\q} &= -\Gamma^\dag \left(P \frac{1}{2}\Lambda P^\dag\right)^* \Gamma\\
&= \left( \frac{1}{\sqrt{2}}\Gamma^\dag P^* C \right) (-C\Lambda C) \left( \frac{1}{\sqrt{2}}\Gamma^\dag P^* C \right)^\dag.
\end{aligned}
\end{equation}
and thus we make the identification
\begin{equation}
U_- = \frac{1}{\sqrt{2}}\Gamma_{-\q}^\dag P^*C.
\end{equation}
Here we make explicit the momentum dependence in the $\Gamma$-matrix for clarity.

\section{Expression for spin correlators}\label{ap: spin correlators}

As was suggested Baskaran et.al.~\cite{BaskaranExact} the calculation of spin correlators in the Kitaev model can be mapped onto a quantum quench problem. This is one of the key steps that allows us to make use of machinery developed in the context of the X-ray edge problem~\cite{Nozieres}. Below we outline the steps in re-expressing and solving the problem as was originally done in 2D for the hyperhoneycomb by some of the authors~\cite{JohannesThesis,PRL}.

This mapping relies on the static nature of the $\mathbb{Z}_2$ gauge field and is facilitated by the definition of complex bond fermions
\begin{equation}
\hat{\chi}^{\phantom{0}}_{\la j k \ra_a} = \frac{1}{2}(\hat{b}^a_j + i \hat{b}^a_k),
\end{equation}
where we enforce $j < k$. In terms of these fermions the bond operators represent the occupation numbers for bond fermions $\hat{u}_{jk} = 2\hat{\chi}^{\dag}_{\la j k \ra_a}\hat{\chi}^{\phantom{0}}_{\la j k \ra_a}-1$. Two spin operators on the bond $jk$ can be then expressed as
\begin{equation}
\hat{\sigma}^a_j = i ( \hat{\chi}^{\phantom{0}}_{\la j k \ra_a} + \hat{\chi}^{\dag}_{\la j k \ra_a}) \hat{c}_j,  \ \hat{\sigma}^a_k =( \hat{\chi}^{\phantom{0}}_{\la jk \ra_a} - \hat{\chi}^{\dag}_{\la jk \ra_a}) \hat{c}_k.
\end{equation}
One can see that the effect of a spin operator is to flip the direction of the bond it is associated with, which in turn changes the flux through the adjacent loops. Figure~\ref{fig: loops} shows examples of the fluxes that are changed by flipping a single bond for our three lattices. To get back to the ground state flux sector with a single bond flip we must flip back the same bond meaning our correlators are ultra-short ranged.

By inverting the relationship between bond fermions and bond operators we are able to remove the bond fermions from the expression of spin correlators for a given gauge~\cite{BaskaranExact}, leaving them in the gauge invariant form
\begin{equation}\label{eq: S in G}
S^{aa}_{jk}(t)  = \left\{\begin{array}{cc} -i u_{jk} \la M_0 | e^{i\hat{H}_0t} \hat{c}_j e^{-i(\hat{H}_0 + \hat{V}_{jk})t} \hat{c}_k | M_0 \ra, & \la j k \ra \\ \la M_0 |\; e^{i\hat{H}_0t} \hat{c}_j e^{-i(\hat{H}_0 + \hat{V}_{jk})t} \hat{c}_j \; | M_0 \ra, & j=k,\end{array}\right.
\end{equation}
where $\hat{V}_{jk} = -i u_{jk} J_{a_{jk}} \hat{c}_j \hat{c}_k$ and $\hat{H}_0$ is the Hamiltonian \eqref{eq: H Majorana} for the hopping Majorana fermions with gauge $\{u_{jk}\}$.

\subsection{Expression in terms of Green Functions}\label{ap: spin correlators GF}

To be able to use the integral equation approach described in~\cite{PRL,Smith} we need to re-express the spin correlators in terms of Green's functions for complex fermions. Here, the first step is to use the interaction representation and put the `free' time dependence into the fermions, i.e.
\begin{equation}
\begin{aligned}
S^{aa}_{jk}(t) &\propto \la M_0 |\;  \hat{c}_j(t) e^{i\hat{H}_0t} e^{-i(\hat{H}_0 + \hat{V}_{jk})t} \hat{c}_k(0) \; | M_0 \ra\\
&= \la M_0 |\;  \hat{c}_j(t)  \hat{c}_k(0) \hat{S}(t,0) \; | M_0 \ra
\end{aligned}
\end{equation}
where the S-matrix is defined as
\begin{equation}
\hat{S}(t,0) = e^{i\hat{H}_0t} e^{-i(\hat{H}_0 + \hat{V}_{jk})t} = \mathbb{T}\exp\left\{ -\int_0^t d\tau \; \hat{V}_{jk}(\tau) \right\}
\end{equation}
due to quench potential $\hat{V}_{jk}(\tau) = -i u_{jk} J_a \hat{c}_j(\tau) \hat{c}_k(\tau)$.

In order to express these correlators in terms of fermionic Green's functions we combine Majorana `matter' fermions into complex fermions. We define the complex fermion along the bond involved in the spin correlator, e.g. for the j-k bond we define
\begin{equation}
\hat{f} = \frac{1}{2}(\hat{c}_j + i\hat{c}_k).
\end{equation}
Note that one has to define complex fermions for each type of correlator. The way to deal with this is explained in Appendix~\ref{ap: H momentum diagonalisation}.

The scattering potential can be written in terms of complex fermions as
\begin{equation}
\hat{V}_{jk}(t) = -2u_{jk}J_a \left[ \hat{f}^\dag(t) \hat{f}(t) - \frac{1}{2} \right].
\end{equation}
Now we can express nearest neighbour correlators in terms of complex fermions as 
\begin{widetext}
\begin{equation}
\begin{aligned}
\la 0 | \; \hat{\sigma}^a_j(t) \hat{\sigma}^a_k (0) \;| 0 \ra &= u_{jk} \left[ \la M_0 | \; \hat{f}(t) \hat{f}^\dag(0) \hat{S}(t,0) \;|M_0\ra - \la M_0 | \; \hat{f}^\dag(t) \hat{f}(0) \hat{S}(t,0) \;|M_0\ra \right]\\
&= u_{jk} \left[ \la M_0 | \; \mathbb{T} \hat{f}(t) \hat{f}^\dag(0) \hat{S}(t,0) \;|M_0\ra + \la M_0 | \; \mathbb{T} \hat{f}(0) \hat{f}^\dag(t) \hat{S}(t,0) \;|M_0\ra \right],
\end{aligned}
\end{equation}
\end{widetext}
where $\mathbb{T}$ denotes time-ordering. One would generally expect to also have contributions from anomalous Green's functions $-i\la M_0 | \; \mathbb{T} \hat{f}(t) \hat{f}(0) \hat{S}(t,0) \;|M_0\ra$. However, our definition of the complex matter fermions above ensures that the anomalous contributions always vanish.

In terms of Green functions
\begin{equation}\label{eq: GF def}
\begin{aligned}
G(t,0) &= -i \la M_0 | \; \mathbb{T} \hat{f}(t) \hat{f}^\dag(0) \hat{S}(t,0) \;|M_0\ra,\\
 G^{neg}(0,t) &= -i \la M_0 | \; \mathbb{T} \hat{f}(0) \hat{f}^\dag(t) \hat{S}(t,0) \;|M_0\ra,
\end{aligned}
\end{equation}
these nearest-neighbour spin correlators can be written as
\begin{equation}\label{eq: spin GF nn}
S^{aa}_{jk}(t) = i u_{jk} \left[ G(t,0) + G^{neg}(0,t) \right] = S^{aa}_{kj}(t).
\end{equation}
Similarly for the same-site correlators we obtain
\begin{equation}\label{eq: spin GF ss}
S^{aa}_{jj}(t) = i\left[ G(t,0) - G^{neg}(0,t) \right] = S^{aa}_{kk}(t).
\end{equation}

Equations \eqref{eq: spin GF nn} and \eqref{eq: spin GF ss} reveal an interesting effect of the gauge transformation. If we change the gauge then $u_{jk} \rightarrow -u_{jk}$ for some bond in the lattice. If we remain in the same flux sector then this cannot change the correlation functions. We can thus see that this gauge transformation has the effect of interchanging the roles of the positive and negative time Green functions (more precisely, $G(t,0) \rightarrow -G^{neg}(0,t)$). This boils down to a change in the bare GF $G_0(\omega) \rightarrow -G_0(-\omega)$ and a change in the scattering potential $\hat{V}_{jk}(t) \rightarrow -\hat{V}_{jk}(t)$. Once again we see that the dynamics of the Majorana fermions depends significantly on the choice of gauge but the gauge invariant expressions \eqref{eq: spin GF nn} and \eqref{eq: spin GF ss} compensate in just the right way to keep physical quantities invariant.

\begin{figure*}[t!]
\centering
\subfigimg[width=0.32\textwidth]{\large  \hspace*{-0pt} (a)}{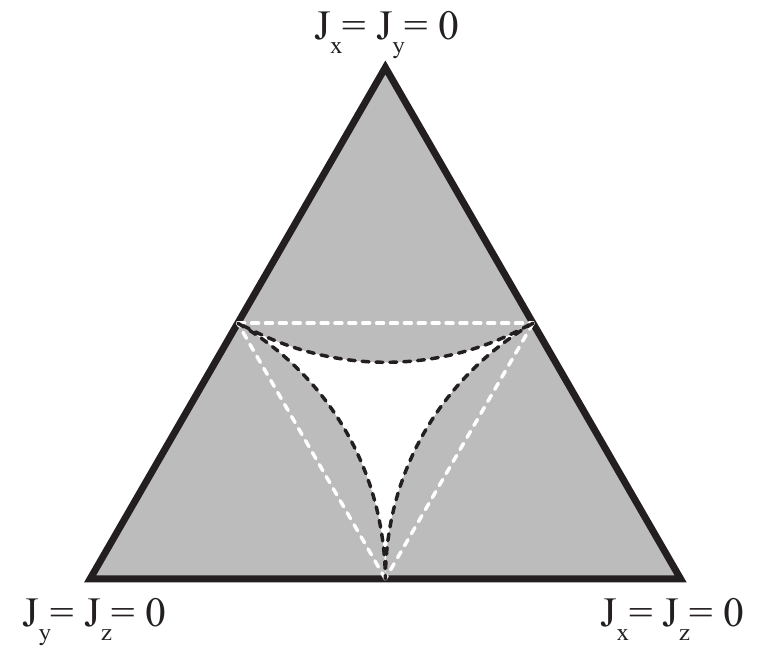}
\subfigimg[width=0.32\textwidth]{\large \hspace*{-0pt} (b)}{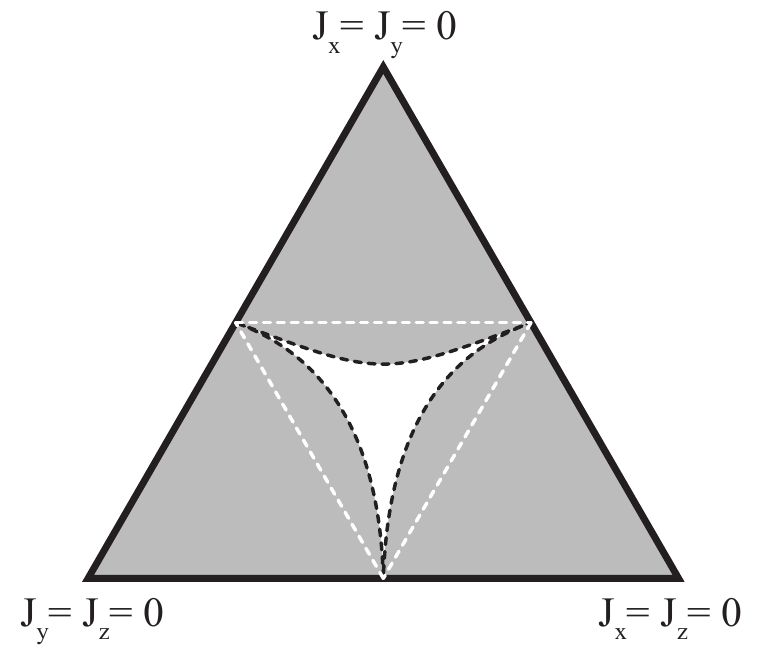}
\subfigimg[width=0.32\textwidth]{\large \hspace*{-0pt} (c)}{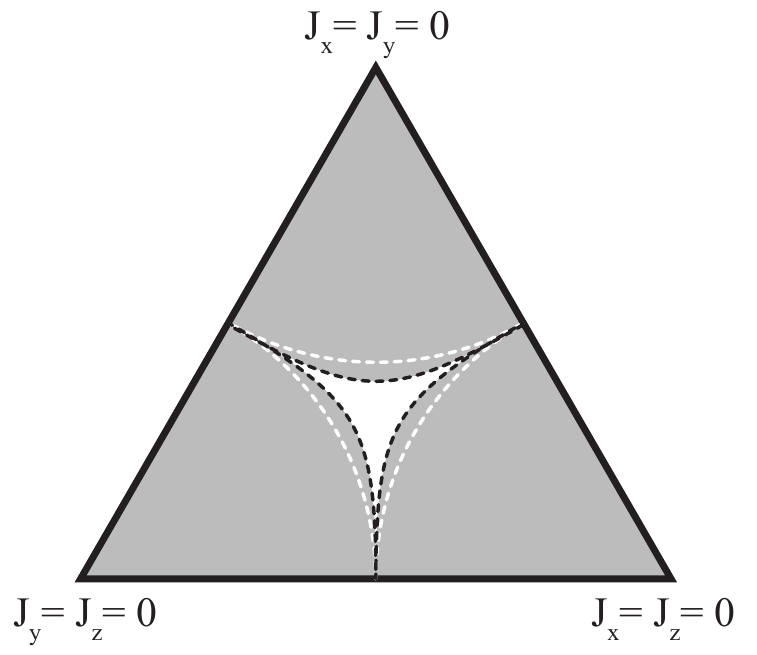}
\caption{Dynamical phase diagram for our three lattices~\cite{Knolle2015}. The response in the shaded region has a delta function contribution whereas the central region does not. The boundary of the static phase diagram is indicated by a dashed white line.}\label{fig: dynamical phase diagram}
\end{figure*}

We are now in a position to write out the full dynamical structure factor in terms of Green functions. Recall the definition of the structure factor:
\begin{equation}
S(\mathbf{q},\omega) = \frac{1}{N} \int^\infty_{-\infty} dt\; e^{i\omega t} \sum_{a,b} \sum_{j,k} e^{-i\mathbf{q}\cdot\mathbf{r}_{jk}} S^{aa}_{jk}(t),
\end{equation}
where $\mathbf{r}_{jk} = \mathbf{r}_j - \mathbf{r}_k$ are the vectors between neighbouring lattice sites. Let us first consider the summation of terms for a single bond. For nearest neighbours we have
\begin{equation}
e^{-i\mathbf{q}\cdot\mathbf{r}_{jk}} S^{aa}_{jk} (t) + e^{-i\mathbf{q}\cdot\mathbf{r}_{kj}} S^{aa}_{kj} (t) = 2 \cos(\mathbf{q}\cdot\mathbf{r}_{jk}) S^{aa}_{jk}(t),
\end{equation}
and for the same site correlators we get
\begin{equation}
S^{aa}_{jj}(t) + S^{aa}_{kk}(t) = 2 S^{aa}_{jj}(t).
\end{equation}
We can then sum over all lattice sites to get
\begin{widetext}
\begin{equation}
\begin{aligned}
S(\mathbf{q},\omega) &=  \int^\infty_{-\infty} dt\; e^{i\omega t} \sum_{\substack{\text{bonds in}\\ \text{primitive cell}}} 2 \cos(\mathbf{q}\cdot\mathbf{r}_{jk}) S^{aa}_{jk}(t) +  2 S^{aa}_{jj}(t)\\
&= \sum_{\substack{\text{bonds in}\\ \text{primitive cell}}} 2 i \Big\{ \big[u_{jk}\cos(\mathbf{q}\cdot\mathbf{r}_{jk}) + 1 \big] G(\omega) + \big[u_{jk}\cos(\mathbf{q}\cdot\mathbf{r}_{jk}) - 1 \big] G^{\textrm{neg}}(\omega) \Big\},
\end{aligned}
\end{equation}
\end{widetext}
where
\begin{equation}
G(\omega) = \int^\infty_{-\infty} dt\; e^{i\omega t} G(t,0),
\end{equation}
and
\begin{equation}
G^{\textrm{neg}}(\omega) = \int^\infty_{-\infty} dt\; e^{i\omega t} G^{neg}(0,t).
\end{equation}

\subsection{Dynamical Phase Diagram}\label{ap: dynamical phase diagram}

As discussed in Refs.~\cite{PRL,Knolle2015}, by looking at the Lehmann representation of the spin correlators we find that we have either only an  odd number of excitations in this expansion, or only an even number. Whether we have an odd or even number is determined by the relative parity of the ground state $|M_0 \ra$ of the Hamiltonian $\hat{H}$ in Eq. \eqref{eq: H Majorana} with fixed $u_{jk}$ and $| M_F \ra$ of the Hamiltonian with a single bond flipped, i.e. the one involved in the spin correlator. The overlap $\la M_F | M_0 \ra$ is zero if the states are of opposite parity and non-zero otherwise which allows us to determine the relative parities numerically.

If a spin correlator can be written in terms of only even numbers of excitations the Lehmann expansion includes a `zero particle' term which corresponds to a delta function contribution to the structure factor. Figure~\ref{fig: dynamical phase diagram} shows the dynamical phase diagram for our three lattices. The shaded regions are those where there exists a correlator that has a delta-function contribution, and in the central unshaded region all correlators have only odd numbers of excitations, which is the case we study in this paper.

For the hyperoctagon and hyperhexagon lattices we find that the dynamical phase diagram is symmetric in $J_a$ due to lattice symmetry as discussed in Ref.~\cite{HermannsZoo}. While all bonds are equivalent for the hyperoctagon lattice, which leads to a purely radial response, the hyperhexagon has two distinct type of bond and thus the more complicated momentum dependence that we have observed. In contrast, for the hyperhoneycomb lattice we find that while the x- and y-bonds are equivalent, the z-bonds are distinct (the two z-bonds are still related to each other by symmetry). This in turn leads to asymmetries in the dynamical phase diagram.
Note that for all three lattices the boundaries of the dynamical and the static phase diagrams (see Fig.~\ref{fig: phase diagram}) are different, and in all these cases there exists a gapless spin-liquid with a delta function contribution to the DSF

\subsection{The Exact Integral Equation Solution}

The calculation of Green functions, and hence the DSF, is simplified by the fact that they can be split into connected and loop contributions $G(t,0) = G_c(t,0)L(t,0)$~\cite{Abrikosov}. The connected GFs then satisfy the Dyson equation
\begin{equation}\label{eq: Gc pos}
G_c(t,t') = G_0(t,t') - 2J_a u_{jk} \int^t_{t'} d\tau G_0(t,\tau) G_c(\tau,t'),
\end{equation}
and similarly for the negative times
\begin{equation}\label{eq: Gc neg}
G^{neg}_c(t',t) = G_0(t',t) - 2J_a u_{jk} \int^t_{t'} d\tau G_0(\tau,t) G^{neg}_c(t',\tau).
\end{equation}
The loop contributions are given by
\begin{equation}\label{eq: L definition}
L(t,0) = \left\la \mathbb{T} \exp\left\{-i\int^t_0 dt' V(t') \right\} \right\ra.
\end{equation} 
In the above equations $G_0(t,t')$ is the bare GF
\begin{equation}
G_0(t,t') = -i \big\la \mathbb{T} f(t) f^\dag(t') \big\ra,
\end{equation}
which are calculated for a large, but finite lattice as shown in Appendix~\ref{ap: bare GF}.

Equations \eqref{eq: Gc pos} and \eqref{eq: Gc neg} are singular Fredholm integral equations of the second kind. These were solved numerically exactly in the context of the honeycomb Kitaev model in Ref.~\cite{JohannesThesis,Smith,PRL}. In their current form they are not suitable for numeric solution. The transformation we perform to render them numerically tractable can be summarised in two main steps:
\begin{itemize}
\item Using the analytic properties of the bare GF $G_0(t-\tau)$ and introducing normalised GF $\varphi^{(neg)} = G^{(neg)}(\omega,t)/G_0(\omega)$ the integral equations can be restated in the form
\begin{equation}\label{eq: K IES}
\hat{K}_2 \varphi^{(neg)} = f
\end{equation}
where $\hat{K}_2$ is a singular integral operator with Cauchy-type kernel. Importantly the kernel of these integral operators have finite support due to a factor of $\Im m[G^A_0(\omega)]$ which is proportional to the finite-bandwidth Majorana DOS.
\item If we then apply a second integral operator $\hat{K}_1$ to this equation it is possible, following the general prescription of Muskhelishvili~\cite{muskhelishvili2013singular}, to choose $\hat{K}_1$ such that
\begin{equation}
\hat{K}_1\hat{K}_2 \varphi^{(neg)} = \hat{K}_1f
\end{equation}
is non-singular and has the same solution as \eqref{eq: K IES}. We are then left with a non-singular integral equation with finite support that can be solved numerically.
\end{itemize}
The integral equation resulting from this procedure is presented in the supplementary material of Ref.~\cite{PRL} and is derived in detail in the appendix of Ref.~\cite{JohannesThesis}.

\subsection{Sum Rules}

As a check of our DSF calculations we have at our disposal the sum rules
\begin{equation}
S_{jk}^{aa}(t=0) = \frac{1}{2\pi} \int^\infty_{-\infty} d\omega\; S_{jk}^{aa}(\omega).
\end{equation}
The left hand side of this equation is simply the zero time spin correlator
\begin{equation}
\la 0| \hat{\sigma}^a_j (0) \hat{\sigma}^b_k(0) | 0 \ra = \la M_0 | \hat{c}_j \hat{c}_k | M_0 \ra,
\end{equation}
which can be calculated exactly with no non-equilibrium complications. In fact one can immediately see that same site correlators $S_{jj}^{aa}(t=0) = 1$ due to property that $\hat{c}_j^2 = 1$ for Majorana fermions.

Since we can express our spin correlators in terms of Green's functions the sum rules can also used to check the GF directly. Therefore, we must have that at $t=0$
\begin{equation}
\begin{aligned}
G(t=0,0) &= -i(u_{jk}S_{jk}^{aa}(t=0) +1),\\ G^{neg}(0,t=0) &= -i(u_{jk}S_{jk}^{aa}(t=0) - 1).
\end{aligned}
\end{equation}
By Fourier transform we also have that
\begin{equation}
\begin{aligned}
\int^\infty_{-\infty} d\omega\; G(\omega) &= -2\pi i(u_{jk}S_{jk}^{aa}(t=0) +1),\\
\int^\infty_{-\infty} d\omega\; G^{neg}(\omega) &=-2\pi i(u_{jk}S_{jk}^{aa}(t=0) - 1).
\end{aligned}
\end{equation}

We find from these sum rules a maximum error in our computations of $\sim 1\%$ across the three lattices.

\section{Calculating the Bare Green Functions}\label{ap: bare GF}

Having expressed our spin correlators in terms of fermionic Green functions and being equipped to solve the corresponding Dyson equations, all that is left to calculate is the bare GF $G_0(t)$:
\begin{equation}
\begin{aligned}
&-i \la \mathbb{T} \hat{f}(t) \hat{f}^\dag(0) \ra \\
&= -i \left[ \Theta(t) \la \hat{f}(t)\hat{f}^\dag(0) \ra - \Theta(-t) \la \hat{f}^\dag(0)\hat{f}(t) \ra \right]
\end{aligned}
\end{equation}
where $\hat{f}$ is the relevant complex fermion for the bond. By translational invariance we can just consider bonds in the primitive cell at $\mathbf{r} = 0$. Thus the bare GF can be written as
\begin{equation}
-\frac{i}{N} \sum_{\q \in B.Z.} \left[ \Theta(t) \la \hat{f}_\q(t)\hat{f}_{\q}^\dag(0) \ra - \Theta(-t) \la \hat{f}_{\q}^\dag(0)\hat{f}_\q(t) \ra \right].
\end{equation}
Note that the summation is over one momentum variable $\mathbf{q}$ since the correlators are zero unlesss they are over the same momenta.

We will consider the case of a 4$\times$4 momentum space matrix, which is the case of the hyperhoneycomb and hyperoctagon but it can be extended obviously to the Weyl case where instead of two species of complex fermion ($\hat{f}$ and $\hat{g}$) we end up with three ($\hat{f}$, $\hat{g}$ and $\hat{h}$).

Let us consider first $\Lambda = \text{diag}(\lambda_1,\lambda_2,\lambda_3,\lambda_4)$ with $\lambda_1\leq\lambda_2 \leq 0 \leq \lambda_3 \leq\lambda_4$. This is the case for the hyperhoneycomb where we further have that $\lambda_1=-\lambda_4$ and $\lambda_2 = -\lambda_3$.

For each $\q$ the complex fermions $\hat{f}$ and $\hat{g}$ can be related to the diagonalizing fermions $\hat{b}_i$'s by
\begin{equation}
\left( \begin{array}{c} \hat{f}_\q \\ \hat{f}^\dag_{-\q} \\ \hat{g}_\q \\ \hat{g}^\dag_{-\q} \end{array} \right) = U_+ \left( \begin{array}{c} \hat{b}^\dag_{1,\q} \\ \hat{b}^\dag_{2,\q}\\ \hat{b}_{3,\q} \\ \hat{b}_{4,\q}\end{array} \right).
\end{equation}
Dropping the $+$ symbol on the $U$ matrix $\hat{f}_{\q}$ and $\hat{f}^\dag_{\q}$ can be written out as
\begin{subequations}\label{eq: f to b}
\begin{align}
\hat{f}_\q &= U_{11} \hat{b}^\dag_{1,\q} + U_{12} \hat{b}^\dag_{2,\q}  + U_{13} \hat{b}_{3,\q}  + U_{14} \hat{b}_{4,\q} \label{eq: f1} \\
\hat{f}^\dag_\q &= U_{11}^* \hat{b}_{1,\q} + U_{12}^* \hat{b}_{2,\q}  + U_{13}^* \hat{b}^\dag_{3,\q}  + U_{14}^* \hat{b}^\dag_{4,\q} \label{eq: f3}
\end{align}
\end{subequations}
where by using the symmetrized Hamiltonian \eqref{eq: H symmetrized} these hold for all $\q \in B.Z.$.
As shown in \eqref{eq: fermion t dependence} the $\hat{b}$ fermions have the time dependence
\begin{equation}\label{eq: b t dependence}
\hat{b}_{i,\q}(t) = \hat{b}_{i,\q}e^{-2i|\lambda_i| t}
\end{equation}
Plugging \eqref{eq: f to b} and \eqref{eq: b t dependence} into the $\la f_\q(t) f^\dag_\q(0) \ra$ we get
\begin{equation}
\la f_\q(t) f^\dag_\q(0) \ra = |U_{13}|^2 e^{-2i|\lambda_3| t} + |U_{14}|^2 e^{-2i|\lambda_4| t},
\end{equation}
and similarly,
\begin{equation}
\la f^\dag_\q(0) f_\q(t) \ra = |U_{11}|^2 e^{2i|\lambda_1| t} + |U_{12}|^2 e^{2i|\lambda_2| t},
\end{equation}
Moving to frequency space we have
\begin{widetext}
\begin{equation}\label{eq: HHC full bare GF}
\begin{aligned}
G_0(\omega) &= -i \int^\infty_0 dt e^{i(\omega +i\delta)t} \la \hat{f}(t) f^\dag(0) \ra + i \int^0_{-\infty} dt e^{i(\omega -i\delta)t} \la \hat{f}^\dag(0) f(t) \ra\\
& = \frac{1}{N} \sum_{\q \in B.Z.}\left[ \frac{|U_{11}|^2}{\omega + 2|\lambda_1| -i\delta} + \frac{|U_{12}|^2}{\omega  + 2|\lambda_2| - i\delta} + \frac{|U_{13}|^2}{\omega - 2|\lambda_3| + i\delta} + \frac{|U_{14}|^2}{\omega - 2|\lambda_4| +i\delta} \right],
\end{aligned}
\end{equation}
\end{widetext}
where the $U$ matrix elements in the sum have an implicit $\q$ dependence and there is an implicit limit $\delta \rightarrow 0$. This GF is the bare GF for the bond associated with $\hat{f}$. To get those associated with $\hat{g}$ (or $\hat{h}$) we simply exchange $U_{1,i} \rightarrow U_{3,i}$ (or $U_{1,i} \rightarrow U_{5,i}$).

The hyperoctagon case is slightly different because $\Lambda$ does not have the same form (with $\pm \lambda_{1/2}$) and generally consists of four distinct eigenvalues $\Lambda = \text{diag}(\lambda_1,\lambda_2,\lambda_3,\lambda_4)$ with $\lambda_1\leq\lambda_2 \leq \lambda_3 \leq\lambda_4$. To see how this affects things let us consider the case of three negative eigenvalues $\lambda_1,\lambda_2,\lambda_3$ and one positive $\lambda_4$. In that case we have
\begin{subequations}\label{eq: f to b HO}
\begin{align}
\hat{f}_\q &= U_{11} \hat{b}^\dag_{1,\q} + U_{12} \hat{b}^\dag_{2,\q}  + U_{13} \hat{b}^\dag_{3,\q}  + U_{14} \hat{b}_{4,\q} \label{eq: f1 HO} \\
\hat{f}^\dag_\q &= U_{11}^* \hat{b}_{1,\q} + U_{12}^* \hat{b}_{2,\q}  + U_{13}^* \hat{b}_{3,\q}  + U_{14}^* \hat{b}^\dag_{4,\q}.\label{eq: f3 HO}
\end{align}
\end{subequations}
Using \eqref{eq: f to b HO} and \eqref{eq: b t dependence} we get
\begin{equation}\label{eq: f correlator1 HO}
\la f_\q(t) f^\dag_\q(0) \ra =|U_{14}|^2 e^{-2i|\lambda_4| t},
\end{equation}
and
\begin{equation}\label{eq: f correlator2 HO}
\la f^\dag_\q(0) f_\q(t) \ra = |U_{11}|^2 e^{2i|\lambda_1| t} + |U_{12}|^2 e^{2i|\lambda_2| t} + |U_{13}|^2 e^{2i|\lambda_3| t}.
\end{equation}
Following the same steps as for obtaining \eqref{eq: HHC full bare GF} we find that in frequency space the bare GF can generally be written as
\begin{equation}
G_0(\omega) = \frac{1}{N} \sum_{\q \in B.Z.}\sum_j \frac{|U_{1j}|^2}{\omega - 2\lambda_j +i\delta\,\text{sign}(\lambda_j)}
\end{equation}
where $\Lambda = \text{diag}(\lambda_1,\lambda_2,\cdots)$ and $\lambda_1 \leq \lambda_2 \leq \cdots$. This is now a completely general expression for the bare GFs and applies to all three lattices and for any number of positive/negative signs in $\Lambda$.

\bibliography{references}

\begin{thebibliography}{35}%
\makeatletter
\providecommand \@ifxundefined [1]{%
 \@ifx{#1\undefined}
}%
\providecommand \@ifnum [1]{%
 \ifnum #1\expandafter \@firstoftwo
 \else \expandafter \@secondoftwo
 \fi
}%
\providecommand \@ifx [1]{%
 \ifx #1\expandafter \@firstoftwo
 \else \expandafter \@secondoftwo
 \fi
}%
\providecommand \natexlab [1]{#1}%
\providecommand \enquote  [1]{``#1''}%
\providecommand \bibnamefont  [1]{#1}%
\providecommand \bibfnamefont [1]{#1}%
\providecommand \citenamefont [1]{#1}%
\providecommand \href@noop [0]{\@secondoftwo}%
\providecommand \href [0]{\begingroup \@sanitize@url \@href}%
\providecommand \@href[1]{\@@startlink{#1}\@@href}%
\providecommand \@@href[1]{\endgroup#1\@@endlink}%
\providecommand \@sanitize@url [0]{\catcode `\\12\catcode `\$12\catcode
  `\&12\catcode `\#12\catcode `\^12\catcode `\_12\catcode `\%12\relax}%
\providecommand \@@startlink[1]{}%
\providecommand \@@endlink[0]{}%
\providecommand \url  [0]{\begingroup\@sanitize@url \@url }%
\providecommand \@url [1]{\endgroup\@href {#1}{\urlprefix }}%
\providecommand \urlprefix  [0]{URL }%
\providecommand \Eprint [0]{\href }%
\providecommand \doibase [0]{http://dx.doi.org/}%
\providecommand \selectlanguage [0]{\@gobble}%
\providecommand \bibinfo  [0]{\@secondoftwo}%
\providecommand \bibfield  [0]{\@secondoftwo}%
\providecommand \translation [1]{[#1]}%
\providecommand \BibitemOpen [0]{}%
\providecommand \bibitemStop [0]{}%
\providecommand \bibitemNoStop [0]{.\EOS\space}%
\providecommand \EOS [0]{\spacefactor3000\relax}%
\providecommand \BibitemShut  [1]{\csname bibitem#1\endcsname}%
\let\auto@bib@innerbib\@empty
\bibitem [{\citenamefont {Kitaev}(2006)}]{Kitaev}%
  \BibitemOpen
  \bibfield  {author} {\bibinfo {author} {\bibfnamefont {A.~Y.}\ \bibnamefont
  {Kitaev}},\ }\href@noop {} {\bibfield  {journal} {\bibinfo  {journal} {Ann.
  Phys. (Amsterdam)}\ }\textbf {\bibinfo {volume} {321}},\ \bibinfo {pages} {2}
  (\bibinfo {year} {2006})}\BibitemShut {NoStop}%
\bibitem [{\citenamefont {Mandal}\ and\ \citenamefont
  {Surendran}(2009)}]{MandalKitaev}%
  \BibitemOpen
  \bibfield  {author} {\bibinfo {author} {\bibfnamefont {S.}~\bibnamefont
  {Mandal}}\ and\ \bibinfo {author} {\bibfnamefont {N.}~\bibnamefont
  {Surendran}},\ }\href {\doibase 10.1103/PhysRevB.79.024426} {\bibfield
  {journal} {\bibinfo  {journal} {Phys. Rev. B.}\ }\textbf {\bibinfo {volume}
  {79}},\ \bibinfo {pages} {024426} (\bibinfo {year} {2009})}\BibitemShut
  {NoStop}%
\bibitem [{\citenamefont {Kimchi}\ \emph {et~al.}(2014)\citenamefont {Kimchi},
  \citenamefont {Analytis},\ and\ \citenamefont {Vishwanath}}]{Kimchi}%
  \BibitemOpen
  \bibfield  {author} {\bibinfo {author} {\bibfnamefont {I.}~\bibnamefont
  {Kimchi}}, \bibinfo {author} {\bibfnamefont {J.~G.}\ \bibnamefont
  {Analytis}}, \ and\ \bibinfo {author} {\bibfnamefont {A.}~\bibnamefont
  {Vishwanath}},\ }\href {\doibase 10.1103/PhysRevB.90.205126} {\bibfield
  {journal} {\bibinfo  {journal} {Phys. Rev. B}\ }\textbf {\bibinfo {volume}
  {90}},\ \bibinfo {pages} {205126} (\bibinfo {year} {2014})}\BibitemShut
  {NoStop}%
\bibitem [{\citenamefont {O'Brien}\ \emph {et~al.}(2016)\citenamefont
  {O'Brien}, \citenamefont {Hermanns},\ and\ \citenamefont
  {Trebst}}]{HermannsZoo}%
  \BibitemOpen
  \bibfield  {author} {\bibinfo {author} {\bibfnamefont {K.}~\bibnamefont
  {O'Brien}}, \bibinfo {author} {\bibfnamefont {M.}~\bibnamefont {Hermanns}}, \
  and\ \bibinfo {author} {\bibfnamefont {S.}~\bibnamefont {Trebst}},\ }\href
  {\doibase 10.1103/PhysRevB.93.085101} {\bibfield  {journal} {\bibinfo
  {journal} {Phys. Rev. B}\ }\textbf {\bibinfo {volume} {93}},\ \bibinfo
  {pages} {085101} (\bibinfo {year} {2016})}\BibitemShut {NoStop}%
\bibitem [{\citenamefont {Jackeli}\ and\ \citenamefont
  {Khaliullin}(2009)}]{Jackeli}%
  \BibitemOpen
  \bibfield  {author} {\bibinfo {author} {\bibfnamefont {G.}~\bibnamefont
  {Jackeli}}\ and\ \bibinfo {author} {\bibfnamefont {G.}~\bibnamefont
  {Khaliullin}},\ }\href {\doibase 10.1103/PhysRevLett.102.017205} {\bibfield
  {journal} {\bibinfo  {journal} {Phys. Rev. Lett.}\ }\textbf {\bibinfo
  {volume} {102}},\ \bibinfo {pages} {017205} (\bibinfo {year}
  {2009})}\BibitemShut {NoStop}%
\bibitem [{\citenamefont {Singh}\ and\ \citenamefont
  {Gegenwart}(2010)}]{Singh2010}%
  \BibitemOpen
  \bibfield  {author} {\bibinfo {author} {\bibfnamefont {Y.}~\bibnamefont
  {Singh}}\ and\ \bibinfo {author} {\bibfnamefont {P.}~\bibnamefont
  {Gegenwart}},\ }\href {\doibase 10.1103/PhysRevB.82.064412} {\bibfield
  {journal} {\bibinfo  {journal} {Phys. Rev. B}\ }\textbf {\bibinfo {volume}
  {82}},\ \bibinfo {pages} {064412} (\bibinfo {year} {2010})}\BibitemShut
  {NoStop}%
\bibitem [{\citenamefont {Singh}\ \emph {et~al.}(2012)\citenamefont {Singh},
  \citenamefont {Manni}, \citenamefont {Reuther}, \citenamefont {Berlijn},
  \citenamefont {Thomale}, \citenamefont {Ku}, \citenamefont {Trebst},\ and\
  \citenamefont {Gegenwart}}]{Singh2012}%
  \BibitemOpen
  \bibfield  {author} {\bibinfo {author} {\bibfnamefont {Y.}~\bibnamefont
  {Singh}}, \bibinfo {author} {\bibfnamefont {S.}~\bibnamefont {Manni}},
  \bibinfo {author} {\bibfnamefont {J.}~\bibnamefont {Reuther}}, \bibinfo
  {author} {\bibfnamefont {T.}~\bibnamefont {Berlijn}}, \bibinfo {author}
  {\bibfnamefont {R.}~\bibnamefont {Thomale}}, \bibinfo {author} {\bibfnamefont
  {W.}~\bibnamefont {Ku}}, \bibinfo {author} {\bibfnamefont {S.}~\bibnamefont
  {Trebst}}, \ and\ \bibinfo {author} {\bibfnamefont {P.}~\bibnamefont
  {Gegenwart}},\ }\href {\doibase 10.1103/PhysRevLett.108.127203} {\bibfield
  {journal} {\bibinfo  {journal} {Phys. Rev. Lett.}\ }\textbf {\bibinfo
  {volume} {108}},\ \bibinfo {pages} {127203} (\bibinfo {year}
  {2012})}\BibitemShut {NoStop}%
\bibitem [{\citenamefont {Plumb}\ \emph {et~al.}(2014)\citenamefont {Plumb},
  \citenamefont {Clancy}, \citenamefont {Sandilands}, \citenamefont {Shankar},
  \citenamefont {Hu}, \citenamefont {Burch}, \citenamefont {Kee},\ and\
  \citenamefont {Kim}}]{Plumb2014}%
  \BibitemOpen
  \bibfield  {author} {\bibinfo {author} {\bibfnamefont {K.~W.}\ \bibnamefont
  {Plumb}}, \bibinfo {author} {\bibfnamefont {J.~P.}\ \bibnamefont {Clancy}},
  \bibinfo {author} {\bibfnamefont {L.~J.}\ \bibnamefont {Sandilands}},
  \bibinfo {author} {\bibfnamefont {V.~V.}\ \bibnamefont {Shankar}}, \bibinfo
  {author} {\bibfnamefont {Y.~F.}\ \bibnamefont {Hu}}, \bibinfo {author}
  {\bibfnamefont {K.~S.}\ \bibnamefont {Burch}}, \bibinfo {author}
  {\bibfnamefont {H.-Y.}\ \bibnamefont {Kee}}, \ and\ \bibinfo {author}
  {\bibfnamefont {Y.-J.}\ \bibnamefont {Kim}},\ }\href {\doibase
  10.1103/PhysRevB.90.041112} {\bibfield  {journal} {\bibinfo  {journal} {Phys.
  Rev. B}\ }\textbf {\bibinfo {volume} {90}},\ \bibinfo {pages} {041112}
  (\bibinfo {year} {2014})}\BibitemShut {NoStop}%
\bibitem [{\citenamefont {Sears}\ \emph {et~al.}(2015)\citenamefont {Sears},
  \citenamefont {Songvilay}, \citenamefont {Plumb}, \citenamefont {Clancy},
  \citenamefont {Qiu}, \citenamefont {Zhao}, \citenamefont {Parshall},\ and\
  \citenamefont {Kim}}]{Sears2015}%
  \BibitemOpen
  \bibfield  {author} {\bibinfo {author} {\bibfnamefont {J.~A.}\ \bibnamefont
  {Sears}}, \bibinfo {author} {\bibfnamefont {M.}~\bibnamefont {Songvilay}},
  \bibinfo {author} {\bibfnamefont {K.~W.}\ \bibnamefont {Plumb}}, \bibinfo
  {author} {\bibfnamefont {J.~P.}\ \bibnamefont {Clancy}}, \bibinfo {author}
  {\bibfnamefont {Y.}~\bibnamefont {Qiu}}, \bibinfo {author} {\bibfnamefont
  {Y.}~\bibnamefont {Zhao}}, \bibinfo {author} {\bibfnamefont {D.}~\bibnamefont
  {Parshall}}, \ and\ \bibinfo {author} {\bibfnamefont {Y.-J.}\ \bibnamefont
  {Kim}},\ }\href {\doibase 10.1103/PhysRevB.91.144420} {\bibfield  {journal}
  {\bibinfo  {journal} {Phys. Rev. B}\ }\textbf {\bibinfo {volume} {91}},\
  \bibinfo {pages} {144420} (\bibinfo {year} {2015})}\BibitemShut {NoStop}%
\bibitem [{\citenamefont {Majumder}\ \emph {et~al.}(2015)\citenamefont
  {Majumder}, \citenamefont {Schmidt}, \citenamefont {Rosner}, \citenamefont
  {Tsirlin}, \citenamefont {Yasuoka},\ and\ \citenamefont
  {Baenitz}}]{Majumdar2015}%
  \BibitemOpen
  \bibfield  {author} {\bibinfo {author} {\bibfnamefont {M.}~\bibnamefont
  {Majumder}}, \bibinfo {author} {\bibfnamefont {M.}~\bibnamefont {Schmidt}},
  \bibinfo {author} {\bibfnamefont {H.}~\bibnamefont {Rosner}}, \bibinfo
  {author} {\bibfnamefont {A.~A.}\ \bibnamefont {Tsirlin}}, \bibinfo {author}
  {\bibfnamefont {H.}~\bibnamefont {Yasuoka}}, \ and\ \bibinfo {author}
  {\bibfnamefont {M.}~\bibnamefont {Baenitz}},\ }\href {\doibase
  10.1103/PhysRevB.91.180401} {\bibfield  {journal} {\bibinfo  {journal} {Phys.
  Rev. B}\ }\textbf {\bibinfo {volume} {91}},\ \bibinfo {pages} {180401}
  (\bibinfo {year} {2015})}\BibitemShut {NoStop}%
\bibitem [{\citenamefont {Sandilands}\ \emph {et~al.}(2015)\citenamefont
  {Sandilands}, \citenamefont {Tian}, \citenamefont {Plumb}, \citenamefont
  {Kim},\ and\ \citenamefont {Burch}}]{Sandilands2015}%
  \BibitemOpen
  \bibfield  {author} {\bibinfo {author} {\bibfnamefont {L.~J.}\ \bibnamefont
  {Sandilands}}, \bibinfo {author} {\bibfnamefont {Y.}~\bibnamefont {Tian}},
  \bibinfo {author} {\bibfnamefont {K.~W.}\ \bibnamefont {Plumb}}, \bibinfo
  {author} {\bibfnamefont {Y.-J.}\ \bibnamefont {Kim}}, \ and\ \bibinfo
  {author} {\bibfnamefont {K.~S.}\ \bibnamefont {Burch}},\ }\href {\doibase
  10.1103/PhysRevLett.114.147201} {\bibfield  {journal} {\bibinfo  {journal}
  {Phys. Rev. Lett.}\ }\textbf {\bibinfo {volume} {114}},\ \bibinfo {pages}
  {147201} (\bibinfo {year} {2015})}\BibitemShut {NoStop}%
\bibitem [{\citenamefont {Banerjee}\ \emph {et~al.}(2016)\citenamefont
  {Banerjee}, \citenamefont {Bridges}, \citenamefont {Yan}, \citenamefont
  {Aczel}, \citenamefont {Li}, \citenamefont {Stone}, \citenamefont {Granroth},
  \citenamefont {Lumsden}, \citenamefont {Yiu}, \citenamefont {Knolle},
  \citenamefont {Bhattacharjee}, \citenamefont {Kovrizhin}, \citenamefont
  {Moessner}, \citenamefont {Tennant}, \citenamefont {Mandrus},\ and\
  \citenamefont {Nagler}}]{Banerjee2015}%
  \BibitemOpen
  \bibfield  {author} {\bibinfo {author} {\bibfnamefont {A.}~\bibnamefont
  {Banerjee}}, \bibinfo {author} {\bibfnamefont {C.~A.}\ \bibnamefont
  {Bridges}}, \bibinfo {author} {\bibfnamefont {J.~Q.}\ \bibnamefont {Yan}},
  \bibinfo {author} {\bibfnamefont {A.~A.}\ \bibnamefont {Aczel}}, \bibinfo
  {author} {\bibfnamefont {L.}~\bibnamefont {Li}}, \bibinfo {author}
  {\bibfnamefont {M.~B.}\ \bibnamefont {Stone}}, \bibinfo {author}
  {\bibfnamefont {G.~E.}\ \bibnamefont {Granroth}}, \bibinfo {author}
  {\bibfnamefont {M.~D.}\ \bibnamefont {Lumsden}}, \bibinfo {author}
  {\bibfnamefont {Y.}~\bibnamefont {Yiu}}, \bibinfo {author} {\bibfnamefont
  {J.}~\bibnamefont {Knolle}}, \bibinfo {author} {\bibfnamefont
  {S.}~\bibnamefont {Bhattacharjee}}, \bibinfo {author} {\bibfnamefont {D.~L.}\
  \bibnamefont {Kovrizhin}}, \bibinfo {author} {\bibfnamefont {R.}~\bibnamefont
  {Moessner}}, \bibinfo {author} {\bibfnamefont {D.~A.}\ \bibnamefont
  {Tennant}}, \bibinfo {author} {\bibfnamefont {D.~G.}\ \bibnamefont
  {Mandrus}}, \ and\ \bibinfo {author} {\bibfnamefont {S.~E.}\ \bibnamefont
  {Nagler}},\ }\href {http://dx.doi.org/10.1038/nmat4604} {\bibfield  {journal}
  {\bibinfo  {journal} {Nat Mater}\ }\textbf {\bibinfo {volume} {advance online
  publication}},\  (\bibinfo {year} {2016})}\BibitemShut {NoStop}%
\bibitem [{\citenamefont {Modic}\ \emph {et~al.}(2014)\citenamefont {Modic},
  \citenamefont {Smidt}, \citenamefont {Kimchi}, \citenamefont {Breznay},
  \citenamefont {Biffin}, \citenamefont {Choi}, \citenamefont {Johnson},
  \citenamefont {Coldea}, \citenamefont {Watkins-Curry}, \citenamefont
  {McCandless}, \citenamefont {Chan}, \citenamefont {Gandara}, \citenamefont
  {Islam}, \citenamefont {Vishwanath}, \citenamefont {Shekhter}, \citenamefont
  {McDonald},\ and\ \citenamefont {Analytis}}]{Modic}%
  \BibitemOpen
  \bibfield  {author} {\bibinfo {author} {\bibfnamefont {K.~A.}\ \bibnamefont
  {Modic}}, \bibinfo {author} {\bibfnamefont {T.~E.}\ \bibnamefont {Smidt}},
  \bibinfo {author} {\bibfnamefont {I.}~\bibnamefont {Kimchi}}, \bibinfo
  {author} {\bibfnamefont {N.~P.}\ \bibnamefont {Breznay}}, \bibinfo {author}
  {\bibfnamefont {A.}~\bibnamefont {Biffin}}, \bibinfo {author} {\bibfnamefont
  {S.}~\bibnamefont {Choi}}, \bibinfo {author} {\bibfnamefont {R.~D.}\
  \bibnamefont {Johnson}}, \bibinfo {author} {\bibfnamefont {R.}~\bibnamefont
  {Coldea}}, \bibinfo {author} {\bibfnamefont {P.}~\bibnamefont
  {Watkins-Curry}}, \bibinfo {author} {\bibfnamefont {G.~T.}\ \bibnamefont
  {McCandless}}, \bibinfo {author} {\bibfnamefont {J.~Y.}\ \bibnamefont
  {Chan}}, \bibinfo {author} {\bibfnamefont {F.}~\bibnamefont {Gandara}},
  \bibinfo {author} {\bibfnamefont {Z.}~\bibnamefont {Islam}}, \bibinfo
  {author} {\bibfnamefont {A.}~\bibnamefont {Vishwanath}}, \bibinfo {author}
  {\bibfnamefont {A.}~\bibnamefont {Shekhter}}, \bibinfo {author}
  {\bibfnamefont {R.~D.}\ \bibnamefont {McDonald}}, \ and\ \bibinfo {author}
  {\bibfnamefont {J.~G.}\ \bibnamefont {Analytis}},\ }\href
  {http://dx.doi.org/10.1038/ncomms5203} {\bibfield  {journal} {\bibinfo
  {journal} {Nat Commun}\ }\textbf {\bibinfo {volume} {5}} (\bibinfo {year}
  {2014})}\BibitemShut {NoStop}%
\bibitem [{\citenamefont {Takayama}\ \emph {et~al.}(2015)\citenamefont
  {Takayama}, \citenamefont {Kato}, \citenamefont {Dinnebier}, \citenamefont
  {Nuss}, \citenamefont {Kono}, \citenamefont {Veiga}, \citenamefont {Fabbris},
  \citenamefont {Haskel},\ and\ \citenamefont {Takagi}}]{Takayama}%
  \BibitemOpen
  \bibfield  {author} {\bibinfo {author} {\bibfnamefont {T.}~\bibnamefont
  {Takayama}}, \bibinfo {author} {\bibfnamefont {A.}~\bibnamefont {Kato}},
  \bibinfo {author} {\bibfnamefont {R.}~\bibnamefont {Dinnebier}}, \bibinfo
  {author} {\bibfnamefont {J.}~\bibnamefont {Nuss}}, \bibinfo {author}
  {\bibfnamefont {H.}~\bibnamefont {Kono}}, \bibinfo {author} {\bibfnamefont
  {L.~S.}\ \bibnamefont {Veiga}}, \bibinfo {author} {\bibfnamefont
  {G.}~\bibnamefont {Fabbris}}, \bibinfo {author} {\bibfnamefont
  {D.}~\bibnamefont {Haskel}}, \ and\ \bibinfo {author} {\bibfnamefont
  {H.}~\bibnamefont {Takagi}},\ }\href {\doibase
  10.1103/PhysRevLett.114.077202} {\bibfield  {journal} {\bibinfo  {journal}
  {Phys. Rev. Lett.}\ }\textbf {\bibinfo {volume} {114}},\ \bibinfo {pages}
  {077202} (\bibinfo {year} {2015})}\BibitemShut {NoStop}%
\bibitem [{\citenamefont {Kim}\ \emph {et~al.}(2015)\citenamefont {Kim},
  \citenamefont {Lee},\ and\ \citenamefont {Kim}}]{Kim}%
  \BibitemOpen
  \bibfield  {author} {\bibinfo {author} {\bibfnamefont {H.-S.}\ \bibnamefont
  {Kim}}, \bibinfo {author} {\bibfnamefont {E.~K.-H.}\ \bibnamefont {Lee}}, \
  and\ \bibinfo {author} {\bibfnamefont {Y.~B.}\ \bibnamefont {Kim}},\ }\href
  {http://stacks.iop.org/0295-5075/112/i=6/a=67004} {\bibfield  {journal}
  {\bibinfo  {journal} {EPL (Europhysics Letters)}\ }\textbf {\bibinfo {volume}
  {112}},\ \bibinfo {pages} {67004} (\bibinfo {year} {2015})}\BibitemShut
  {NoStop}%
\bibitem [{\citenamefont {{Yamaji}}\ \emph {et~al.}(2016)\citenamefont
  {{Yamaji}}, \citenamefont {{Suzuki}}, \citenamefont {{Yamada}}, \citenamefont
  {{Suga}}, \citenamefont {{Kawashima}},\ and\ \citenamefont
  {{Imada}}}]{Yamaji2016}%
  \BibitemOpen
  \bibfield  {author} {\bibinfo {author} {\bibfnamefont {Y.}~\bibnamefont
  {{Yamaji}}}, \bibinfo {author} {\bibfnamefont {T.}~\bibnamefont {{Suzuki}}},
  \bibinfo {author} {\bibfnamefont {T.}~\bibnamefont {{Yamada}}}, \bibinfo
  {author} {\bibfnamefont {S.-i.}\ \bibnamefont {{Suga}}}, \bibinfo {author}
  {\bibfnamefont {N.}~\bibnamefont {{Kawashima}}}, \ and\ \bibinfo {author}
  {\bibfnamefont {M.}~\bibnamefont {{Imada}}},\ }\href@noop {} {\bibfield
  {journal} {\bibinfo  {journal} {ArXiv e-prints}\ } (\bibinfo {year}
  {2016})},\ \Eprint {http://arxiv.org/abs/1601.05512} {arXiv:1601.05512
  [cond-mat.str-el]} \BibitemShut {NoStop}%
\bibitem [{\citenamefont {{Nasu}}\ \emph {et~al.}(2016)\citenamefont {{Nasu}},
  \citenamefont {{Knolle}}, \citenamefont {{Kovrizhin}}, \citenamefont
  {{Motome}},\ and\ \citenamefont {{Moessner}}}]{Nasu2016}%
  \BibitemOpen
  \bibfield  {author} {\bibinfo {author} {\bibfnamefont {J.}~\bibnamefont
  {{Nasu}}}, \bibinfo {author} {\bibfnamefont {J.}~\bibnamefont {{Knolle}}},
  \bibinfo {author} {\bibfnamefont {D.~L.}\ \bibnamefont {{Kovrizhin}}},
  \bibinfo {author} {\bibfnamefont {Y.}~\bibnamefont {{Motome}}}, \ and\
  \bibinfo {author} {\bibfnamefont {R.}~\bibnamefont {{Moessner}}},\
  }\href@noop {} {\bibfield  {journal} {\bibinfo  {journal} {ArXiv e-prints}\ }
  (\bibinfo {year} {2016})},\ \Eprint {http://arxiv.org/abs/1602.05277}
  {arXiv:1602.05277 [cond-mat.str-el]} \BibitemShut {NoStop}%
\bibitem [{\citenamefont {Anderson}(1973)}]{AndersonRVB}%
  \BibitemOpen
  \bibfield  {author} {\bibinfo {author} {\bibfnamefont {P.~W.}\ \bibnamefont
  {Anderson}},\ }\href
  {http://www.sciencedirect.com/science/article/pii/0025540873901670}
  {\bibfield  {journal} {\bibinfo  {journal} {Mat. Res. Bull.}\ }\textbf
  {\bibinfo {volume} {8}},\ \bibinfo {pages} {155} (\bibinfo {year}
  {1973})}\BibitemShut {NoStop}%
\bibitem [{\citenamefont {Wen}(1990)}]{WenTopological}%
  \BibitemOpen
  \bibfield  {author} {\bibinfo {author} {\bibfnamefont {X.~G.}\ \bibnamefont
  {Wen}},\ }\href {\doibase 10.1142/S0217979290000139} {\bibfield  {journal}
  {\bibinfo  {journal} {Int. J. Mod. Phys. B}\ }\textbf {\bibinfo {volume}
  {04}},\ \bibinfo {pages} {239} (\bibinfo {year} {1990})}\BibitemShut
  {NoStop}%
\bibitem [{\citenamefont {Knolle}\ \emph {et~al.}(2014)\citenamefont {Knolle},
  \citenamefont {Kovrizhin}, \citenamefont {Chalker},\ and\ \citenamefont
  {Moessner}}]{PRL}%
  \BibitemOpen
  \bibfield  {author} {\bibinfo {author} {\bibfnamefont {J.}~\bibnamefont
  {Knolle}}, \bibinfo {author} {\bibfnamefont {D.~L.}\ \bibnamefont
  {Kovrizhin}}, \bibinfo {author} {\bibfnamefont {J.~T.}\ \bibnamefont
  {Chalker}}, \ and\ \bibinfo {author} {\bibfnamefont {R.}~\bibnamefont
  {Moessner}},\ }\href {http://link.aps.org/doi/10.1103/PhysRevLett.112.207203}
  {\bibfield  {journal} {\bibinfo  {journal} {Phys. Rev. Lett.}\ }\textbf
  {\bibinfo {volume} {112}},\ \bibinfo {pages} {207203} (\bibinfo {year}
  {2014})}\BibitemShut {NoStop}%
\bibitem [{\citenamefont {Knolle}\ \emph {et~al.}(2015)\citenamefont {Knolle},
  \citenamefont {Kovrizhin}, \citenamefont {Chalker},\ and\ \citenamefont
  {Moessner}}]{Knolle2015}%
  \BibitemOpen
  \bibfield  {author} {\bibinfo {author} {\bibfnamefont {J.}~\bibnamefont
  {Knolle}}, \bibinfo {author} {\bibfnamefont {D.~L.}\ \bibnamefont
  {Kovrizhin}}, \bibinfo {author} {\bibfnamefont {J.~T.}\ \bibnamefont
  {Chalker}}, \ and\ \bibinfo {author} {\bibfnamefont {R.}~\bibnamefont
  {Moessner}},\ }\href {\doibase 10.1103/PhysRevB.92.115127} {\bibfield
  {journal} {\bibinfo  {journal} {Phys. Rev. B}\ }\textbf {\bibinfo {volume}
  {92}},\ \bibinfo {pages} {115127} (\bibinfo {year} {2015})}\BibitemShut
  {NoStop}%
\bibitem [{\citenamefont {Hermanns}\ and\ \citenamefont
  {Trebst}(2014)}]{HermannsQSL}%
  \BibitemOpen
  \bibfield  {author} {\bibinfo {author} {\bibfnamefont {M.}~\bibnamefont
  {Hermanns}}\ and\ \bibinfo {author} {\bibfnamefont {S.}~\bibnamefont
  {Trebst}},\ }\href {\doibase 10.1103/PhysRevB.89.235102} {\bibfield
  {journal} {\bibinfo  {journal} {Phys. Rev. B}\ }\textbf {\bibinfo {volume}
  {89}},\ \bibinfo {pages} {235102} (\bibinfo {year} {2014})}\BibitemShut
  {NoStop}%
\bibitem [{\citenamefont {Lee}\ \emph {et~al.}(2014)\citenamefont {Lee},
  \citenamefont {Schaffer}, \citenamefont {Bhattacharjee},\ and\ \citenamefont
  {Kim}}]{Lee}%
  \BibitemOpen
  \bibfield  {author} {\bibinfo {author} {\bibfnamefont {E.~K.-H.}\
  \bibnamefont {Lee}}, \bibinfo {author} {\bibfnamefont {R.}~\bibnamefont
  {Schaffer}}, \bibinfo {author} {\bibfnamefont {S.}~\bibnamefont
  {Bhattacharjee}}, \ and\ \bibinfo {author} {\bibfnamefont {Y.~B.}\
  \bibnamefont {Kim}},\ }\href@noop {} {\bibfield  {journal} {\bibinfo
  {journal} {Phys. Rev. B.}\ }\textbf {\bibinfo {volume} {89}},\ \bibinfo
  {pages} {045117} (\bibinfo {year} {2014})}\BibitemShut {NoStop}%
\bibitem [{\citenamefont {Smith}\ \emph {et~al.}(2015)\citenamefont {Smith},
  \citenamefont {Knolle}, \citenamefont {Kovrizhin}, \citenamefont {Chalker},\
  and\ \citenamefont {Moessner}}]{Smith}%
  \BibitemOpen
  \bibfield  {author} {\bibinfo {author} {\bibfnamefont {A.}~\bibnamefont
  {Smith}}, \bibinfo {author} {\bibfnamefont {J.}~\bibnamefont {Knolle}},
  \bibinfo {author} {\bibfnamefont {D.~L.}\ \bibnamefont {Kovrizhin}}, \bibinfo
  {author} {\bibfnamefont {J.~T.}\ \bibnamefont {Chalker}}, \ and\ \bibinfo
  {author} {\bibfnamefont {R.}~\bibnamefont {Moessner}},\ }\href {\doibase
  10.1103/PhysRevB.92.180408} {\bibfield  {journal} {\bibinfo  {journal} {Phys.
  Rev. B}\ }\textbf {\bibinfo {volume} {92}},\ \bibinfo {pages} {180408}
  (\bibinfo {year} {2015})}\BibitemShut {NoStop}%
\bibitem [{\citenamefont {Hermanns}\ \emph {et~al.}(2015)\citenamefont
  {Hermanns}, \citenamefont {Trebst},\ and\ \citenamefont
  {Rosch}}]{Hermanns2015}%
  \BibitemOpen
  \bibfield  {author} {\bibinfo {author} {\bibfnamefont {M.}~\bibnamefont
  {Hermanns}}, \bibinfo {author} {\bibfnamefont {S.}~\bibnamefont {Trebst}}, \
  and\ \bibinfo {author} {\bibfnamefont {A.}~\bibnamefont {Rosch}},\ }\href
  {\doibase 10.1103/PhysRevLett.115.177205} {\bibfield  {journal} {\bibinfo
  {journal} {Phys. Rev. Lett.}\ }\textbf {\bibinfo {volume} {115}},\ \bibinfo
  {pages} {177205} (\bibinfo {year} {2015})}\BibitemShut {NoStop}%
\bibitem [{\citenamefont {Lieb}(1994)}]{Lieb}%
  \BibitemOpen
  \bibfield  {author} {\bibinfo {author} {\bibfnamefont {E.~H.}\ \bibnamefont
  {Lieb}},\ }\href {\doibase 10.1103/PhysRevLett.73.2158} {\bibfield  {journal}
  {\bibinfo  {journal} {Phys. Rev. Lett.}\ }\textbf {\bibinfo {volume} {73}},\
  \bibinfo {pages} {2158} (\bibinfo {year} {1994})}\BibitemShut {NoStop}%
\bibitem [{\citenamefont {Baskaran}\ \emph {et~al.}(2007)\citenamefont
  {Baskaran}, \citenamefont {Mandal},\ and\ \citenamefont
  {Shankar}}]{BaskaranExact}%
  \BibitemOpen
  \bibfield  {author} {\bibinfo {author} {\bibfnamefont {G.}~\bibnamefont
  {Baskaran}}, \bibinfo {author} {\bibfnamefont {S.}~\bibnamefont {Mandal}}, \
  and\ \bibinfo {author} {\bibfnamefont {R.}~\bibnamefont {Shankar}},\ }\href
  {\doibase 10.1103/PhysRevLett.98.247201} {\bibfield  {journal} {\bibinfo
  {journal} {Phys. Rev. Lett.}\ }\textbf {\bibinfo {volume} {98}},\ \bibinfo
  {pages} {247201} (\bibinfo {year} {2007})}\BibitemShut {NoStop}%
\bibitem [{\citenamefont {Lovesey}(1986)}]{Lovesey}%
  \BibitemOpen
  \bibfield  {author} {\bibinfo {author} {\bibfnamefont {S.~W.}\ \bibnamefont
  {Lovesey}},\ }\href@noop {} {\emph {\bibinfo {title} {The Theory of Neutron
  Scattering from Condensed Matter}}},\ Vol.~\bibinfo {volume} {II}\ (\bibinfo
  {publisher} {Clarendon Press},\ \bibinfo {year} {1986})\BibitemShut {NoStop}%
\bibitem [{\citenamefont {Gogolin}\ \emph {et~al.}(1998)\citenamefont
  {Gogolin}, \citenamefont {Nersesyan},\ and\ \citenamefont
  {Tsvelik}}]{Bosonization}%
  \BibitemOpen
  \bibfield  {author} {\bibinfo {author} {\bibfnamefont {A.~O.}\ \bibnamefont
  {Gogolin}}, \bibinfo {author} {\bibfnamefont {A.~A.}\ \bibnamefont
  {Nersesyan}}, \ and\ \bibinfo {author} {\bibfnamefont {A.~M.}\ \bibnamefont
  {Tsvelik}},\ }\href@noop {} {\emph {\bibinfo {title} {Bosonization and
  Strongly Correlated Systems}}}\ (\bibinfo  {publisher} {Cambridge University
  Press},\ \bibinfo {address} {Cambridge, UK},\ \bibinfo {year}
  {1998})\BibitemShut {NoStop}%
\bibitem [{\citenamefont {Nozi\`eres}\ and\ \citenamefont
  {De~Dominicis}(1969)}]{Nozieres}%
  \BibitemOpen
  \bibfield  {author} {\bibinfo {author} {\bibfnamefont {P.}~\bibnamefont
  {Nozi\`eres}}\ and\ \bibinfo {author} {\bibfnamefont {C.~T.}\ \bibnamefont
  {De~Dominicis}},\ }\href {\doibase 10.1103/PhysRev.178.1097} {\bibfield
  {journal} {\bibinfo  {journal} {Phys. Rev.}\ }\textbf {\bibinfo {volume}
  {178}},\ \bibinfo {pages} {1097} (\bibinfo {year} {1969})}\BibitemShut
  {NoStop}%
\bibitem [{\citenamefont {Tikhonov}\ and\ \citenamefont
  {Feigel'man}(2010)}]{Tikhonov2010}%
  \BibitemOpen
  \bibfield  {author} {\bibinfo {author} {\bibfnamefont {K.~S.}\ \bibnamefont
  {Tikhonov}}\ and\ \bibinfo {author} {\bibfnamefont {M.~V.}\ \bibnamefont
  {Feigel'man}},\ }\href {\doibase 10.1103/PhysRevLett.105.067207} {\bibfield
  {journal} {\bibinfo  {journal} {Phys. Rev. Lett.}\ }\textbf {\bibinfo
  {volume} {105}},\ \bibinfo {pages} {067207} (\bibinfo {year}
  {2010})}\BibitemShut {NoStop}%
\bibitem [{\citenamefont {Knolle}(2016)}]{JohannesThesis}%
  \BibitemOpen
  \bibfield  {author} {\bibinfo {author} {\bibfnamefont {J.}~\bibnamefont
  {Knolle}},\ }\href {\doibase 10.1007/978-3-319-23953-8} {\emph {\bibinfo
  {title} {Dynamics of a Quantum Spin Liquid}}}\ (\bibinfo  {publisher}
  {Springer International Publishing},\ \bibinfo {year} {2016})\BibitemShut
  {NoStop}%
\bibitem [{\citenamefont {Blaizot}\ and\ \citenamefont
  {Ripka}(1986)}]{BlaizotRipka}%
  \BibitemOpen
  \bibfield  {author} {\bibinfo {author} {\bibfnamefont {J.-P.}\ \bibnamefont
  {Blaizot}}\ and\ \bibinfo {author} {\bibfnamefont {G.}~\bibnamefont
  {Ripka}},\ }\href@noop {} {\emph {\bibinfo {title} {Quantum theory of finite
  systems}}}\ (\bibinfo  {publisher} {MIT press, Cambridge Mass.},\ \bibinfo
  {year} {1986})\BibitemShut {NoStop}%
\bibitem [{\citenamefont {Abrikosov}\ \emph {et~al.}(1963)\citenamefont
  {Abrikosov}, \citenamefont {Gorkov},\ and\ \citenamefont
  {Dzyaloshinski}}]{Abrikosov}%
  \BibitemOpen
  \bibfield  {author} {\bibinfo {author} {\bibfnamefont {A.~A.}\ \bibnamefont
  {Abrikosov}}, \bibinfo {author} {\bibfnamefont {L.~P.}\ \bibnamefont
  {Gorkov}}, \ and\ \bibinfo {author} {\bibfnamefont {I.~E.}\ \bibnamefont
  {Dzyaloshinski}},\ }\href@noop {} {\emph {\bibinfo {title} {Methods of
  quantum field theory in statistical physics}}}\ (\bibinfo  {publisher}
  {Prentice-Hall, Inc.},\ \bibinfo {address} {Englewood Cliffs, N.J.},\
  \bibinfo {year} {1963})\BibitemShut {NoStop}%
\bibitem [{\citenamefont {Muskhelishvili}(2013)}]{muskhelishvili2013singular}%
  \BibitemOpen
  \bibfield  {author} {\bibinfo {author} {\bibfnamefont {N.}~\bibnamefont
  {Muskhelishvili}},\ }\href {https://books.google.co.uk/books?id=cOtrLCS32AMC}
  {\emph {\bibinfo {title} {Singular Integral Equations: Boundary Problems of
  Function Theory and Their Application to Mathematical Physics}}},\ Dover
  Books on Mathematics\ (\bibinfo  {publisher} {Dover Publications},\ \bibinfo
  {year} {2013})\BibitemShut {NoStop}%
\end{thebibliography}%

\end{document}